\documentclass{article}

\usepackage{arxiv}

\usepackage[utf8]{inputenc} 
\usepackage[T1]{fontenc}    
\usepackage{hyperref}       
\usepackage{url}            
\usepackage{booktabs}       
\usepackage{amsfonts}       
\usepackage{nicefrac}       
\usepackage{microtype}      
\usepackage{lipsum}
\usepackage{graphicx}
\usepackage{braket}
\usepackage{bbm}
\usepackage{caption}
\usepackage{subcaption}
\usepackage{amsmath}
\usepackage{amssymb}
\usepackage[table]{xcolor}
\graphicspath{ {./images/} }

\title{Graph Neural Networks for Source Detection: A Review and Benchmark Study}

\author{
 Martin Sterchi \\
  University of Applied Sciences and Arts Northwestern Switzerland FHNW\\
  and University of Zürich \\
  Switzerland \\
  \texttt{martin.sterchi@fhnw.ch} \\
   \And
 Nathan Brack \\
  University of Applied Sciences and Arts Northwestern Switzerland FHNW\\
  and University of Zürich \\
  Switzerland \\
  \And
 Lorenz Hilfiker \\
  University of Applied Sciences and Arts Northwestern Switzerland FHNW\\
  Switzerland \\
}

\begin{document}
\maketitle
\begin{abstract}
The source detection problem arises when an epidemic process unfolds over a contact network, and the objective is to identify its point of origin, i.e., the source node. Research on this problem began with the seminal work of Shah and Zaman in 2010, who formally defined it and introduced the notion of {\itshape rumor centrality}. With the emergence of Graph Neural Networks (GNNs), several studies have proposed GNN-based approaches to source detection. However, there is room to strengthen methodological clarity and reproducibility across these works. As a result, it remains unclear whether GNNs truly outperform more traditional source detection methods across comparable settings. In this paper, we first systematically review existing GNN-based methods for source detection, clearly outlining the specific settings each addresses and the architectures they employ. We then reproduce and benchmark four representative GNN architectures against a diverse set of traditional and MLP-based baselines under controlled, comparable conditions. We also investigate key questions surrounding this problem, including how detectability evolves over time, how performance scales with training set size, and how sensitive methods are to uncertainty in observation timing and epidemic parameters. Our experiments show that GNNs substantially outperform all other methods we test across a variety of network topologies. Although we initially set out to challenge the notion of GNNs as a solution to source detection, our results instead demonstrate their remarkable effectiveness for this task. To ensure full reproducibility, we release all code and data on GitHub. Finally, we argue that epidemic source detection constitutes a natural and attractive benchmark task for evaluating GNN architectures.
\end{abstract}


\section{Introduction}\label{intro}

Epidemic processes are ubiquitous and extend far beyond biological pathogens such as SARS-CoV-2 in the human domain or the foot-and-mouth disease in the animal domain. With the rise of the internet and social media, the spread of rumors and misinformation has become an acute problem. On a more positive note, the technological revolution based on the introduction of the internet has also contributed to making data about the contact network underlying such epidemic processes more available. For example, RFID sensors and the Bluetooth technology allow collecting human face-to-face contacts~\cite{Vanhems:2013,Stopczynski:2014}, industrial animal movements can be collected from farmers or transport companies~\cite{Noremark:2011,Sterchi:2019}, often using digital reporting tools, and online contact network data can be scraped from the internet~\cite{Rocha:2010,Leskovec:2012}, or are sometimes provided to the public by social media or other companies~\cite{Makice:2009,Klein:2020,Brady:2021}. The availability of such data has prompted many interesting research questions, one of them being the problem of source detection, famously introduced by Shah and Zaman~\cite{Shah:2010}.

The aim of source detection is to identify one or multiple source nodes of an epidemic process, given the underlying network of contacts and some observed state about the process (e.g., the current epidemic state of some or all nodes in the network). This is generally a hard problem for which closed-form solutions only exist for simplified problem settings~\cite{Shah:2010}. After a plethora of solutions have been proposed in the years following the seminal introduction of the source detection problem by Shah and Zaman in 2010~\cite{Shah:2010}, Dong et al.~\cite{Dong:2019} and, a bit later, Shah et al.~\cite{Shah:2020} were the first to propose the use of Graph Neural Networks~(GNNs) for the source detection problem. The fundamental idea is that a GNN is trained on a static network of contacts using a large number of simulated epidemic outcomes as training data (learning step). The trained GNN can then be used to predict the source for a new epidemic outcome on this network (inference step). As Dong et al.~\cite{Dong:2019} point out, GNNs seem to be well suited to tackle the source detection problem as the node embeddings resulting from a GNN aggregate neighborhood information which contains the key information to determine whether a node is a likely source or not: a source node would have many infected nodes in the close neighborhood, but a decreasing density of infected nodes in further out neighborhoods. If the embeddings, learned by the GNN, can encode this information in a meaningful way, then GNNs should be a sensible approach to source detection.

While research on GNNs in the context of epidemics is currently a very active research area~\cite{Liu:2024}, the problem of using GNNs for source detection is rather understudied. We counted a total of (at least) seven papers proposing GNN-based source detection methods~\cite{Dong:2019,Shah:2020,Sha:2021,Shu:2021,Guo:2021,Haddad:2023,Ru:2023} at the time of writing this paper. The seven papers address varied problem settings: while some consider the {\itshape single-source} problem~\cite{Shah:2020,Shu:2021,Guo:2021,Ru:2023}, others focus on {\itshape multi-source} problems~\cite{Dong:2019,Haddad:2023}, and one paper covers both problems~\cite{Sha:2021}. Or, as another example, Ru et al.~\cite{Ru:2023} assume a temporal contact network while all other works assume static contact networks. This makes it hard or even impossible to compare the different approaches. Another issue with the proposed works is that they are difficult to reproduce for two reasons: first, some of the papers lack the exact details of their GNN architectures (e.g., number of layers), and second, only few authors~\cite{Sha:2021,Haddad:2023} provide the code for their methods and experiments. Finally, the precision and rigor of methodological presentations vary across the literature, and some claims warrant closer scrutiny. For example, several papers~\cite{Dong:2019,Shah:2020,Shu:2021,Guo:2021,Ru:2023} claim that, unlike more conventional methods, GNN-based source detection is independent of the spreading model and its parameters or even model-agnostic and fail to acknowledge that all GNN-based approaches are trained on simulated spreading scenarios for which, at least implicitly, strong assumptions about the spreading process are made.

This paper brings order to the understudied area of using GNNs for source detection. We approach this as a systematic survey and benchmark study, critically reviewing and formally describing all existing GNN-based source detection methods, and evaluating four GNN-based methods form the literature against conventional source detection approaches under controlled, comparable conditions. After an exposition of the general problem setting to be studied (Section~\ref{sec:problem}), we review prior research (traditional and GNN-based) in this subfield (Section~\ref{sec:rel-work}) and provide a formal description of the different GNN architectures that have been proposed for source detection (Section~\ref{sec:methods}). We then outline how we implement the four GNN architectures from related work and more generally outline the experimental setting (Section~\ref{sec:setting}). We aim to address several important questions that, in our view, remain unresolved (Section~\ref{sec:experiments}). The central question we examine is whether GNN-based source detectors truly outperform conventional source detection methods. In addition, we investigate how source detectability changes over time, how detection performance scales with increasing training set sizes, and how critical it is to know the exact time elapsed between the start of the epidemic and the observation of the snapshot of node states as well as the exact epidemic parameters. Section~\ref{sec:complexity} elaborates on the computational complexity of the GNN-based approach and some of the benchmarks. Before we conclude in Section~\ref{conclusion}, we present a real-world use case in which we apply a learned GNN to an observed snapshot of the 2009 H1N1 pandemic (Section~\ref{sec:real-world}).

Our main contributions can be summarized as follows:
\begin{itemize}
\item A systematic survey of GNN-based source detection methods, classifying them according to the problem settings addressed and the architectural design choices made.
\item A comprehensive benchmark showing whether GNN-based source detection truly outperforms conventional methods.
\item Insights into source detectability over time, training set size scaling, and sensitivity to observation timing and misspecification of spreading model parameters.
\item A real-world application to the 2009 H1N1 pandemic.
\end{itemize}

\section{Problem Formulation}\label{sec:problem}

In this section, we first present the preliminaries on network representations and epidemic processes. We then formalize the specific problem setting considered in this work. Finally, we discuss the distinction between the single-source and multi-source variants of the problem.

\subsection{Preliminaries}

In this paper, we focus on the source detection problem on {\itshape static} networks, as do all the related papers except for Ru et al.~\cite{Ru:2023}. We denote such networks as $G(V,E)$ with $V$ being the set of nodes and $E$ the set of edges. The network contains a total of $N=|V|$ nodes. We write an edge as $(v,u)\in E$, with $v,u \in V$. Note that edges can be directed or undirected, but, in line with related work, we focus on the case of undirected edges. The network’s adjacency matrix is denoted by $\mathbf{A}\in \mathbb{R}^{N\times N}$, where $A_{ij}=1$ if $(i,j)\in E$ and $A_{ij}=0$ otherwise. For each node $v$, we denote by $\mathcal{N}(v)$ the set of its immediate neighbors.

The observed epidemic state of each node is the result of an underlying (stochastic) propagation process that evolves in continuous time. At any time $t$, a node $v \in V$ is in some epidemic state that we denote as $X_v(t)$. The most prominent such propagation model, and therefore the one we will focus on here, is the {\itshape Susceptible-Infectious-Recovered} (SIR) model in which any node $v\in V$ is in one of three possible states, i.e., $X_v(t)\in \{S,I,R\}$. In the most straightforward version of the SIR model, there are two possible state transitions, both following a Poisson process, and all events are independent of each other~\cite{Kiss:2017}. The state transitions are (i) a susceptible node becoming infected at a rate $k\beta$ for $k$ infected neighbors, and (ii) an infected node recovering at a constant rate $\mu$.

\subsection{Our Problem Setting}

We assume that the described SIR process originates at some time $t_0$ from a single initially infected node in the network ({\itshape single-source} problem). At time $t_1=t_0+T$, we observe a complete snapshot of the epidemic, i.e., the states of all nodes in the network. Formally, we represent this snapshot (i.e., the observed {\itshape evidence}) as $E_{t_1}=\{(v,x_v(t_1)): v \in V\}$. The majority of the related work assumes a similar observation process. An important decision in setting up the precise problem is whether the start of the epidemic, $t_0$, and, by implication, its duration at the time of observation, $T$, are assumed to be known. Throughout this work, we primarily assume that $T$ is known. Nonetheless, we also include results that examine how the detection performance is affected when $T$ is not precisely known.

The task is to infer the source node $v^*$, given the network $G(V,E)$ and the observed snapshot of the epidemic $E_{t_1}$ at time $t_1$. $G(V,E)$ and $E_{t_1}$ form an attributed graph in which each node is assigned its epidemic state at $t_1$. Our objective is to develop a model that takes this attributed graph as input and outputs a prediction of the source node. Consequently, the task can be framed as a {\itshape graph prediction} problem, specifically as a multi-class classification problem with $N$ possible classes, one for each node in the network.

A natural way to approach this problem using GNNs is to treat it as a supervised learning task, where a large set of simulated epidemic scenarios is used as training and validation data. Each training instance consists of the adjacency matrix $\mathbf{A}$, the observed snapshot which is typically encoded as a node feature matrix $\mathbf{X}\in \mathbb{R}^{N\times 3}$ containing the one-hot encoded node states at time $t_1$, and the corresponding label (true source node). Crucially, the reliance on simulated data implies that some knowledge of the underlying spreading process is available.

\begin{figure}[ht]
  \centering
  \includegraphics[width=\linewidth]{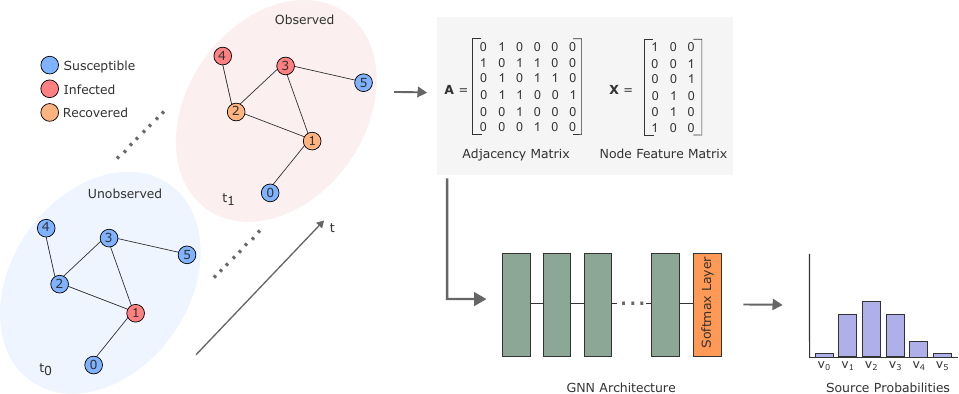}
  \caption{Schematic representation of the source detection problem. An epidemic originates from node 1 and evolves over time. At observation time $t_1$ , a full snapshot of node states is available, but the identity of the source is unknown. The GNN receives as input the graph adjacency matrix and a node feature matrix containing the one-hot encoded epidemic states, and outputs a probability distribution over nodes indicating the most likely source.}
  \label{fig:problem}
\end{figure}

Figure~\ref{fig:problem} schematically illustrates the overall source inference process once the model has been trained.

\subsection{Single-Source vs. Multi-Source Problem}

As stated above, this paper focuses on the single-source problem. Some related work, however, considers the multi-source setting, which has important implications for the design of the GNN’s output layer and the corresponding loss function.

For the single-source case, the output layer typically uses a softmax activation together with the cross-entropy loss, defined as
\begin{equation}
\label{eq:mc-cross-entropy}
    E_m(\mathbf{\Theta}) = -\sum_{v\in V}  y_{v,m}\cdot\text{ln}\left(\hat{y}_{v,m}(\mathbf{\Theta})\right)
\end{equation}
where $y_{v,m}=1$ if node $v$ is the true source in the $m$-th training instance and $y_{v,m}=0$ otherwise. The term $\hat{y}_{v,m}$ denotes the softmax output of the model for node $v$ in instance $m$, and $\mathbf{\Theta}$ represents all learnable model parameters.

In contrast, the multi-source problem can be formulated as a multi-label binary classification task, where the GNN predicts, for each node, whether it belongs to the set of sources. This formulation implicitly assumes independence among source nodes. Instead of a softmax activation, each output neuron applies a sigmoid function, and training is performed using the multi-label binary cross-entropy loss:
\begin{equation}
\label{eq:binary-ml-cross-entropy}
    E_m(\mathbf{\Theta}) = -\sum_{v\in V}  y_{v,m}\cdot\text{ln}\left(\hat{y}_{v,m}(\mathbf{\Theta})\right) + (1-y_{v,m})\cdot\text{ln}\left(1-\hat{y}_{v,m}(\mathbf{\Theta})\right),
\end{equation}
where $y_{v,m}=1$ for each node $v$ that is a true source in instance $m$ and $y_{v,m}=0$ otherwise, and $\hat{y}_{v,m}$ denotes the sigmoid output of the model for node $v$.

In the following section, we discuss the related work in greater detail, highlighting how the corresponding problem formulations differ from the one adopted in this paper.

\section{Source Detection Methods}\label{sec:rel-work}

In this section, we first introduce traditional approaches to (single-) source detection that mostly build on methods from physics and statistics. We then discuss recent studies that employ GNNs for source detection.

\subsection{Traditional Source Detection}

The field of source detection on networks was coined by Shah and Zaman in 2010~\cite{Shah:2010} who proposed a maximum likelihood source estimator (called {\itshape rumor center}). Their analysis focused on a simplified setting involving undirected regular tree graphs and a {\itshape Susceptible-Infectious} (SI) spreading model, and they proposed an approximate solution based on breadth-first search (BFS) trees for general networks. Zhu and Ying~\cite{Zhu:2016} later proposed an estimator applicable to SIR processes, called {\itshape Jordan center}, which identifies the source as the node minimizing the maximum distance to any node in the infected subgraph. Pinto et al.~\cite{Pinto:2012} examined a different problem setting in which a subset of nodes acts as observers that record infection arrival times. On tree graphs, the source can then be estimated under a distributional assumption on propagation delays, while for general networks a BFS-based heuristic is applied. Paluch et al.~\cite{Paluch:2018} improved the computational efficiency of Pinto et al.'s method, and Xu et al.~\cite{Xu:2019} proposed a simpler correlation-based estimator that selects the node maximizing the correlation between infection arrival times and shortest-path distances from the candidate source to the observers.

Since closed-form solutions only exist for simplified problem settings, a substantial portion of the literature on source detection uses Monte-Carlo simulations of epidemic processes. Beginning with the work by Agaskar and Lu~\cite{Agaskar:2013} who sample arrival times which they use to determine the most likely source, simulation-based approaches have also been used by Antulov-Fantulin et al.~\cite{Antulov-Fantulin:2015}, Dutta et al.~\cite{Dutta:2018}, and Sterchi et al.~\cite{Sterchi:2023}. Both Antulov-Fantulin et al. and Dutta et al. compare simulated outcomes with the observed epidemic snapshot to infer source likelihoods. Sterchi et al., in contrast, use simulations to estimate node-state probabilities conditional on each possible source, from which a (factorized) likelihood is derived under a node independence assumption. A related line of research~\cite{Lokhov:2014,Rocha:2016} replaces stochastic simulations with deterministic update equations to compute node-state probabilities, again relying on independence assumptions. Another probabilistically motivated approach by Braunstein and Ingrosso~\cite{Braunstein:2016} employs loopy belief propagation to approximate the posterior distribution of infection times for all nodes, identifying the source as the node with the highest marginal probability of initial infection.

A topological perspective on source detection is offered by Brockmann and Helbing~\cite{Brockmann:2013} and Rozenshtein et al.~\cite{Rozenshtein:2016}. Brockmann and Helbing address a metapopulation problem setting and introduce the concept of {\itshape effective distance}, derived from fluxes between subpopulations. Their source estimate corresponds to the node producing the most concentric infection pattern in terms of shortest distances between infected nodes (using the effective distances). Rozenshtein et al.~\cite{Rozenshtein:2016} proposed a truly model-agnostic formulation, identifying the source as the node that minimizes the cost of a Steiner tree spanning all infected nodes.

\subsection{GNN-based Source Detection}
In this subsection, we review the related work that employs GNN architectures for the source detection problem, classifying methods into single- and multi-source approaches. We begin with the two multi-source approaches, one of which represents the earliest known GNN-based source detector, and then proceed to single-source contributions. Our discussion is limited to studies that consider the spread of biological pathogens as a representative use case. A number of related works instead focus uniquely on rumor propagation, which entails a somewhat different problem formulation; these works are briefly summarized at the end of this subsection. Moreover, we note that Li et al.~\cite{Li:2021} also propose a GCN-based framework for source detection. However, due to several technical and methodological issues, we do not include this work in our detailed discussion. Finally, we conclude with a short summary highlighting the open questions that emerge from the reviewed literature.

\subsubsection{Multi-Source Approaches}
To the best of our knowledge, Dong et al.~\cite{Dong:2019} were the first to apply GNNs to the source detection problem on networks. Specifically, they address a multi-source setting on static networks, assuming that a full snapshot of all node states is observed at some time after the outbreak and that the number of sources is known a priori. Their approach, termed {\itshape Graph Convolutional Networks based Source Identification} (GCNSI), is motivated by rumor spreading in social networks and guided by two principles: {\itshape source prominence}, the notion that true sources tend to be surrounded by many infected nodes, and {\itshape rumor centrality}, the idea that distant nodes are less likely to be infected. These principles align naturally with the message-passing paradigm of GNNs, in which nodes learn representations based on information exchanged with their neighbors.

Methodologically, Dong et al.~\cite{Dong:2019} first apply a label-propagation method~\cite{Wang:2017} to augment initial node representations beyond simple one-hot encoded epidemic states. However, the benefit of this feature augmentation remains unclear, as GNNs inherently learn higher-order node representations. The study does not include a quantitative analysis of its effect. Moreover, the approach is benchmarked only against two earlier multi-source methods~\cite{Prakash:2012,Wang:2017} and is evaluated in  simplified epidemic scenarios, where each node is either infected or not. While such a binary assumption offers simplicity and may be appropriate for rumor propagation, disease-spreading contexts often involve additional states such as {\itshape exposed} or {\itshape recovered}. Some of the reported results deserve closer attention, for instance, the model’s performance appears to improve with larger numbers of true sources, which, presumably, is an artifact of the evaluation metric. Finally, although the authors describe GCNSI as model-agnostic (meaning it does not require explicit prior knowledge of the spreading dynamics), the approach relies on specifying a spreading model to generate the training data. This notion of spreading model agnosticism is a recurring claim throughout the subsequent literature that we argue is somewhat misleading. Table~\ref{tab:problems} summarizes the problem settings addressed by Dong et al.~\cite{Dong:2019} and the other studies reviewed in this subsection along four dimensions.

\begin{table*}
  \caption{The problem settings addressed in related work. Four dimensions are considered: the network representation used, the epidemic spreading model applied, the number of sources assumed (single or multiple), and the type of node state snapshot observed. The contributions are listed in chronological order. All abbreviations are introduced in the main text.}
  \label{tab:problems}
  \centering
  \begin{tabular}{ll|lllll}
    \toprule
    Authors & Model & Network & Epidemics & Single/Multi & Snapshot\\
    \midrule
    Dong et al.~\cite{Dong:2019} & GCNSI & Undirected, static & SI, SIR, SIS, IC, LT & Multi & Full\\
    Shah et al.~\cite{Shah:2020} & - & Undirected, static & SIR, SEIR & Single & Full\\
    Shu et al.~\cite{Shu:2021} & MCGNN & Undirected, static & Heterogeneous SI & Single & Full\\
    Guo et al.~\cite{Guo:2021} & IGCN & Undirected, static & SI, SIR, IC & Single & Full\\
    Sha et al.~\cite{Sha:2021} & SD-STGCN & Undirected, static & SIR, SEIR, delay SIR & Both & Multiple full\\
    Had. and Fig.~\cite{Haddad:2023} & - & Undirected, static & SI & Multi & Full\\
    Ru et al.~\cite{Ru:2023} & BN & Directed/undir., temporal & SIR & Single & Full\\
    \bottomrule
  \end{tabular}
\end{table*}

Haddad and Figueiredo~\cite{Haddad:2023} investigate the same multi-source detection problem as Dong et al.~\cite{Dong:2019}, but differ in two key aspects. First, they employ the GraphSAGE architecture instead of the GCN architecture used in~\cite{Dong:2019}. Second, they design additional node features to enrich each node $v$’s initial representation: (i) the fraction of infected nodes within its $k$-hop neighborhood, and (ii) the fraction of infected nodes among the neighbors of $v$ at distance $k$, for $k = 1, \dots, K$, where $K$ is a hyperparameter. In contrast to related work, Haddad and Figueiredo train and test their models on potentially different networks which is an unusual choice that departs from the standard assumption in source detection that the underlying network is fully known and fixed during inference. As in~\cite{Dong:2019}, the number of true sources is assumed to be known a priori, a simplification that makes the task more tractable.

\subsubsection{Single-Source Approaches}
In contrast to the previous two contributions, Shah et al.~\cite{Shah:2020} address the single-source detection problem. The general setting remains similar, assuming a full snapshot of node states is available at inference time $t_1$. Unlike Dong et al.~\cite{Dong:2019} and Haddad and Figueiredo~\cite{Haddad:2023}, Shah et al. do not augment the initial node features and instead use only one-hot encoded node states as input. The training data are generated from large numbers of simulated epidemic realizations using SIR (or SEIR) dynamics. Shah et al. evaluate several GNN architectures, including variants of Graph Convolutional Networks (GCNs) and Graph Attention Networks (GATs), but find no substantial performance differences between them. While their approach clearly outperforms the benchmark method of Lokhov et al.~\cite{Lokhov:2014}, the exceptionally long inference times reported for the benchmark method raise some doubts regarding the robustness of the results.

Shu et al.~\cite{Shu:2021} propose a {\itshape Multi-Channel GNN}~(MCGNN) framework. The overall setting is similar to that in~\cite{Shah:2020}, except that they consider a heterogeneous SI model in which transmission probabilities may vary across edges. As the name suggests, their model comprises two channels: the first applies GCN layers to process node features, while the second constructs a line graph and computes various node properties within that graph. This second channel is intended to enhance performance by incorporating edge-level information, which the authors identify as the main improvement over prior methods. The outputs of both channels are concatenated and fed into the model’s output layer. As in related work~\cite{Dong:2019,Haddad:2023}, feature augmentation plays a central role. In both channels, several network-based measures (primarily centrality metrics) are computed for each node and used as input features. Additionally, the first (node) channel is enriched with established source detection indicators such as rumor centrality~\cite{Shah:2010} and Jordan centrality~\cite{Zhu:2016}. 

One aspect of their approach worth noting is the formulation of the learning objective. Specifically, the single-source problem is treated as a multi-label classification task, with an output layer containing $N$ sigmoid-activated units trained using independent binary cross-entropy losses. As the single-source problem requires a probability distribution over all nodes, a softmax output with categorical cross-entropy loss may be a more appropriate formulation.

Guo et al.~\cite{Guo:2021} address the same problem setting as Shah et al.~\cite{Shah:2020} but propose a distinct architecture, the {\itshape Infected Graph Convolutional Network}~(IGCN). After an initial GCN layer, the model applies three lightweight neighborhood aggregation layers containing only four trainable weights. These weights depend on the epidemic states of the focal node and its neighbors and are used during message aggregation. Since the authors consider only two possible states (infected vs. not infected), there are four state combinations per node pair, requiring exactly four parameters. Notably, the authors account for uncertainty in the underlying spreading dynamics by simulating the training data from three different spreading models with varying parameter settings.

Sha et al.~\cite{Sha:2021} attempt to address some of the critiques raised during the review process of Shah et al.~\cite{Shah:2020} on OpenReview~\cite{Openreview:2020}, and accordingly modify their problem setting. However, this modification appears to stem from a misunderstanding of one reviewer’s comment. The reviewer questioned the feasibility of a learning-based approach (apparently unaware that training data can be generated through simulations) by noting that ``we only have access to one snapshot for a given graph and learning is impossible''~\cite{Openreview:2020}. Sha et al. interpret this remark as motivation to modify the problem setting so that, instead of a single full snapshot, multiple full snapshots of the epidemic at different time steps are available for inference.

In addition to this revised observation model, the authors adopt more realistic epidemic dynamics, including SEIR and delayed SIR models with non-exponential distributions of the infectious periods, and they evaluate their method on a broader range of both synthetic and empirical networks. Because multiple snapshots are observed, the node feature input becomes a three-dimensional tensor, with the third dimension representing time. To process such data, Sha et al. propose a {\itshape Spatio-Temporal GCN architecture for Source Detection}~(SD-STGCN), originally developed for traffic forecasting~\cite{Yu:2018}. Each block of the SD-STGCN consists of a Convolutional Neural Network (CNN) layer applied along the temporal dimension, a GCN layer aggregating neighborhood information, and another CNN layer for temporal refinement. Overall, the paper presents a thorough evaluation, including comparisons with baseline GCNs, experiments on different observation schemes (randomly sampled vs. consecutive snapshots), and training across varying epidemic parameters to account for uncertainty in the underlying dynamics.

While all previously discussed studies assume static networks, Ru et al.~\cite{Ru:2023} address the single-source detection problem on {\itshape temporal} networks, which they conceptually treat as sequences of static snapshots. As in prior work, a full snapshot of a SIR epidemic is assumed to be available at inference time. Ru et al. explicitly incorporate uncertainty about the epidemic’s start time $t_0$ and duration $T$: their method requires an additional input specifying a time interval within which the outbreak is assumed to have begun, which is used during the simulation of training data.

The core idea of their approach, termed the {\itshape Backtracking Network}~(BN), is to represent the network as static by including all edges that appeared across time, but to incorporate temporality through edge features. Specifically, the model takes as input both the one-hot encoded epidemic states of nodes and an edge feature matrix encoding the temporal activity patterns of the edges. It employs two interdependent convolution-style update equations: one for computing edge embeddings and another for computing node embeddings. They then compare their BN approach to several baselines, including Antulov-Fantulin et al.’s soft margin estimator~(SME)~\cite{Antulov-Fantulin:2015}. We note, however, that the description of SME differs from the original formulation, and with only 200 simulations per source node, the comparison may be unreliable and potentially favor BN.

\subsubsection{Other Works}
Cheng et al.~\cite{Cheng:2024} address a multi-source problem setting that is heavily focused on the information spreading use case. As a consequence, they assume that for a fraction of the nodes the node states, but also the times when nodes received the information, and potentially the propagating nodes are observed. Their method is then a mix of feature augmentation based on a spectral decomposition of the infected subgraph Laplacian and a simple attention mechanism similar to GAT. Another work focused on information spreading is that of Ling et al.~\cite{Ling:2022}, who apply a Variational Autoencoder to the problem of multi-source detection. Their problem setting differs from ours in that they observe, for each node, a probability indicating whether the information has reached it. Given these probabilities and the underlying graph, they aim to infer both the number and the identities of the source nodes.

\subsubsection{Summary}
Our review of the GNN-based source detection literature reveals several open questions and directions for improvement. First, we argue that these approaches should not be described as model-agnostic, since in all cases the training data are simulated under explicit assumptions about the underlying spreading dynamics, and the learned models inevitably encode those assumptions. Second, although prior work does compare GNN-based methods against baselines, important benchmark models, such as the SME by Antulov-Fantulin et al.~\cite{Antulov-Fantulin:2015}, are often omitted or implemented in possibly incorrect ways. Consequently, the true potential of GNN-based approaches relative to classical methods remains uncertain. Third, while several studies emphasize feature augmentation, few provide a quantitative analysis of its actual effect on model performance. Fourth, the influence of training set size is rarely discussed. Given that training data are generated through simulations, it remains unclear whether systematic scaling laws exist, i.e., how model performance improves with increasing training data. Finally, only a small number of works relax the assumption that the duration $T$ between the outbreak and the observed snapshot and/or the exact spreading parameters are known, even though these parameters are typically uncertain in realistic scenarios.

\section{GNN Architectures for Source Detection}\label{sec:methods}

In this section, we offer a more formal exposition of the GNN architectures proposed for the source detection problem. For each method, we outline the node-level update equations, detail any specific processing steps applied to the input features, and describe the structure of the output layer together with the corresponding loss function.

\subsection{Graph Convolutional Network (GCN)}\label{sec:gcn}

The most commonly employed architecture for source detection is the Graph Convolutional Network (GCN)~\cite{Dong:2019,Shah:2020,Shu:2021}. Dong et al.~\cite{Dong:2019} adopt the standard GCN formulation proposed by Kipf and Welling~\cite{Kipf:2017}. Each node $v$ is initialized with its feature vector $\mathbf{h}_v^{(0)}=\mathbf{x}_v\in\mathbb{R}^{m}$, where $m$ denotes the number of node features. The layer-wise update rule is given by
\begin{equation}
\label{eq:updateDong}
    \mathbf{h}_v^{(l)} = \text{ReLU}\left(
    \mathbf{W}_l 
    \sum_{u \in \{\mathcal{N}(v) \cup \{v\}\}}
    \frac{\mathbf{h}_u^{(l-1)}}{\sqrt{\tilde{d}_v\,\tilde{d}_u}}
    \right), 
    \qquad l=1,\dots,L,
\end{equation}
where $\tilde{d}_v$ denotes the degree of node $v$ augmented by one. The symmetric normalization applied here has been shown to perform well in practice~\cite{Kipf:2017}. The neighborhood aggregation includes the node itself, and $\mathbf{W}_l \in \mathbb{R}^{s_l \times s_{l-1}}$ projects the embeddings to dimension $s_l$, followed by a ReLU activation. Typically, after the first layer projects the input features to the hidden dimension, the number of hidden units is kept constant, i.e., $s_{l-1}=s_l$ for all $l\geq2$.

For the output layer, consistent with the multi-label nature of the multi-source problem, a dense layer with sigmoid activation is used:
\begin{equation}
\label{eq:outputDong}
    \hat{y}_v = \sigma\!\left(\mathbf{w}^\intercal \mathbf{h}_v^{(L)}\right),
\end{equation}
where $\mathbf{w}\in\mathbb{R}^{s_L}$. The model is trained using binary cross-entropy loss with L2 regularization. Table~\ref{tab:hyperparams} summarizes the hyperparameter choices and training details for this and the other approaches discussed below.

\begin{table*}
  \caption{Overview of design choices and learning configurations used by the different contributions. A dash (-) indicates that no information has been reported. Note that the methods by Sha et al.~\cite{Sha:2021} and Ru et al.~\cite{Ru:2023} are conceptually quite different to the others and hyperparameter settings should thus not be directly compared.}
  \label{tab:hyperparams}
  \centering
  \begin{tabular}{l|lllllll}
    \toprule
    Authors & \# Layers & \# Hidden Units & Learning Rate & Dropout Rate & \# Epochs & Optim.\\
    \midrule
    Dong et al.~\cite{Dong:2019} & 5--10 & 512 or 1024 & 0.001--0.005 & 0.1 or 0.3 & 900--4000 & ADAM\\
    Shah et al.~\cite{Shah:2020} & 10 (GAT: 5) & 128 & 0.003 (GAT: 0.004) & 0.265 & 150 & ADAM\\
    Shu et al.~\cite{Shu:2021} & - & - & - & - & - & -\\
    Sha et al.~\cite{Sha:2021} & 2--4 & 36 & 0.001 & - & 3 & RMSProp\\
    Guo et al.~\cite{Guo:2021} & 3 (+1 GCN) & 512 & 0.001 & 0.1 & 50 & ADAM\\
    Had. and Fig.~\cite{Haddad:2023} & 3 & 128 & 0.001 & 0.1 & 500 (early stop.) & ADAM\\
    Ru et al.~\cite{Ru:2023} & 1--7 & - & 0.001 & - & 200 & ADAM\\
    \bottomrule
  \end{tabular}
\end{table*}

Shah et al.~\cite{Shah:2020} use an architecture similar to that of Dong et al.~\cite{Dong:2019}, with several notable modifications. Before applying the GCN layers, the node feature vectors are linearly projected to match the hidden layer dimension (required for dimensional compatibility with the residual connections described below),
\begin{equation}
    \mathbf{h}_v^{(0)} = \mathbf{U}\mathbf{x}_v \in \mathbb{R}^{s_0}, \qquad \mathbf{U}\in \mathbb{R}^{s_0\times m}.
\end{equation}
The authors introduce three main architectural differences: (i) inclusion of a bias term $\mathbf{b}_l\in\mathbb{R}^{s_l}$, (ii) residual connections between layers instead of self-inclusion in the aggregation, and (iii) the use of batch normalization (BatchNorm). The resulting layer update is given by
\begin{equation}
\label{eq:updateShah}
    \mathbf{h}_v^{(l)} = \mathbf{h}_v^{(l-1)} + 
    \text{LeakyReLU}\!\left(
    \text{BatchNorm}\!\left(
    \mathbf{W}_l 
    \sum_{u \in \mathcal{N}(v)} 
    \frac{\mathbf{h}_u^{(l-1)}}{\sqrt{d_v\,d_u}}
    + \mathbf{b}_l
    \right)\right),
    \qquad l=1,\dots,L.
\end{equation}
Although alternative aggregation schemes were tested, the results were largely invariant to this choice. Their output layer is defined as
\begin{equation}
\label{eq:outputShah}
    \hat{y}_v = \mathbf{p}^\intercal \text{ReLU}\!\left(\mathbf{Q}\mathbf{h}_v^{(L)}\right),
\end{equation}
where $\mathbf{Q}\in\mathbb{R}^{s_L\times s_L}$ and $\mathbf{p}\in\mathbb{R}^{s_L}$ transform the final node embedding into the model prediction. The authors do not appear to apply a softmax activation in the output layer, despite the fact that the task is single-source detection. Moreover, no information regarding the loss function is provided.

Shu et al.~\cite{Shu:2021} also build on a GCN backbone similar to Dong et al.~\cite{Dong:2019}, with the addition of a bias term in the layer update:
\begin{equation}
\label{eq:updateShu}
    \mathbf{h}_v^{(l)} = \sigma\!\left( 
    \mathbf{W}_l 
    \sum_{u \in \{\mathcal{N}(v)\cup\{v\}\}} 
    \frac{\mathbf{h}_u^{(l-1)}}{\sqrt{\tilde{d}_v\,\tilde{d}_u}}
    + \mathbf{b}_l
    \right), \qquad l=1,\dots,L.
\end{equation}
The final GCN embedding $\mathbf{h}_v^{(L)}$ is concatenated with $f$ additional node-level features derived from a line graph; a dense layer followed by a sigmoid activation yields the model output $\hat{y}_v$. Training is performed with binary cross-entropy and L2 regularization. As stated above, we believe that using a sigmoid output and independent binary cross-entropy losses is not suited for the single-source formulation considered in the paper. The single-source problem requires a valid probability distribution over all candidate nodes (i.e., mutually exclusive classes), which is more naturally modeled with a softmax output and a categorical (multi-class) cross-entropy loss rather than independent sigmoids. The authors do not report any additional hyperparameter settings.

\subsection{GraphSAGE}
Haddad and Figueiredo~\cite{Haddad:2023} employ the GraphSAGE architecture~\cite{Hamilton:2017} instead of a GCN. Their model consists of $L=3$ GraphSAGE layers that use element-wise max. pooling to aggregate neighbor embeddings:
\begin{equation}
\label{eq:updateHaddad}
    \mathbf{h}_v^{(l)} = \text{ReLU}\!\left(
        \mathbf{W}_l^1 \mathbf{h}_v^{(l-1)} + \mathbf{W}_l^2 \cdot \text{MAX}\!\left(\{\mathbf{h}_u^{(l-1)} : u \in \mathcal{N}(v)\}\right)
        + \mathbf{b}_l
    \right), \qquad l = 1, \dots, L,
\end{equation}
The first two layers use ReLU activations, while the third layer serves directly as the output layer, with dimensions of $\mathbf{W}_3^1$, $\mathbf{W}_3^2$, and $\mathbf{b}_3$ chosen such that $\mathbf{h}_v^{(3)}$ is a scalar.

Since Haddad and Figueiredo address the multi-source problem, they apply a sigmoid activation in the final layer to obtain the probability that node $v$ is a source. Training is based on a weighted binary cross-entropy loss, where source nodes are upweighted to address class imbalance.

\subsection{Graph Attention Network (GAT)}\label{sec:gat}
Besides several GCN variants, Shah et al.~\cite{Shah:2020} also evaluate a Graph Attention Network (GAT)~\cite{Velickovic:2018}. The node update equation is given by
\begin{equation}
\label{eq:updateGAT}
    \mathbf{h}_v^{(l)} = \sigma\!\left(
        \mathbf{W}_l \sum_{u \in \{\mathcal{N}(v) \cup \{v\}\}} a_{vu}\, \mathbf{h}_u^{(l-1)} + \mathbf{b}_l
    \right), \qquad l = 1, \dots, L,
\end{equation}
where $a_{vu}$ denotes the learned attention coefficient between node $v$ and its neighbor $u$.  

The attention mechanism is computed in two steps. First, an unnormalized attention weight is obtained as
\begin{equation}
\label{eq:attention1}
    e_{vu} = \text{LeakyReLU}\!\left(
        \mathbf{a}^\intercal
        \left[
            \mathbf{W}_l\mathbf{h}_v^{(l-1)} \mathbin\Vert 
            \mathbf{W}_l\mathbf{h}_u^{(l-1)}
        \right]
    \right),
\end{equation}
where $\mathbf{a} \in \mathbb{R}^{2s_l}$ ensures that $e_{vu}$ is a scalar. These attention scores are then normalized across all neighbors using a softmax:
\begin{equation}
\label{eq:attention2}
    a_{vu} = 
    \frac{\exp(e_{vu})}
         {\sum_{w \in \{\mathcal{N}(v) \cup \{v\}\}} \exp(e_{vw})}.
\end{equation}

In practice, attention mechanisms are often used in a {\itshape multi-head} setting, with several attention functions computed in parallel to stabilize training and improve expressiveness. Shah et al.~\cite{Shah:2020}, for instance, use $K=4$ attention heads, combining their outputs via element-wise averaging or summing in the final layer.

\subsection{Infected Graph Convolutional Network (IGCN)}
Guo et al.~\cite{Guo:2021} first augment the node  feature vectors with additional topological descriptors such as betweenness centrality and subsequently normalize all features. This input layer is followed by a standard GCN layer using LeakyReLU activation. The core innovation lies in a sequence of IGCN layers, each defined as
\begin{equation}
\label{eq:updateIGCN}
    \mathbf{h}_v^{(l)} = \sigma\!\left(\sum_{u\in \{\mathcal{N}(v)\cup \{v\}\}} w\left(x_v(t_1), x_u(t_1)\right)\,\cdot \mathbf{h}_u^{(l-1)}\right), \qquad l=1,\dots,L,
\end{equation}
where
\begin{equation}
\label{eq:IGCNweights}
w\left(x_v(t_1), x_u(t_1)\right)=\begin{cases}
    w_{S,S}, & \text{if } x_v(t_1)=S,\, x_u(t_1)=S,\\
    w_{S,I}, & \text{if } x_v(t_1)=S,\, x_u(t_1)=I,\\
    w_{I,S}, & \text{if } x_v(t_1)=I,\, x_u(t_1)=S,\\
    w_{I,I}, & \text{if } x_v(t_1)=I,\, x_u(t_1)=I,
\end{cases}
\end{equation}
represents four learnable weights used to modulate the aggregation based on the epidemic states of nodes $v$ and $u$. No other parameters are learned within these layers. The architecture concludes with a fully connected layer and a softmax activation, producing a probability distribution over all nodes. Model training uses the cross-entropy loss with an additional regularization term.

\subsection{Source Detection Spatial-Temporal Graph Convolutional Network (SD-STGCN)}
Sha et al.~\cite{Sha:2021} extend the spatio-temporal GCN architecture originally proposed by Yu et al.~\cite{Yu:2018} to address the source detection problem with multiple epidemic snapshots. Their model consists of two {\itshape spatio-temporal convolution blocks}, each containing a GCN layer sandwiched between two temporal convolution (CNN-based) layers.  

The model takes as input the adjacency matrix $\mathbf{A}$ and a node feature tensor $\mathbf{X} \in \mathbb{R}^{k \times N \times m}$, where $k$ is the number of epidemic snapshots observed at different times. The temporal layers are implemented as 1-D CNNs followed by Gated Linear Units (GLUs). A total of 36 CNN kernels with sizes $K_t \in \{2,3,4\}$ are used, applied without padding, thereby reducing the temporal dimension by $K_t - 1$ at each step.  

Within each block, the spatial dependency among nodes is captured by the GCN layer, implemented either as the standard GCN~\cite{Kipf:2017} (using 2--4 layers) or the precursor based on the Chebyshev polynomials approximation~\cite{Defferrard:2016}, both followed by a ReLU activation. In the output layer, a final CNN reduces the temporal dimension to one, after which a dense layer and a softmax activation produce a probability vector $\mathbf{\hat{y}} \in \mathbb{R}^{N}$ over all nodes.

\subsection{Backtracking Network (BN)}
As noted earlier, Ru et al.~\cite{Ru:2023} are the only ones to address the single-source problem on {\itshape temporal networks}, which they model as a sequence of static graphs $G_0, \dots, G_T$. Their GNN takes as input the (one-hot encoded) node states $\mathbf{x}_v \in \mathbb{R}^m$, which are first linearly projected, the adjacency matrix of the aggregated network $\widetilde{G}$ containing all edges that appear in any time slice, $\mathbf{A}_{\widetilde{G}}$, and edge feature vectors $\mathbf{x}_{(v,u)} \in \mathbb{R}^{T-1}$ that indicate, for each time step, whether edge $(v,u)$ is active or not.  

To incorporate temporal edge information, the model alternates between updating {\itshape edge} and {\itshape node} embeddings. First, edge embeddings are updated as
\begin{equation}
\label{eq:updateEdgeBN}
    \mathbf{e}_{(v,u)}^{(l)} =
    \text{ReLU}\!\left(
        \mathbf{W}_l^{1} \left[
            \mathbf{e}_{(v,u)}^{(l-1)} \mathbin\Vert 
            \mathbf{h}_u^{(l-1)}
        \right]
    \right), \qquad l = 1, \dots, L,
\end{equation}
followed by node embedding updates:
\begin{equation}
\label{eq:updateNodeBN}
    \mathbf{h}_v^{(l)} =
    \text{ReLU}\!\left(\mathbf{W}_l^{2} \mathbf{h}_v^{(l-1)}\right)
    + \sum_{u \in \mathcal{N}(v)} \mathbf{e}_{(v,u)}^{(l)}.
\end{equation}

A final dense layer maps each node embedding to a scalar, and a softmax activation yields a probability distribution over all nodes, representing the likelihood of each being the source. Accordingly, the model is trained using the multi-class cross-entropy loss.

\section{Experimental Setting}\label{sec:setting}

In this section, we first introduce the exact specifications and training hyperparameters of the four GNN architectures from related work used for comparisons with traditional source detection methods. We then describe the network datasets and SIR process used to train the four models. This section is intentionally comprehensive to ensure that our results and training process can be readily reproduced. All source code is publicly available on GitHub (\url{https://github.com/martinSter/gnn-source-detection}).

\subsection{Architectures}\label{sec:design}

We implement four GNN architectures from related work that are compatible with the problem formulation used in this paper: (i)~Dong et al.'s GCN-based architecture~\cite{Dong:2019}, described in equation~\ref{eq:updateDong}; (ii)~Shah et al.'s modified GCN architecture~\cite{Shah:2020}, described in equation~\ref{eq:updateShah}; (iii)~Guo et al.'s IGCN architecture~\cite{Guo:2021}, described in equations~\ref{eq:updateIGCN} and \ref{eq:IGCNweights}; and (iv)~Haddad and Figueiredo's GraphSAGE-based architecture~\cite{Haddad:2023}, described in equation~\ref{eq:updateHaddad}. Other proposed architectures were not considered here because they apply to temporal networks~\cite{Ru:2023}, apply to a vastly different problem setting with several observed snapshots~\cite{Sha:2021}, or provide too few details for reproducing the architecture~\cite{Shu:2021}. GAT~(Subsection~\ref{sec:gat}) is excluded from the comparison on the basis that Shah et al.~\cite{Shah:2020} find GCN-based architectures to perform at least as well across several network types.

Dong et al.~\cite{Dong:2019} employ a standard GCN architecture as outlined in Subsection~\ref{sec:gcn}. Instead of a sigmoid activation after the final dense layer, we use the log-softmax activation to take into account the single-source nature of our problem setting. Similarly, for the architecture proposed by Shah et al.~\cite{Shah:2020}, we replace their output layer (equation~\ref{eq:outputShah}) with a dense layer projecting the embeddings for each node to a scalar followed by a log-softmax activation. For Guo et al.'s IGCN model~\cite{Guo:2021}, we make two modifications. First, we extend the model from the SI to the SIR case by allowing 9 ($3^2$~possible pairs of node states) instead of 4 learnable weights in the IGCN layer. Second, we normalize the neighborhood aggregation outlined in equation~\ref{eq:updateIGCN} by node $v$'s degree. Omitting this normalization resulted in unstable training and substantially degraded detection performance, possibly due to scale explosion. Finally, Haddad and Figueiredo's GraphSAGE model~\cite{Haddad:2023} similarly requires replacing the sigmoid activations in the output layer with a log-softmax activation to account for the single-source setting.

\begin{figure}[ht]
\centering
\includegraphics[width=0.6\linewidth]{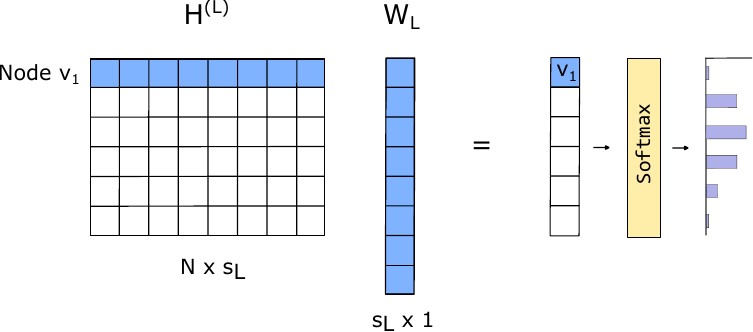}
\caption{Schematic depiction of the output layer for an example with $s_L=8$ hidden units. Note that this does not reflect the embedding sizes used in our architectures. Unlike many graph learning architectures, no global pooling operation is applied here to aggregate node embeddings $\mathbf{H}^{(L)}$ across nodes. Instead, each node's embedding is projected individually to a scalar score.}
\label{fig:outputlayer}
\end{figure}

The output layer's weight matrix is shaped such that, for each node $v$, a linear transformation of the preceding embeddings yields a scalar value representing the predicted likelihood of node $v$ being the epidemic source (Figure~\ref{fig:outputlayer}). The log-softmax activation produces a normalized probability distribution across all nodes in the graph. Unlike in other graph-level prediction tasks, we do not apply global pooling operations, since any aggregation across nodes would, to some degree, eliminate the node-specific information that was learned across the message-passing layers.

Crucially, we deviate in some cases from the model parameters used in the related work in order to ensure a fair comparison across architectures in terms of model expressiveness. We set the number of message-passing layers and the number of hidden units to 5 and 128, respectively, for each architecture. No feature augmentation is performed, i.e., all models receive the one-hot encoded node states as input features.

\subsection{Training}
For the learning hyperparameters, the related work (Table~\ref{tab:hyperparams}) and the GNN design space analysis of You et al.~\cite{You:2020} show broad consensus regarding their choice. Specifically, we set the learning rate to 0.001, employ the ADAM optimizer with a small weight decay of 0.0005, and use a batch size of 128 to ensure stable gradient updates given our relatively small graphs. The maximum number of training epochs is set to 500, but we apply early stopping with a patience of five epochs. The loss function is the negative log-likelihood loss, which aligns with the log-softmax output activation. For all four models, message-passing layers are followed by dropout with a dropout rate of 0.1.

\subsection{Network Data and SIR Process}\label{sec:network-data}
We train our model on six static networks, most of which are well established in the network science literature. Their key structural characteristics, along with the epidemic parameters used in our experiments, are summarized in Table~\ref{tab:networks}.

\begin{table*}
    \caption{Overview of empirical networks. The middle part of the table reports the number of nodes, the number of edges, the average degree, the average shortest path length, the diameter, and the average local clustering coefficient for each network. The right part lists the infection rate $\beta$ and the duration $T$ until the snapshot was observed. Note that the Workplace and Highschool networks are originally temporal networks and we simply use their static projections.}
    \label{tab:networks}
    \centering
    \begin{tabular}{l|rrrrrr|rr}
        \toprule
        \textbf{Name} & $|V|$ & $|E|$ & $\braket{k}$ & $\braket{l}$ & $\text{diam}(G)$ & $C$ & $\beta$ & $T$ \\
        \midrule
        Karate~\cite{Zachary:1977} & 34 & 77 & 4.53 & 2.42 & 5 & 0.54 & 1.300 & 0.85 \\
        Iceland~\cite{Haraldsdottir:1992} & 75 & 114 & 3.04 & 3.20 & 6 & 0.29 & 5.100 & 0.34 \\
        Dolphin~\cite{Lusseau:2003} & 62 & 159 & 5.13 & 3.36 & 8 & 0.26 & 0.900 & 2.20 \\
        Fraternity~\cite{Killworth:1976,Sha:2021} & 58 & 967 & 33.34 & 1.42 & 3 & 0.75 & 0.073 & 3.50 \\
        Workplace~\cite{Genois:2015} & 92 & 755 & 16.41 & 1.96 & 3 & 0.43 & 0.165 & 3.50 \\
        Highschool~\cite{Mastrandrea:2015} & 327 & 5818 & 35.58 & 2.16 & 4 & 0.50 & 0.065 & 7.50 \\
        \bottomrule
    \end{tabular}
\end{table*}

Without loss of generality, we fix the recovery rate to $\mu = 1$. The infection rate $\beta$ is chosen such that the basic reproduction number $R_0 \approx 2$, ensuring comparable epidemic intensity across networks. The duration until the snapshot is observed, $T$, is selected such that, on average, roughly 40\% of nodes are infected at the observation time.

For each experiment, epidemic realizations are simulated using the SIR parameters specified in Table~\ref{tab:networks}. For the large majority of experiments, we generate 500 simulated outbreaks per potential source node, thus ensuring a balanced dataset. The resulting data are split in a stratified manner into 70\% training and 30\% validation data.

\section{Experiments}\label{sec:experiments}

In this section, we present the results that address our core research questions. We first turn to our central question: the performance of GNN-based source detection methods relative to a number of heuristic and probabilistic benchmarks. Second, we perform an ablation study investigating the importance of different architectural features. Third, we examine how detectability evolves as the time between outbreak and observation of the snapshot increases. Finally, we analyze how performance scales with training set size for three selected approaches, including the best-performing GNN architecture, and assess the importance of knowing the precise time interval between the outbreak and the observed snapshot as well as the infection rate.

We evaluate all results using top-5 accuracy, defined as the proportion of experiments in which the true source appears among the five highest-ranked nodes returned by a method. In the subsection comparing GNNs with heuristic and probabilistic benchmarks, we additionally report alternative evaluation metrics, highlighting that the choice of performance measure for source detection is itself a nontrivial consideration.

\subsection{Comparison with Benchmarks}\label{sec:benchmark-results}
In this subsection, we address the central question of this paper: do GNNs truly outperform traditional source-detection methods? The suitability of GNNs for this task has been questioned~\cite{Openreview:2020}, and the existing literature remains inconclusive regarding their true potential in source detection. A major reason is that prior work (focused on the single-source setting) has primarily compared GNNs with heuristic approaches (e.g., the Jordan center) or with the dynamic message passing (DMP) method introduced by Lokhov et al.~\cite{Lokhov:2014}. Only one study~\cite{Ru:2023} incorporates the soft margin estimator (SME), a notable contribution by Antulov-Fantulin et al.~\cite{Antulov-Fantulin:2015}. Moreover, none of the existing studies includes a simple MLP baseline to isolate the contribution of message passing.

\subsubsection{Benchmark Methods}
Our simplest baseline, referred to as Random, selects an estimated source uniformly at random from all nodes in the infected subgraph. We then consider two purely topological heuristics: the Jordan center~\cite{Zhu:2016} and the node with the highest betweenness centrality within the infected subgraph (Betweenness). Although we also evaluated degree- and closeness-based heuristics, betweenness centrality consistently yielded the strongest performance among them.

To incorporate probabilistic methods into the comparison, we evaluate the soft margin estimator~(SME)~\cite{Antulov-Fantulin:2015} as well as the Monte Carlo mean field~(MCMF) approach~\cite{Sterchi:2023} (see Appendix for details). The SME has the appealing property that, as the number of simulations per potential source increases, it converges to direct Monte Carlo estimation and should, in principle, recover the true likelihood of each source node eventually. In practice, however, achieving this regime is computationally infeasible except for very small graphs. MCMF is conceptually similar to dynamic message passing (DMP)~\cite{Lokhov:2014} but can be more computationally efficient, alleviating some of the scalability limitations reported for DMP~\cite{Shah:2020}. With a sufficiently large number of simulations, MCMF also tends to yield more accurate estimates of node state probabilities than DMP. Importantly, SME and MCMF are both computed using the same set of simulated outbreaks that serve as training data for the GNNs, ensuring that all methods operate on the same data substrate.

Finally, we compare the four GNN variants to two MLP architectures (see Appendix for details): MLP$_{\text{node}}$, which processes each node's state independently, and MLP$_{\text{snapshot}}$, which takes a flattened representation of the complete snapshot as input. Because  MLP$_{\text{node}}$ does not share information across nodes, it can only distinguish the three node states and assigns identical scores to all nodes in the same state. MLP$_{\text{snapshot}}$, by contrast, processes the complete snapshot and can therefore learn which configurations of node states are indicative of a particular source.

\subsubsection{Results}
To evaluate the different source detection methods, we simulate an additional 100 outbreak scenarios per node, using the same epidemic parameters as those used for generating the training data. Several performance metrics are available for assessing the methods. Given our fully balanced experimental design, accuracy-based metrics commonly used in related work~\cite{Shah:2020,Sha:2021,Guo:2021,Ru:2023} are appropriate. Specifically, we report the previously introduced top-5 accuracy. In addition, we compute the average error distance~\cite{Shu:2021}, which measures the mean shortest-path distance between the true and the estimated MAP source. As a third metric, we use the average reciprocal rank~\cite{Sha:2021}, where the reciprocal of the true source’s rank ensures that higher values correspond to better performance. Our fourth metric, the average 90\% credible set size~\cite{Paluch:2020}, applies only to methods that output probability distributions and quantifies the sharpness of those distributions by measuring the number of nodes required to accumulate at least 90\% posterior probability mass. Finally, we evaluate performance using the resistance score, a novel proper scoring rule for distributions on graphs that balances rewarding accuracy with avoiding overconfidence and generally encourages an honest uncertainty quantification (lower values indicate better performance)~\cite{Hilfiker:2026}. For each network, we trained the four GNN variants three times, using different random seeds for both the training data and weight initialization. The results reported in Tables~\ref{tab:dataset_model_metrics1}~and~\ref{tab:dataset_model_metrics2} are averaged over these three runs.

\begin{table}[ht]
\small
\centering
\caption{Performance of different source detection methods across the first three network datasets. Test sets are generated by simulating 100 outbreaks from each node in the network. All performance measures are averaged over three runs, each using a different random seed. The 95\% confidence intervals shown in parentheses reflect the uncertainty across these runs. For each metric (except CSS), the best-performing method is highlighted in bold. The performance of the four GNN variants is highlighted in red.}
\begin{tabular}{l|l|rrrrr}
\toprule
\textbf{Dataset} & \textbf{Model} & \textbf{Top-5 accuracy} & \textbf{Error dist.} & \textbf{Rec. rank} & \textbf{|90\% CSS|} & \textbf{Resistance} \\
\midrule
Karate & Random & 39.37\% ($\pm 0.88\%$) & 1.3209 ($\pm 0.0163$) & 0.3478 ($\pm 0.0076$) & - & -\\
& Jordan        & 52.68\% ($\pm 0.37\%$) & 1.0799 ($\pm 0.0042$) & 0.4034 ($\pm 0.0000$) & - & -\\
& Betweenness   & 55.19\% ($\pm 0.11\%$) & 0.9111 ($\pm 0.0030$) & 0.4273 ($\pm 0.0000$) & - & -\\
& SME           & 60.67\% ($\pm 0.22\%$) & 1.0825 ($\pm 0.0151$) & 0.4770 ($\pm 0.0053$) & 6.1575 ($\pm 0.1141$) & 0.3004 ($\pm 0.0035$)\\
& MCMF           & 65.61\% ($\pm 0.16\%$) & \textbf{0.8996} ($\pm 0.0037$) & 0.5141 ($\pm 0.0018$) & 3.5992 ($\pm 0.0648$) & 0.2489 ($\pm 0.0010$) \\
& MLP$_{\text{node}}$ & 60.01\% ($\pm 0.34\%$) & 1.1825 ($\pm 0.0094$) & 0.4289 ($\pm 0.0000$) & 11.9274 ($\pm 0.0129$) & 0.2357 ($\pm 0.0001$) \\
& MLP$_{\text{snapshot}}$ & 67.90\% ($\pm 0.84\%$) & 1.0374 ($\pm 0.0060$) & 0.5204 ($\pm 0.0020$) & 8.0804 ($\pm 0.1256$) & 0.2338 ($\pm 0.0005$) \\
& \cellcolor[HTML]{BC8F8F}IGCN~\cite{Guo:2021} & \cellcolor[HTML]{BC8F8F}72.83\% ($\pm 0.15\%$) & \cellcolor[HTML]{BC8F8F}0.9745 ($\pm 0.0107$) & \cellcolor[HTML]{BC8F8F}0.5448 ($\pm 0.0008$) & \cellcolor[HTML]{BC8F8F}9.3911 ($\pm 0.1120$) & \cellcolor[HTML]{BC8F8F}0.2173 ($\pm 0.0002$) \\
& \cellcolor[HTML]{BC8F8F}GCN~\cite{Dong:2019} & \cellcolor[HTML]{BC8F8F}68.67\% ($\pm 0.27\%$) & \cellcolor[HTML]{BC8F8F}1.0020 ($\pm 0.0249$) & \cellcolor[HTML]{BC8F8F}0.4984 ($\pm 0.0011$) & \cellcolor[HTML]{BC8F8F}12.6557 ($\pm 0.2816$) & \cellcolor[HTML]{BC8F8F}0.2393 ($\pm 0.0019$) \\
& \cellcolor[HTML]{BC8F8F}GCN~\cite{Shah:2020} & \cellcolor[HTML]{BC8F8F}\textbf{72.87\%} ($\pm 0.03\%$) & \cellcolor[HTML]{BC8F8F}0.9333 ($\pm 0.0352$) & \cellcolor[HTML]{BC8F8F}\textbf{0.5498} ($\pm 0.0012$) & \cellcolor[HTML]{BC8F8F}8.8102 ($\pm 0.0844$) & \cellcolor[HTML]{BC8F8F}\textbf{0.2159} ($\pm 0.0003$) \\
& \cellcolor[HTML]{BC8F8F}GraphSAGE~\cite{Haddad:2023} & \cellcolor[HTML]{BC8F8F}71.89\% ($\pm 0.51\%$) & \cellcolor[HTML]{BC8F8F}0.9808 ($\pm 0.0286$) & \cellcolor[HTML]{BC8F8F}0.5420 ($\pm 0.0012$) & \cellcolor[HTML]{BC8F8F}9.4744 ($\pm 0.4722$) & \cellcolor[HTML]{BC8F8F}0.2189 ($\pm 0.0013$) \\
\midrule
Iceland & Random & 26.51\% ($\pm 0.19\%$) & 1.9537 ($\pm 0.0039$) & 0.2426 ($\pm 0.0010$) & - & -\\
& Jordan        & 35.84\% ($\pm 0.14\%$) & 1.4410 ($\pm 0.0064$) & 0.2838 ($\pm 0.0000$) & - & -\\
& Betweenness   & 37.39\% ($\pm 0.09\%$) & \textbf{1.2484} ($\pm 0.0004$) & 0.2986 ($\pm 0.0000$) & - & -\\
& SME           & 46.00\% ($\pm 0.26\%$) & 1.5094 ($\pm 0.0137$) & 0.3697 ($\pm 0.0014$) & 13.7520 ($\pm 0.0971$) & 0.6795 ($\pm 0.0040$) \\
& MCMF          & 50.95\% ($\pm 0.35\%$) & 1.3609 ($\pm 0.0068$) & 0.4001 ($\pm 0.0003$) & 4.9706 ($\pm 0.0249$) & 0.6079 ($\pm 0.0021$) \\
& MLP$_{\text{node}}$ & 41.50\% ($\pm 0.09\%$) & 1.7609 ($\pm 0.0063$) & 0.3070 ($\pm 0.0000$) & 27.3003 ($\pm 0.0777$) & 0.5895 ($\pm 0.0004$) \\
& MLP$_{\text{snapshot}}$ & 53.07\% ($\pm 0.21\%$) & 1.5442 ($\pm 0.0162$) & 0.4080 ($\pm 0.0030$) & 18.1261 ($\pm 0.4196$) & 0.5632 ($\pm 0.0021$) \\
& \cellcolor[HTML]{BC8F8F}IGCN~\cite{Guo:2021} & \cellcolor[HTML]{BC8F8F}55.77\% ($\pm 0.09\%$) & \cellcolor[HTML]{BC8F8F}1.5523 ($\pm 0.0062$) & \cellcolor[HTML]{BC8F8F}0.4199 ($\pm 0.0007$) & \cellcolor[HTML]{BC8F8F}19.7947 ($\pm 0.1129$) & \cellcolor[HTML]{BC8F8F}0.5315 ($\pm 0.0000$) \\
& \cellcolor[HTML]{BC8F8F}GCN~\cite{Dong:2019} & \cellcolor[HTML]{BC8F8F}55.94\% ($\pm 0.21\%$) & \cellcolor[HTML]{BC8F8F}1.3819 ($\pm 0.0189$) & \cellcolor[HTML]{BC8F8F}0.4150 ($\pm 0.0027$) & \cellcolor[HTML]{BC8F8F}21.6268 ($\pm 1.8725$) & \cellcolor[HTML]{BC8F8F}0.5410 ($\pm 0.0024$) \\
& \cellcolor[HTML]{BC8F8F}GCN~\cite{Shah:2020} & \cellcolor[HTML]{BC8F8F}\textbf{57.95\%} ($\pm 0.39\%$) & \cellcolor[HTML]{BC8F8F}1.3850 ($\pm 0.0776$) & \cellcolor[HTML]{BC8F8F}\textbf{0.4311} ($\pm 0.0016$) & \cellcolor[HTML]{BC8F8F}19.0114 ($\pm 0.4112$) & \cellcolor[HTML]{BC8F8F}\textbf{0.5262} ($\pm 0.0009$) \\
& \cellcolor[HTML]{BC8F8F}GraphSAGE~\cite{Haddad:2023} & \cellcolor[HTML]{BC8F8F}55.54\% ($\pm 0.60\%$) & \cellcolor[HTML]{BC8F8F}1.5269 ($\pm 0.0706$) & \cellcolor[HTML]{BC8F8F}0.4177 ($\pm 0.0022$) & \cellcolor[HTML]{BC8F8F}21.3116 ($\pm 0.5715$) & \cellcolor[HTML]{BC8F8F}0.5405 ($\pm 0.0019$) \\
\midrule
Dolphin & Random & 33.08\% ($\pm 0.30\%$) & 1.7722 ($\pm 0.0117$) & 0.3065 ($\pm 0.0019$) & - & -\\
& Jordan        & 41.63\% ($\pm 0.13\%$) & 1.4484 ($\pm 0.0042$) & 0.3506 ($\pm 0.0000$) & - & -\\
& Betweenness   & 43.05\% ($\pm 0.04\%$) & 1.3902 ($\pm 0.0009$) & 0.3712 ($\pm 0.0000$) & - & -\\
& SME           & 45.37\% ($\pm 0.52\%$) & 1.4762 ($\pm 0.0079$) & 0.3943 ($\pm 0.0022$) & 13.1542 ($\pm 0.1956$) & 0.2826 ($\pm 0.0012$) \\
& MCMF           & 51.25\% ($\pm 0.37\%$) & 1.2215 ($\pm 0.0122$) & 0.4269 ($\pm 0.0007$) & 5.0600 ($\pm 0.1093$) & 0.2708 ($\pm 0.0016$) \\
& MLP$_{\text{node}}$ & 44.59\% ($\pm 0.33\%$) & 1.6118 ($\pm 0.0139$) & 0.3397 ($\pm 0.0000$) & 19.4132 ($\pm 0.4241$) & 0.2408 ($\pm 0.0004$) \\
& MLP$_{\text{snapshot}}$ & 55.94\% ($\pm 0.22\%$) & 1.2967 ($\pm 0.0286$) & 0.4514 ($\pm 0.0021$) & 13.3406 ($\pm 0.1511$) & 0.2300 ($\pm 0.0021$) \\
& \cellcolor[HTML]{BC8F8F}IGCN~\cite{Guo:2021} & \cellcolor[HTML]{BC8F8F}58.30\% ($\pm 0.08\%$) & \cellcolor[HTML]{BC8F8F}1.4481 ($\pm 0.0067$) & \cellcolor[HTML]{BC8F8F}0.4676 ($\pm 0.0004$) & \cellcolor[HTML]{BC8F8F}15.5202 ($\pm 0.4495$) & \cellcolor[HTML]{BC8F8F}0.2175 ($\pm 0.0001$) \\
& \cellcolor[HTML]{BC8F8F}GCN~\cite{Dong:2019} & \cellcolor[HTML]{BC8F8F}57.81\% ($\pm 0.21\%$) & \cellcolor[HTML]{BC8F8F}1.2116 ($\pm 0.0323$) & \cellcolor[HTML]{BC8F8F}0.4523 ($\pm 0.0023$) & \cellcolor[HTML]{BC8F8F}18.2811 ($\pm 0.0569$) & \cellcolor[HTML]{BC8F8F}0.2282 ($\pm 0.0008$) \\
& \cellcolor[HTML]{BC8F8F}GCN~\cite{Shah:2020} & \cellcolor[HTML]{BC8F8F}\textbf{60.33\%} ($\pm 0.10\%$) & \cellcolor[HTML]{BC8F8F}\textbf{1.1632} ($\pm 0.0162$) & \cellcolor[HTML]{BC8F8F}\textbf{0.4829} ($\pm 0.0013$) & \cellcolor[HTML]{BC8F8F}14.5358 ($\pm 0.1117$) & \cellcolor[HTML]{BC8F8F}\textbf{0.2126} ($\pm 0.0003$) \\
& \cellcolor[HTML]{BC8F8F}GraphSAGE~\cite{Haddad:2023} & \cellcolor[HTML]{BC8F8F}59.80\% ($\pm 0.09\%$) & \cellcolor[HTML]{BC8F8F}1.2511 ($\pm 0.0429$) & \cellcolor[HTML]{BC8F8F}0.4777 ($\pm 0.0004$) & \cellcolor[HTML]{BC8F8F}16.2347 ($\pm 0.4023$) & \cellcolor[HTML]{BC8F8F}0.2158 ($\pm 0.0006$) \\
\bottomrule
\end{tabular}
\label{tab:dataset_model_metrics1}
\end{table}

\begin{table}[ht]
\small
\centering
\caption{Performance of different source detection methods across the second set of network datasets. The description provided for Table~\ref{tab:dataset_model_metrics1} applies here as well.}
\begin{tabular}{l|l|rrrrr}
\toprule
\textbf{Dataset} & \textbf{Model} & \textbf{Top-5 accuracy} & \textbf{Error dist.} & \textbf{Rec. rank} & \textbf{|90\% CSS|} & \textbf{Resistance} \\
\midrule
Fraternity & Random & 42.14\% ($\pm 0.50\%$) & 0.8056 ($\pm 0.0044$) & 0.3974 ($\pm 0.0023$) & - & -\\
& Jordan        & 50.35\% ($\pm 0.32\%$) & 0.7761 ($\pm 0.0055$) & 0.4178 ($\pm 0.0000$) & - & -\\
& Betweenness   & 51.95\% ($\pm 0.06\%$) & \textbf{0.6584} ($\pm 0.0021$) & 0.4535 ($\pm 0.0000$) & - & -\\
& SME           & 51.31\% ($\pm 0.17\%$) & 0.7759 ($\pm 0.0025$) & 0.4619 ($\pm 0.0024$) & 6.5087 ($\pm 0.1599$) & 0.0292 ($\pm 0.0003$) \\
& MCMF           & 54.60\% ($\pm 0.18\%$) & 0.6764 ($\pm 0.0054$) & 0.4828 ($\pm 0.0017$) & 6.1986 ($\pm 0.9044$) & 0.0239 ($\pm 0.0008$) \\
& MLP$_{\text{node}}$ & 52.78\% ($\pm 0.14\%$) & 0.7912 ($\pm 0.0041$) & 0.4228 ($\pm 0.0000$) & 16.8509 ($\pm 0.2250$) & 0.0202 ($\pm 0.0000$) \\
& MLP$_{\text{snapshot}}$ & 54.04\% ($\pm 0.17\%$) & 0.7809 ($\pm 0.0072$) & 0.4751 ($\pm 0.0003$) & 16.0714 ($\pm 0.6127$) & 0.0215 ($\pm 0.0001$) \\
& \cellcolor[HTML]{BC8F8F}IGCN~\cite{Guo:2021} & \cellcolor[HTML]{BC8F8F}56.22\% ($\pm 0.47\%$) & \cellcolor[HTML]{BC8F8F}0.8251 ($\pm 0.0294$) & \cellcolor[HTML]{BC8F8F}0.4920 ($\pm 0.0030$) & \cellcolor[HTML]{BC8F8F}16.4714 ($\pm 0.2377$) & \cellcolor[HTML]{BC8F8F}\textbf{0.0199} ($\pm 0.0000$) \\
& \cellcolor[HTML]{BC8F8F}GCN~\cite{Dong:2019} & \cellcolor[HTML]{BC8F8F}26.16\% ($\pm 3.00\%$) & \cellcolor[HTML]{BC8F8F}0.9665 ($\pm 0.0169$) & \cellcolor[HTML]{BC8F8F}0.1947 ($\pm 0.0097$) & \cellcolor[HTML]{BC8F8F}43.1317 ($\pm 1.1168$) & \cellcolor[HTML]{BC8F8F}0.0304 ($\pm 0.0003$) \\
& \cellcolor[HTML]{BC8F8F}GCN~\cite{Shah:2020} & \cellcolor[HTML]{BC8F8F}\textbf{57.05\%} ($\pm 0.35\%$) & \cellcolor[HTML]{BC8F8F}0.7734 ($\pm 0.0243$) & \cellcolor[HTML]{BC8F8F}\textbf{0.4949} ($\pm 0.0070$) & \cellcolor[HTML]{BC8F8F}16.3356 ($\pm 0.2152$) & \cellcolor[HTML]{BC8F8F}0.0200 ($\pm 0.0003$) \\
& \cellcolor[HTML]{BC8F8F}GraphSAGE~\cite{Haddad:2023} & \cellcolor[HTML]{BC8F8F}54.58\% ($\pm 0.21\%$) & \cellcolor[HTML]{BC8F8F}0.8759 ($\pm 0.0183$) & \cellcolor[HTML]{BC8F8F}0.4689 ($\pm 0.0016$) & \cellcolor[HTML]{BC8F8F}17.1501 ($\pm 0.6169$) & \cellcolor[HTML]{BC8F8F}0.0201 ($\pm 0.0000$) \\
\midrule
Workplace & Random & 38.05\% ($\pm 0.25\%$) & 1.1593 ($\pm 0.0058$) & 0.3647 ($\pm 0.0010$) & - & -\\
& Jordan        & 45.20\% ($\pm 0.14\%$) & 1.0174 ($\pm 0.0005$) & 0.3981 ($\pm 0.0000$) & - & -\\
& Betweenness   & 46.21\% ($\pm 0.03\%$) & \textbf{0.9262} ($\pm 0.0013$) & 0.4139 ($\pm 0.0000$) & - & -\\
& SME           & 46.71\% ($\pm 0.31\%$) & 1.1034 ($\pm 0.0032$) & 0.4273 ($\pm 0.0004$) & 6.0358 ($\pm 0.0170$) & 0.0685 ($\pm 0.0002$) \\
& MCMF           & 49.46\% ($\pm 0.13\%$) & 0.9225 ($\pm 0.0049$) & 0.4442 ($\pm 0.0011$) & 3.5631 ($\pm 0.1017$) & 0.0591 ($\pm 0.0002$) \\
& MLP$_{\text{node}}$ & 47.60\% ($\pm 0.46\%$) & 1.1260 ($\pm 0.0090$) & 0.3880 ($\pm 0.0000$) & 25.7391 ($\pm 0.1743$) & 0.0476 ($\pm 0.0000$) \\
& MLP$_{\text{snapshot}}$ & 51.29\% ($\pm 0.11\%$) & 1.0434 ($\pm 0.0084$) & 0.4509 ($\pm 0.0014$) & 24.0596 ($\pm 0.5532$) & 0.0501 ($\pm 0.0002$) \\
& \cellcolor[HTML]{BC8F8F}IGCN~\cite{Guo:2021} & \cellcolor[HTML]{BC8F8F}55.18\% ($\pm 0.67\%$) & \cellcolor[HTML]{BC8F8F}0.9962 ($\pm 0.0630$) & \cellcolor[HTML]{BC8F8F}0.4799 ($\pm 0.0032$) & \cellcolor[HTML]{BC8F8F}23.7295 ($\pm 0.2463$) & \cellcolor[HTML]{BC8F8F}\textbf{0.0465} ($\pm 0.0001$) \\
& \cellcolor[HTML]{BC8F8F}GCN~\cite{Dong:2019} & \cellcolor[HTML]{BC8F8F}45.04\% ($\pm 0.50\%$) & \cellcolor[HTML]{BC8F8F}1.0982 ($\pm 0.0216$) & \cellcolor[HTML]{BC8F8F}0.3489 ($\pm 0.0062$) & \cellcolor[HTML]{BC8F8F}45.2054 ($\pm 0.7457$) & \cellcolor[HTML]{BC8F8F}0.0598 ($\pm 0.0009$) \\
& \cellcolor[HTML]{BC8F8F}GCN~\cite{Shah:2020} & \cellcolor[HTML]{BC8F8F}\textbf{55.64\%} ($\pm 0.10\%$) & \cellcolor[HTML]{BC8F8F}0.9379 ($\pm 0.0271$) & \cellcolor[HTML]{BC8F8F}\textbf{0.4833} ($\pm 0.0005$) & \cellcolor[HTML]{BC8F8F}23.1208 ($\pm 0.3154$) & \cellcolor[HTML]{BC8F8F}0.0466 ($\pm 0.0002$) \\
& \cellcolor[HTML]{BC8F8F}GraphSAGE~\cite{Haddad:2023} & \cellcolor[HTML]{BC8F8F}52.88\% ($\pm 0.26\%$) & \cellcolor[HTML]{BC8F8F}1.0686 ($\pm 0.0359$) & \cellcolor[HTML]{BC8F8F}0.4596 ($\pm 0.0011$) & \cellcolor[HTML]{BC8F8F}25.6172 ($\pm 0.2630$) & \cellcolor[HTML]{BC8F8F}0.0472 ($\pm 0.0000$) \\
\midrule
Highschool & Random & 38.15\% ($\pm 0.08\%$) & 1.2293 ($\pm 0.0020$) & 0.3779 ($\pm 0.0009$) & - & -\\
& Jordan        & 45.80\% ($\pm 0.03\%$) & 1.2036 ($\pm 0.0048$) & 0.4032 ($\pm 0.0000$) & - & -\\
& Betweenness   & 45.99\% ($\pm 0.02\%$) & 1.0292 ($\pm 0.0001$) & 0.4145 ($\pm 0.0000$) & - & -\\
& SME           & 46.06\% ($\pm 0.06\%$) & 1.1971 ($\pm 0.0010$) & 0.4219 ($\pm 0.0002$) & 20.4401 ($\pm 0.4533$) & 0.0273 ($\pm 0.0002$) \\
& MCMF           & 46.36\% ($\pm 0.05\%$) & \textbf{1.0213} ($\pm 0.0101$) & 0.4251 ($\pm 0.0002$) & 1.7003 ($\pm 0.2074$) & 0.0281 ($\pm 0.0009$) \\
& MLP$_{\text{node}}$ & 35.85\% ($\pm 13.27\%$) & 1.4546 ($\pm 0.2750$) & 0.3018 ($\pm 0.1202$) & 112.2050 ($\pm 5.7713$) & 0.0273 ($\pm 0.0081$) \\
& MLP$_{\text{snapshot}}$ & 25.96\% ($\pm 10.07\%$) & 1.6856 ($\pm 0.1207$) & 0.2070 ($\pm 0.0767$) & 112.8184 ($\pm 8.1023$) & 0.0343 ($\pm 0.0022$) \\
& \cellcolor[HTML]{BC8F8F}IGCN~\cite{Guo:2021} & \cellcolor[HTML]{BC8F8F}48.02\% ($\pm 0.15\%$) & \cellcolor[HTML]{BC8F8F}1.1325 ($\pm 0.0385$) & \cellcolor[HTML]{BC8F8F}0.4390 ($\pm 0.0009$) & \cellcolor[HTML]{BC8F8F}94.7303 ($\pm 0.9078$) & \cellcolor[HTML]{BC8F8F}0.0202 ($\pm 0.0000$) \\
& \cellcolor[HTML]{BC8F8F}GCN~\cite{Dong:2019} & \cellcolor[HTML]{BC8F8F}33.93\% ($\pm 9.94\%$) & \cellcolor[HTML]{BC8F8F}1.2655 ($\pm 0.1959$) & \cellcolor[HTML]{BC8F8F}0.2936 ($\pm 0.1146$) & \cellcolor[HTML]{BC8F8F}138.9414 ($\pm 5.3612$) & \cellcolor[HTML]{BC8F8F}0.0277 ($\pm 0.0056$) \\
& \cellcolor[HTML]{BC8F8F}GCN~\cite{Shah:2020} & \cellcolor[HTML]{BC8F8F}\textbf{48.12\%} ($\pm 0.10\%$) & \cellcolor[HTML]{BC8F8F}1.1374 ($\pm 0.0455$) & \cellcolor[HTML]{BC8F8F}\textbf{0.4397} ($\pm 0.0003$) & \cellcolor[HTML]{BC8F8F}95.6440 ($\pm 0.4658$) & \cellcolor[HTML]{BC8F8F}\textbf{0.0201} ($\pm 0.0000$) \\
& \cellcolor[HTML]{BC8F8F}GraphSAGE~\cite{Haddad:2023} & \cellcolor[HTML]{BC8F8F}47.41\% ($\pm 0.01\%$) & \cellcolor[HTML]{BC8F8F}1.1972 ($\pm 0.0262$) & \cellcolor[HTML]{BC8F8F}0.4265 ($\pm 0.0003$) & \cellcolor[HTML]{BC8F8F}106.7379 ($\pm 1.4465$) & \cellcolor[HTML]{BC8F8F}0.0202 ($\pm 0.0000$) \\
\bottomrule
\end{tabular}
\label{tab:dataset_model_metrics2}
\end{table}

The most striking finding is that GNN-based approaches substantially outperform all other methods across most evaluation measures, with the exception of the average error distance, where the betweenness centrality heuristic is often competitive. More generally, probabilistic approaches (including GNNs) consistently outperform heuristic and random baselines.

Among the GNN variants, GCN~\cite{Shah:2020} achieves the best performance in terms of top-5 accuracy, reciprocal rank, and resistance score. Notably, all GNN variants perform well, including IGCN~\cite{Guo:2021} which uses substantially fewer trainable parameters than the other architectures. The one exception is GCN~\cite{Dong:2019}, which exhibits training instability on the Fraternity, Workplace, and Highschool networks. This is likely a consequence of its architectural design rather than a fundamental limitation of GNN-based source detection, since the other GNN variants do not show this behavior.

Among the non-GNN probabilistic baselines, MLP$_{\text{snapshot}}$ ranks second, trailing only the GNN variants and outperforming MCMF on most networks. This is a noteworthy finding: a simple feedforward network with access to the global snapshot (but without any message passing) already captures much of the signal available for source detection. MLP$_{\text{node}}$, by contrast, is not competitive, confirming that node-level state information alone is insufficient without access to the broader infection pattern. On the larger Highschool network, however, both MLP variants exhibit signs of training instability. For MLP$_{\text{snapshot}}$, a likely contributing factor is the mismatch between the high-dimensional input (the flattened snapshot grows linearly in $N$) and the fixed hidden dimension of 128 units, which may be insufficient to represent the global infection pattern on larger networks. MLP$_{\text{node}}$ likely fails for the same reason as on smaller networks, namely the absence of structural information, although instability may be compounded by the larger network size.

The results further underscore the limitations of MCMF, whose probability mass tends to be overly concentrated on a small subset of nodes, as reflected by its small 90\% credible set sizes. When MCMF is wrong, it often assigns very little probability (or none) to the true region of the network. SME, by contrast, may be limited by the relatively small sample size used here (500 simulations per node), as it is expected to surpass mean-field-based approaches when sufficiently large samples are available.

Additional results for the larger Powergrid network ($N=4{,}941$) are presented in the Appendix~\ref{sec:powergrid}. Although the much larger scale of outbreaks on this network makes source detection considerably harder, the GCN still substantially outperforms the random baseline.

\begin{figure}[ht]
  \centering
  \includegraphics[width=1.0\linewidth]{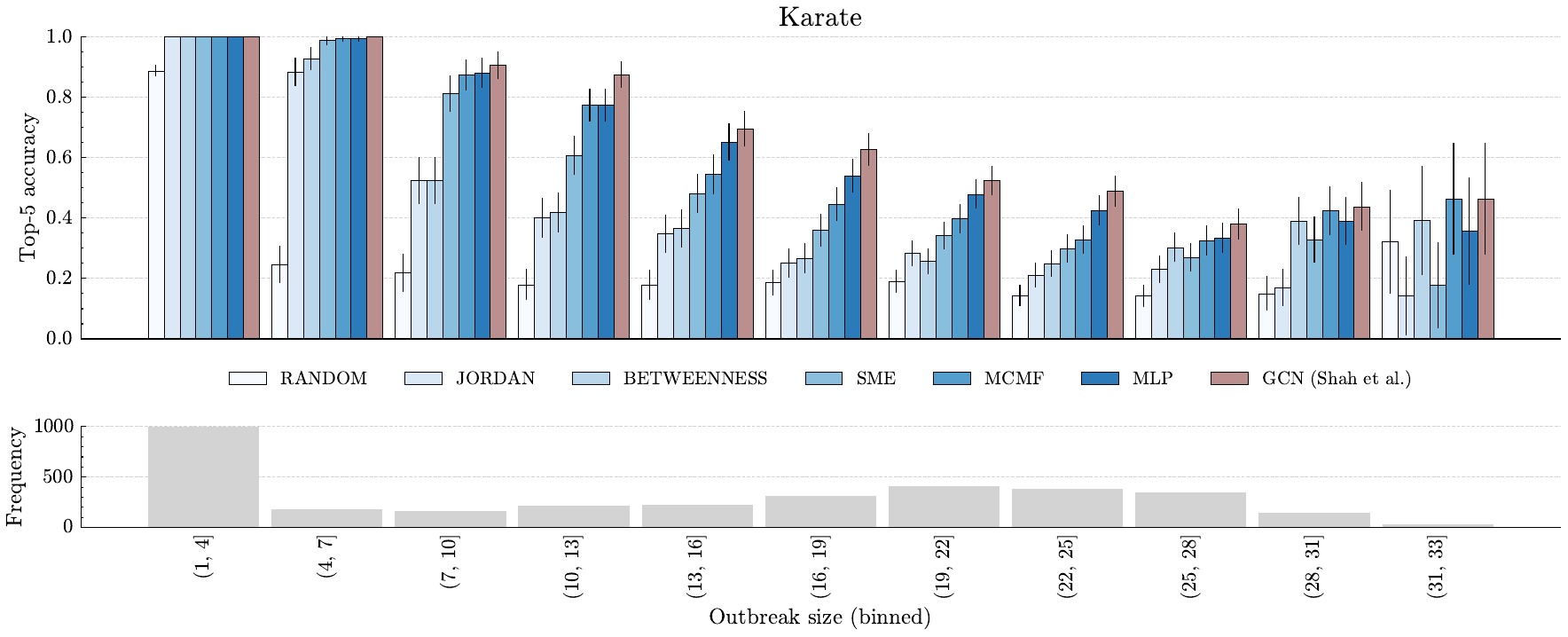}
  \caption{Detection performance by outbreak size for the Karate network. The upper plot reports the top-5 accuracy of the different source detection methods across outbreak categories, defined by outbreak size. Error bars denote 95\% confidence intervals. The lower plot depicts the absolute frequency of outbreaks within each category. The total number of outbreaks is $3{,}400$. MLP results correspond to the MLP$_{\text{snapshot}}$ variant.}
  \label{fig:benchmarks-os}
\end{figure}

Finally, the values reported in Tables~\ref{tab:dataset_model_metrics1}~and~\ref{tab:dataset_model_metrics2} reflect aggregated performances across a wide range of outbreak scenarios. This is further illustrated in Figure~\ref{fig:benchmarks-os}, which plots the top-5 accuracy of the methods for varying outbreak sizes on the Karate network. Notably, GCN~\cite{Shah:2020} attains its primary performance advantages in medium to large outbreak scenarios that are far from trivial.

\subsubsection{Single Scenarios}
\begin{figure}[ht]
  \centering
  \includegraphics[width=0.95\linewidth]{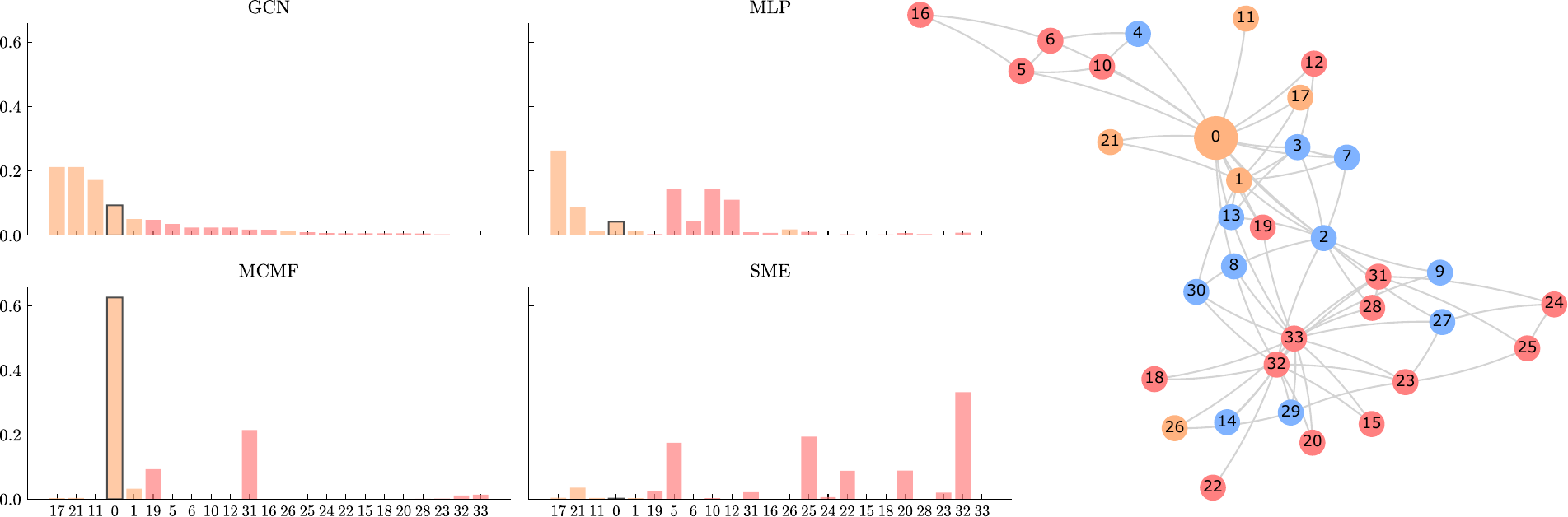}
  \caption{Example outbreak scenario on the Karate network. The plot on the right depicts the network, with node colors indicating epidemic states (blue: susceptible, red: infectious, orange: recovered). The true source node (0) is emphasized by a larger size. The four bar plots on the left show the source predictions of the GCN by Shah et al.~\cite{Shah:2020} (top left), MLP$_{\text{snapshot}}$ (top right), MCMF (bottom left), and SME (bottom right). Nodes on the x-axis are ordered by descending GCN probabilities, and the true source is highlighted with a dark frame. Susceptible nodes are not included in the distributions.}
  \label{fig:example1}
\end{figure}

\begin{figure}[ht]
  \centering
  \includegraphics[width=0.95\linewidth]{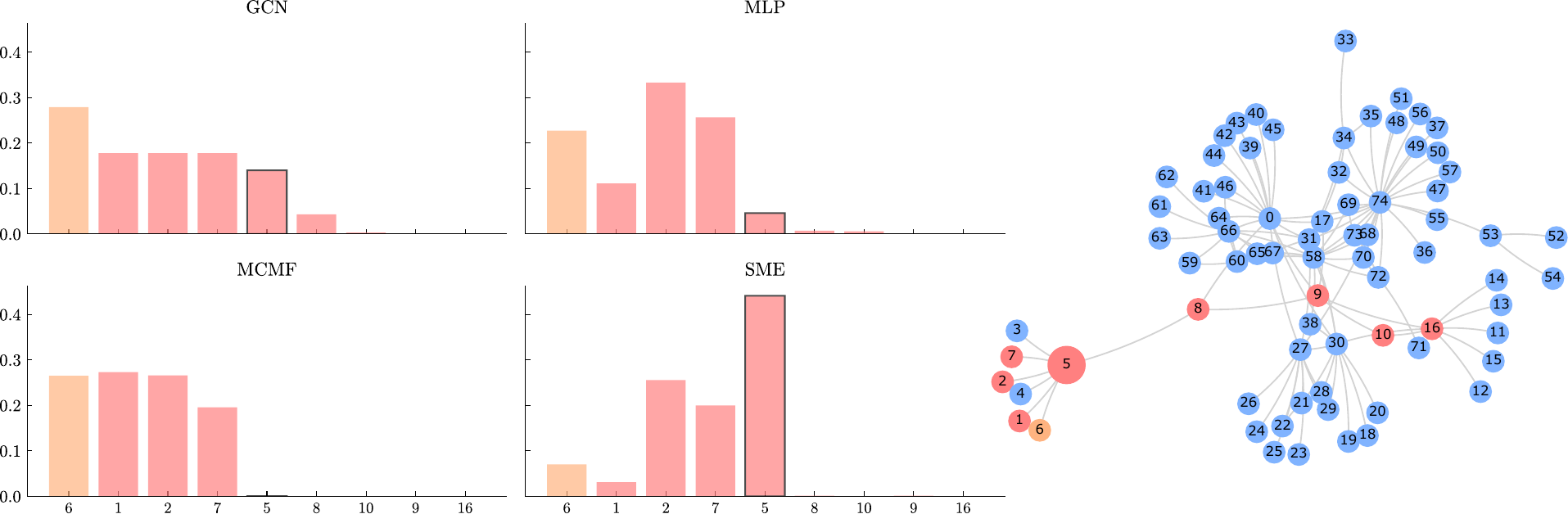}
  \caption{Example outbreak scenario on the Iceland network. The description provided for Figure~\ref{fig:example1} applies here as well.}
  \label{fig:example2}
\end{figure}

To gain deeper insight into the detection quality of a GNN-based approach and its comparison to the MLP, MCMF and SME, we examine two specific outbreak scenarios: one on the Karate network (Figure~\ref{fig:example1}) and another on the Iceland network (Figure~\ref{fig:example2}).

In Figure~\ref{fig:example1}, we observe a relatively large outbreak on the Karate network (23 infected nodes). The true source node~(0) is centrally located and has already recovered, making it a likely source candidate. Accordingly, both the GCN and MCMF assign it a relatively high probability, with MCMF concentrating a large portion of its probability mass on this node, reflecting the fact that the product of node state probabilities computed for node 0 is maximal. However, other plausible source candidates, such as nodes 11, 17, and 21, are assigned virtually no probability by MCMF. Their lower centrality results in node state probability products that are orders of magnitude smaller than those of more central nodes such as 0, 1, 31, and 33. By contrast, the GCN distributes probability mass more smoothly across nodes in the infected subgraph, favoring recovered nodes while accounting for the local density of infectious and recovered neighbors. For example, although node 26 has recovered, it is surrounded by infectious and susceptible nodes only, leading the GCN to assign it lower probability. While MLP$_{\text{snapshot}}$ concentrates a lot of its probability mass on infectious nodes within a cluster close to the true source, the GCN assigns probability more smoothly across the right region of the graph, appropriately favoring recovered nodes. The output of SME is less clear and may reflect instabilities due to the small sample size and the convergence condition (see Appendix).

In Figure~\ref{fig:example2}, we consider a small outbreak on the Iceland network (9 infected nodes). Here, the GCN ranks the true source node~(5) as the fifth most likely source, while MCMF assigns it negligible probability mass. The direct connection of node 5 to the susceptible nodes 3 and 4 makes the product of node state probabilities comparatively small. Interestingly, the GCN correctly assigns equal probability to nodes 1, 2, and 7, which are topologically indistinguishable relative to the rest of the network. MCMF, however, assigns different probabilities to these nodes, likely reflecting small variations in the estimated node state probabilities due to Monte Carlo sampling. SME exhibits a similar discrepancy among nodes 1, 2, and 7, but, unlike MCMF, successfully identifies node 5 as the true source. For MLP$_{\text{snapshot}}$, the most obvious finding is, again, that infectious nodes are assigned an unduly high probability of being the source.

\subsection{Ablation Study}
\begin{figure}[ht]
  \centering
  \includegraphics[width=0.95\linewidth]{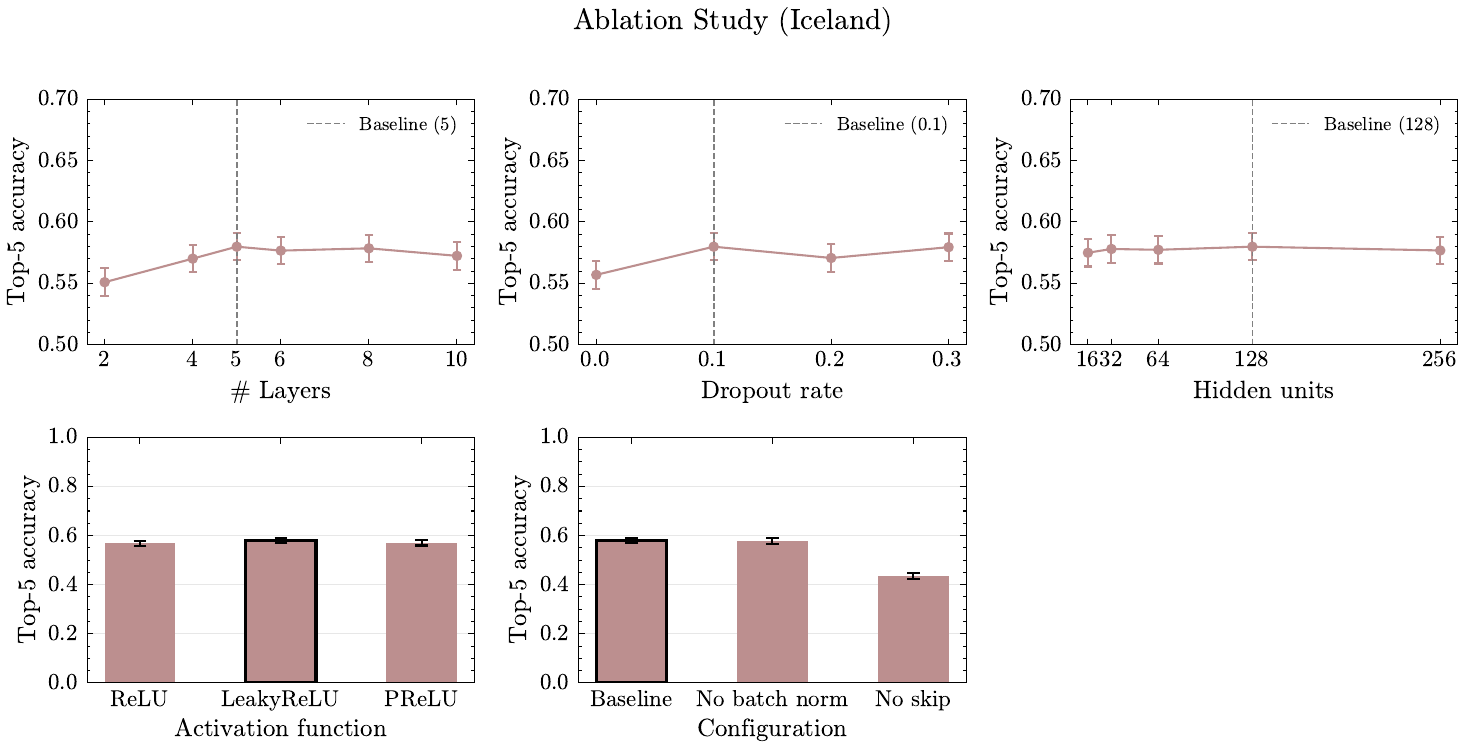}
  \caption{Ablation study results for the Iceland network. All plots show top-5 accuracy evaluated on a test set of 100 simulated outbreaks per node. Error bars represent 95\% confidence intervals. In the top row, the baseline configuration is indicated by a vertical line; in the bottom row, it is marked by a dark border around the corresponding bar}
  \label{fig:iceland-ablation}
\end{figure}

To isolate the effect of individual architectural choices in the best-performing GNN architecture by Shah et al.~\cite{Shah:2020}, we conducted an ablation study varying the number of message-passing layers, the dropout rate, the number of hidden units, the activation function, and the use of batch normalization and residual connections. Figure~\ref{fig:iceland-ablation} shows the results for the Iceland network; results for all other networks are presented in the Appendix.

The most striking finding is that the residual (skip) connection is the only architectural choice that substantially affects detection performance: removing it leads to a marked drop in accuracy, due to irregular and unstable training. This aligns with the original motivation for residual connections, which were introduced to stabilize training in deep networks by smoothing the loss landscape~\cite{He:2015,Li:2017}.

At the same time, most other architectural choices have little consequence for detection performance. Notably, reducing the number of hidden units to as few as 16 consistently falls within the margin of error of larger configurations across all networks. In addition to comparable accuracy, 16 hidden units yields substantially smoother training curves and less overfitting, as illustrated in Figure~\ref{fig:learning-curves}. On the grounds of parsimony, we therefore adopt 16 hidden units as the default for all subsequent experiments.

\subsection{Detectability of Source}
The detectability of the epidemic source has previously been examined by Shah and Zaman~\cite{Shah:2010}, who analyzed their rumor-source estimator on tree graphs. They showed theoretically that, for nontrivial graph structures and the SI model, the probability of correctly identifying the source does not converge to zero as time progresses. Instead, it approaches a strictly positive limit. Antulov-Fantulin et al.~\cite{Antulov-Fantulin:2015} studied detectability from a different angle by computing the normalized Shannon entropy of their source-probability estimates. Their results on lattice graphs suggest that higher infection probabilities, ceteris paribus, lead to greater detectability because outbreak patterns become more distinct across candidate sources, an effect that only manifests when the network is sufficiently large to avoid constraining the spread.

Some GNN-based related works touch on detectability as well, though in indirect ways. For instance, Dong et al.~\cite{Dong:2019} perform inference after 30\% of nodes have become infected, while Haddad and Figueiredo~\cite{Haddad:2023} do so after 20\%. The underlying idea is to choose an epidemic duration that yields non-trivial but not overly extensive outbreaks that may severely hamper detectability. Consistent with this intuition, we set $T$ such that, on average, roughly 40\% of nodes are infected, striking a balance between complexity of the outbreaks and detectability.

\begin{figure}[ht]
    \centering
    \begin{subfigure}[b]{0.32\textwidth}
         \centering
         \includegraphics[width=\textwidth]{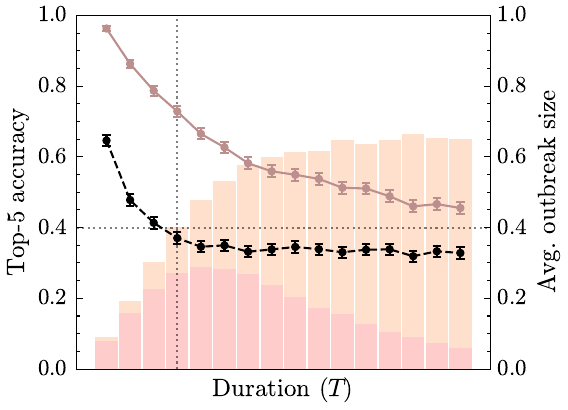}
         \caption{Karate}
         \label{fig:det-karate}
     \end{subfigure}
     \hfill
     \begin{subfigure}[b]{0.32\textwidth}
         \centering
         \includegraphics[width=\textwidth]{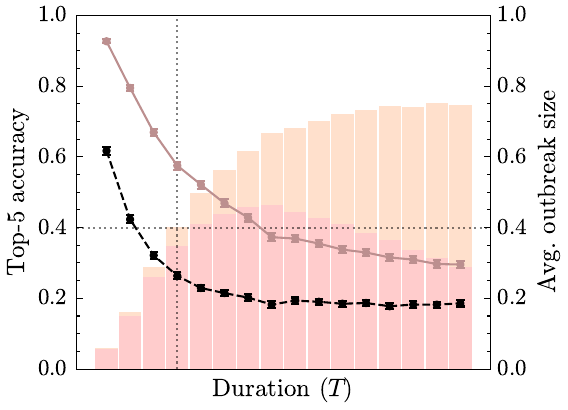}
         \caption{Iceland}
         \label{fig:det-iceland}
     \end{subfigure}
     \hfill
     \begin{subfigure}[b]{0.32\textwidth}
         \centering
         \includegraphics[width=\textwidth]{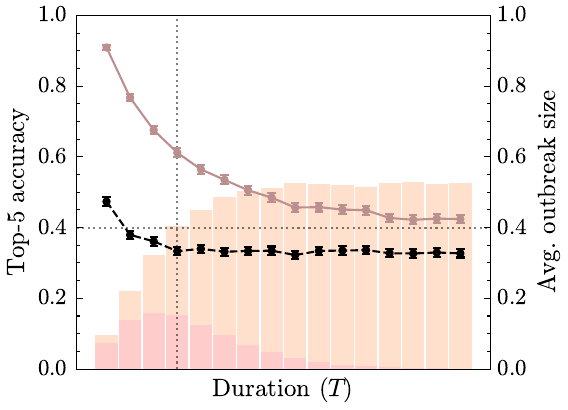}
         \caption{Dolphin}
         \label{fig:det-dolphin}
     \end{subfigure}
     \hfill
     \begin{subfigure}[b]{0.32\textwidth}
         \centering
         \includegraphics[width=\textwidth]{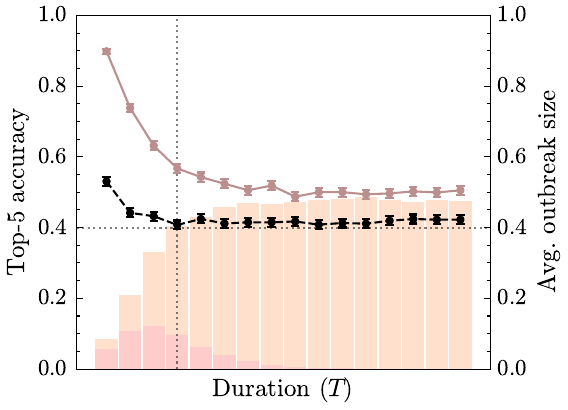}
         \caption{Fraternity}
         \label{fig:det-fraternity}
     \end{subfigure}
     \hfill
     \begin{subfigure}[b]{0.32\textwidth}
         \centering
         \includegraphics[width=\textwidth]{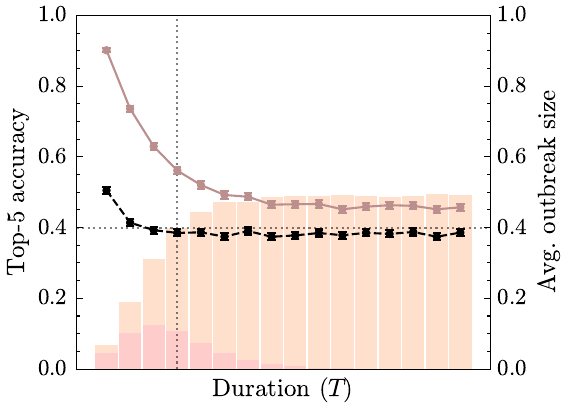}
         \caption{Workplace}
         \label{fig:det-workplace}
     \end{subfigure}
     \hfill
     \begin{subfigure}[b]{0.32\textwidth}
         \centering
         \includegraphics[width=\textwidth]{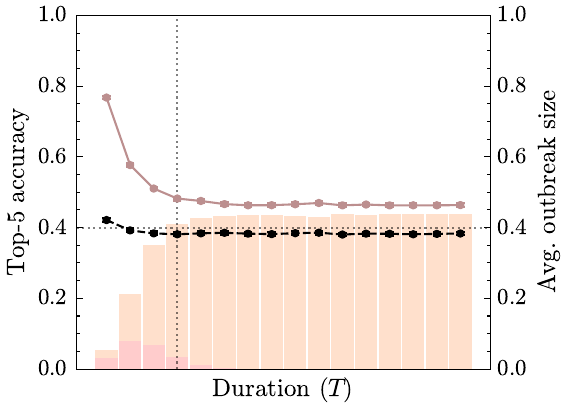}
         \caption{Highschool}
         \label{fig:det-hs2013}
     \end{subfigure}
     \hfill
    \caption{Source detection performance across outbreak durations $T$. Each plot reports top-5 accuracy for the GCN architecture~\cite{Shah:2020} and for a baseline that selects the source uniformly at random from the infected subgraph. Bars indicate the average proportion of infectious and recovered nodes at each duration (secondary y-axis). In each plot, the fourth bar (marked by dotted lines) corresponds to the value of $T$ at which approximately 40\% of all nodes are infected. Units of time are omitted, as the absolute values of durations are not relevant.}
    \label{fig:det}
\end{figure}

To complement prior work, we conduct experiments on our six empirical networks. Keeping $R_0$ fixed (as specified in Table~\ref{tab:networks}), we evaluate the detection performance of the GCN architecture~\cite{Shah:2020} and a simple baseline across increasing values of $T$ (Figure~\ref{fig:det}). The resulting outbreak dynamics differ markedly across the networks despite identical $R_0$ calibration. For example, in the relatively sparse Iceland network (Figure~\ref{fig:det-iceland}), achieving $R_0 \approx 2$ requires a high infection rate, causing outbreaks to propagate quickly and ultimately infect about 75\% of the nodes, on average.

In line with the second law of thermodynamics and prior empirical findings, we expect detectability to deteriorate as $T$ increases. Moreover, Antulov-Fantulin et al.~\cite{Antulov-Fantulin:2015} note that increasing $T$, combined with the modest size of our networks, reduces the distinctiveness of outbreak patterns, making sources harder to distinguish. This expectation is confirmed by the results in Figure~\ref{fig:det}: top-5 accuracy decreases with $T$ for both the GNN and the baseline. More importantly, the GNN consistently converges to a higher accuracy than the baseline across all networks, indicating that the GNN can still yield benefits even when inference is performed relatively late in the outbreak. Finally, the fact that the GNN’s accuracy does not collapse to the random baseline may reflect the theoretical result of Shah and Zaman~\cite{Shah:2010}, who showed that detectability remains bounded away from zero at later times.

\subsection{Training Data}
In this final subsection, we examine three aspects related to the generation of training data that affect not only GNN-based source detection but also MCMF and SME. First, we investigate how detection performance scales with the number of simulations per source node. To the best of our knowledge, no prior work has explored this question. Second, we relax the assumption that the exact time between the outbreak and the observation of the snapshot (denoted by~$T$) is known, and assess how uncertainty in $T$ influences detection performance. Finally, we examine how detection performance is affected by uncertainty in the infection rate $\beta$.

\subsubsection{Scaling the Training Data}
The size of the training (and validation) data, i.e., the number of simulated outbreak scenarios, has received relatively little attention in related work on GNN-based source detection and has often been treated as a negligible hyperparameter. In several studies, the training set size is not reported at all~\cite{Dong:2019,Haddad:2023}, while others adopt widely varying dataset sizes, ranging from $1{,}000$~\cite{Shu:2021} to $20{,}000$~\cite{Shah:2020,Ru:2023} simulated scenarios. Typically, these works further withhold a subset of the simulations for validation and testing. Crucially, however, all prior studies appear to choose the training set size independently of the underlying graph size. In this paper, we argue that training data size should be determined on a per-node basis. Throughout most of our analysis, we therefore use a training/validation set size of 500 simulated scenarios per (source) node. For example, a graph with 50 nodes leads to a total of $25{,}000$ training (and validation) samples, already exceeding the dataset sizes commonly used in related work.

\begin{figure}[ht]
    \centering
    \begin{subfigure}[b]{0.32\textwidth}
         \centering
         \includegraphics[width=\textwidth]{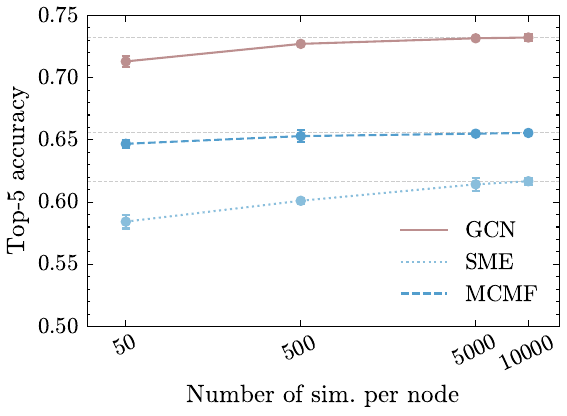}
         \caption{Karate}
         \label{fig:scaling-karate}
     \end{subfigure}
     \hfill
     \begin{subfigure}[b]{0.32\textwidth}
         \centering
         \includegraphics[width=\textwidth]{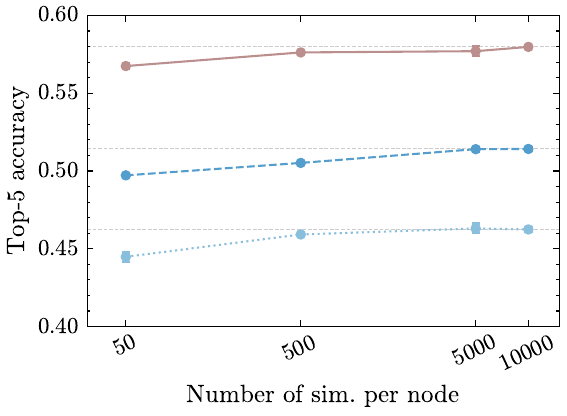}
         \caption{Iceland}
         \label{fig:scaling-iceland}
     \end{subfigure}
     \hfill
     \begin{subfigure}[b]{0.32\textwidth}
         \centering
         \includegraphics[width=\textwidth]{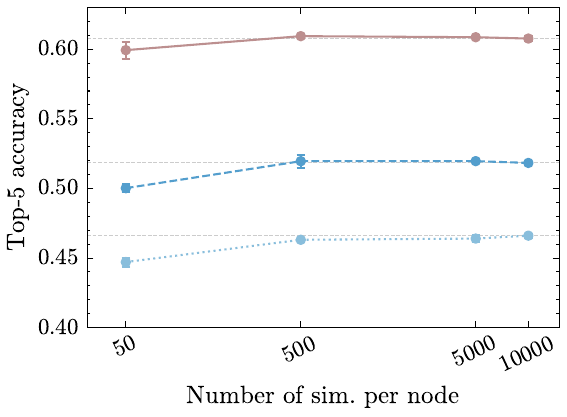}
         \caption{Dolphin}
         \label{fig:scaling-dolphin}
     \end{subfigure}
     \hfill
     \begin{subfigure}[b]{0.32\textwidth}
         \centering
         \includegraphics[width=\textwidth]{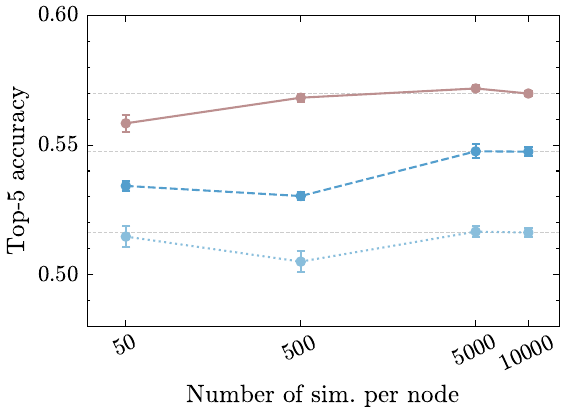}
         \caption{Fraternity}
         \label{fig:scaling-fraternity}
     \end{subfigure}
     \hfill
     \begin{subfigure}[b]{0.32\textwidth}
         \centering
         \includegraphics[width=\textwidth]{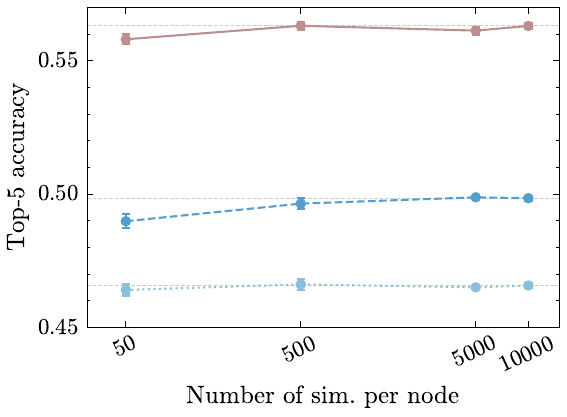}
         \caption{Workplace}
         \label{fig:scaling-workplace}
     \end{subfigure}
     \hfill
     \begin{subfigure}[b]{0.32\textwidth}
         \centering
         \includegraphics[width=\textwidth]{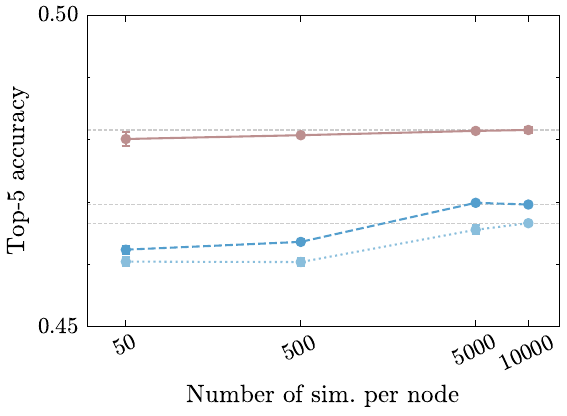}
         \caption{Highschool}
         \label{fig:scaling-highschool}
     \end{subfigure}
     \hfill
    \caption{Detection performance, measured by top-5 accuracy, across all six networks for the three methods GCN~\cite{Shah:2020}, SME, and MCMF. All performance measures are averaged over three runs, each using a different random seed. Results are shown for four sample sizes, corresponding to the number of simulations per node. Horizontal dashed lines indicate the performance achieved with the largest sample size, acting as a visual guide. Error bars denote 95\% confidence intervals, reflecting uncertainty across runs.}
    \label{fig:scaling}
\end{figure}

Figure~\ref{fig:scaling} reports the top-5 accuracy across the six networks as the number of simulations per node increases by orders of magnitude (followed by a final doubling) starting from 50 simulations per node. Overall, detection performance improves moderately with sample size. For the GCN, the most substantial gains are observed when increasing the number of simulations from 50 to 500 per node, with only marginal improvements beyond that point, suggesting that the associated computational cost cannot be justified. For MCMF and SME, further improvements appear possible going from 500 to 5{,}000 simulations per node, though the patterns are less consistent, particularly for SME on the Fraternity network. In principle, however, SME should (at least for SI dynamics) converge to the true likelihood of each candidate source node as the number of simulations grows large, a regime not reached in our experiments because we were not able to choose $n$ large enough and because we consider SIR dynamics.

\subsubsection{Uncertainty in Time until Snapshot Observation}
Another factor that receives little attention in related work is the assumption that the starting time of the outbreak ($t_0$), or, equivalently, the time between the outbreak and the observed snapshot ($T$), is known. This assumption allows us to simulate training data for the same period of time as for the test outbreaks. In practice, however, it is unrealistic: typically only rough estimates of how long a disease has been spreading are available.

A notable exception is the work by Ru et al.~\cite{Ru:2023}, who investigate uncertainty in the knowledge of $t_0$. Their method requires specifying an interval within which the outbreak is assumed to have started, and simulations are generated using a uniformly sampled starting time from this interval. However, their study does not quantify the effect of relaxing the known–$t_0$ assumption on model performance.

In our analysis, we relax this assumption as follows: for each training and validation instance, we sample a value of $T$ uniformly from the interval $(0, 4\,T^*]$, where $T^*$ is the true elapsed time used when generating test instances. This setup mimics a scenario in which knowledge of $T$ and $t_0$ is noisy at best. Moreover, it even introduces systematic bias as the sampled values of $T$ have mean $2T^*$. Note that this implicitly introduces a form of model misspecification: training outbreaks are, on average, observed at a later stage than test outbreaks, producing larger infection patterns similar to those that would arise under a higher infection rate $\beta$.

\begin{figure}[ht]
    \centering
    \begin{subfigure}[b]{0.32\textwidth}
         \centering
         \includegraphics[width=\textwidth]{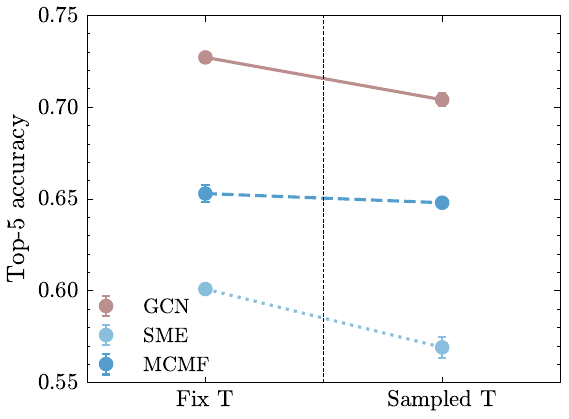}
         \caption{Karate}
         \label{fig:diverse-karate}
     \end{subfigure}
     \hfill
     \begin{subfigure}[b]{0.32\textwidth}
         \centering
         \includegraphics[width=\textwidth]{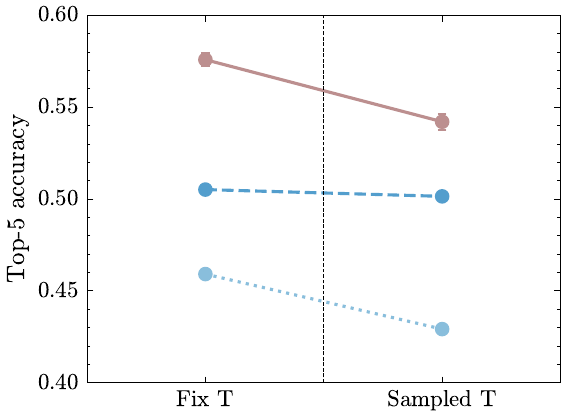}
         \caption{Iceland}
         \label{fig:diverse-iceland}
     \end{subfigure}
     \hfill
     \begin{subfigure}[b]{0.32\textwidth}
         \centering
         \includegraphics[width=\textwidth]{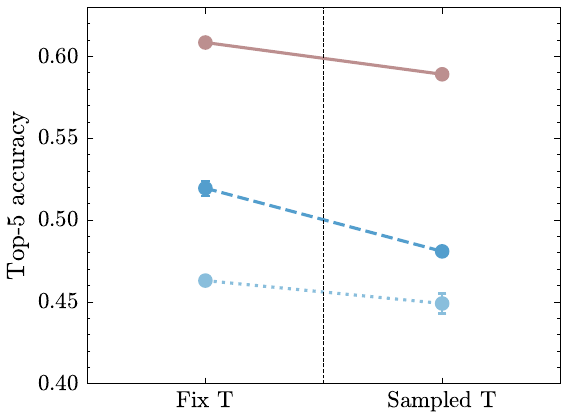}
         \caption{Dolphin}
         \label{fig:diverse-dolphin}
     \end{subfigure}
     \hfill
     \begin{subfigure}[b]{0.32\textwidth}
         \centering
         \includegraphics[width=\textwidth]{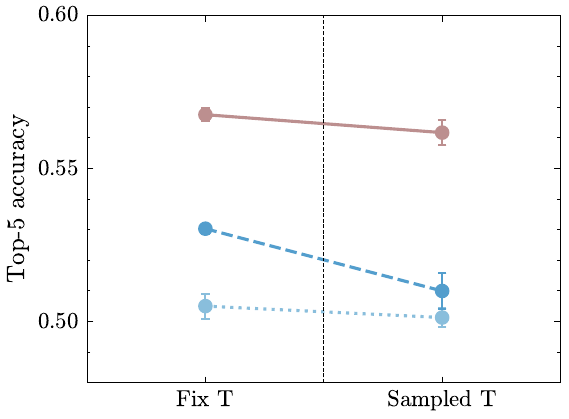}
         \caption{Fraternity}
         \label{fig:diverse-fraternity}
     \end{subfigure}
     \hfill
     \begin{subfigure}[b]{0.32\textwidth}
         \centering
         \includegraphics[width=\textwidth]{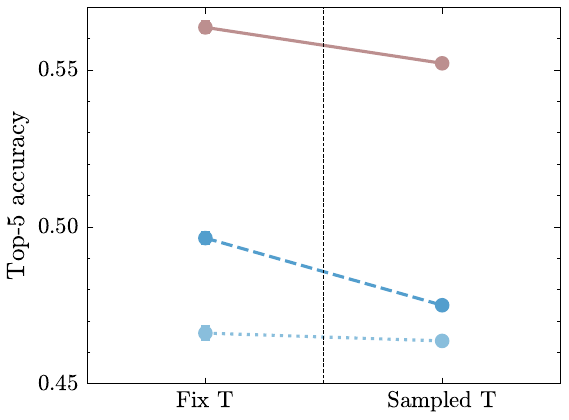}
         \caption{Workplace}
         \label{fig:diverse-workplace}
     \end{subfigure}
     \hfill
     \begin{subfigure}[b]{0.32\textwidth}
         \centering
         \includegraphics[width=\textwidth]{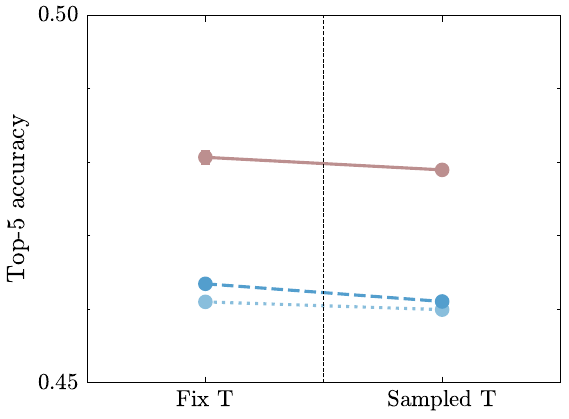}
         \caption{High School}
         \label{fig:diverse-highschool}
     \end{subfigure}
     \hfill
    \caption{Detection performance, measured by top-5 accuracy, across all six networks and for the three methods GCN~\cite{Shah:2020}, SME, and MCMF. All performance measures are averaged over three runs, each using a different random seed. Performance is shown for a fix, known duration $T$ until the snapshot is observed (left) and an uncertain, sampled duration $T$ until the snapshot is observed (right). Error bars represent 95\% confidence intervals, reflecting uncertainty across runs.}
    \label{fig:diverse}
\end{figure}

Figure~\ref{fig:diverse} presents the results. As expected, this uncertainty and bias in $T$ leads to reduced performance across all networks and nearly all methods. On denser networks, such as Fraternity and Highschool, the performance drop for GCN and SME appears negligible. Notably, the qualitative ranking of methods persists and GCN performance remains above benchmark levels even under this relaxation.

\subsubsection{Uncertainty in Infection Rate}

To make the uncertainty in SIR model parameters more explicit, we present results for when the training data are generated with infection rates $\beta$ sampled from a log-normal distribution, while keeping the recovery rate fixed at 1. We set the location parameter to $\mu=\ln\beta$, so that the median of the sampling distribution equals the true infection rate $\beta$, and vary the scale parameter $\sigma \in \{0.0, 0.25, 0.5, 0.75\}$. This represents a different form of model misspecification than in the previous subsection: rather than introducing systematic bias, increasing $\sigma$ introduces variance around the true $\beta$. As a concrete example, for the Karate network with $\beta=1.3$ and $\sigma=0.25$ the 95\% interval of the sampling distribution spans $\beta\cdot \exp(\pm 1.96\sigma) = [0.80, 2.12]$.

\begin{figure}[ht]
    \centering
    \begin{subfigure}[b]{0.32\textwidth}
         \centering
         \includegraphics[width=\textwidth]{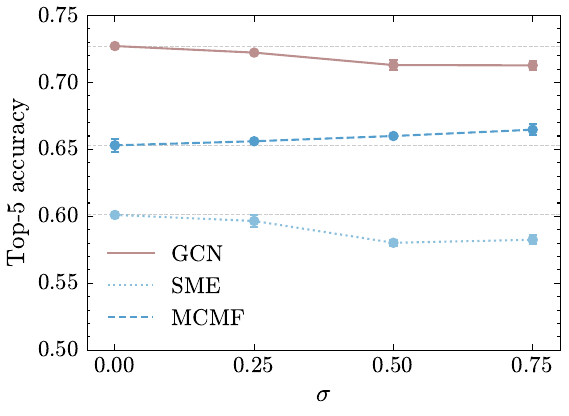}
         \caption{Karate}
         \label{fig:sigmas-karate}
     \end{subfigure}
     \hfill
     \begin{subfigure}[b]{0.32\textwidth}
         \centering
         \includegraphics[width=\textwidth]{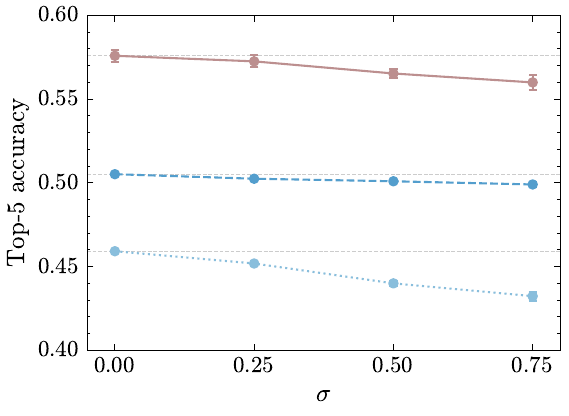}
         \caption{Iceland}
         \label{fig:sigmas-iceland}
     \end{subfigure}
     \hfill
     \begin{subfigure}[b]{0.32\textwidth}
         \centering
         \includegraphics[width=\textwidth]{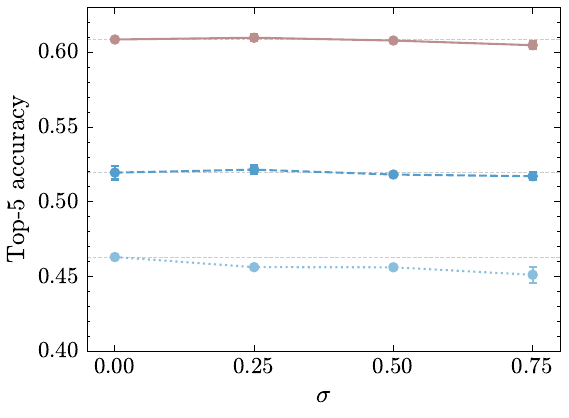}
         \caption{Dolphin}
         \label{fig:sigmas-dolphin}
     \end{subfigure}
     \hfill
     \begin{subfigure}[b]{0.32\textwidth}
         \centering
         \includegraphics[width=\textwidth]{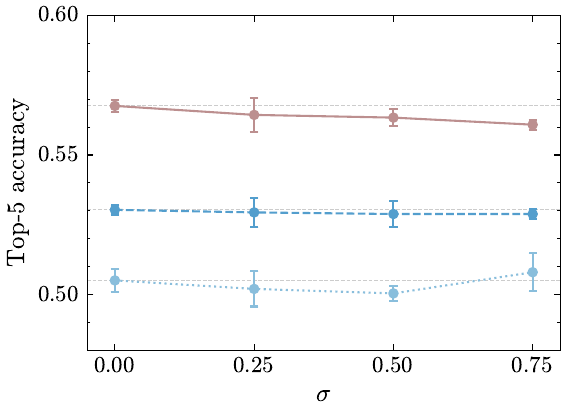}
         \caption{Fraternity}
         \label{fig:sigmas-fraternity}
     \end{subfigure}
     \hfill
     \begin{subfigure}[b]{0.32\textwidth}
         \centering
         \includegraphics[width=\textwidth]{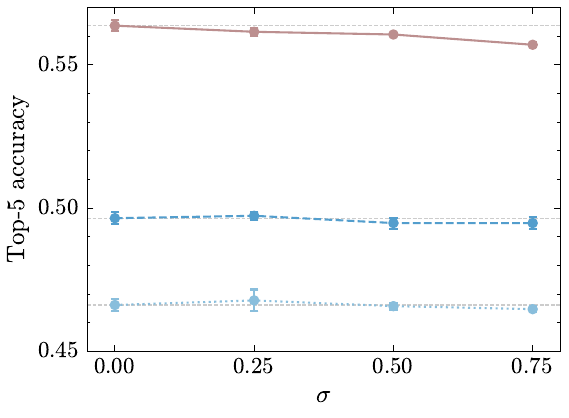}
         \caption{Workplace}
         \label{fig:sigmas-workplace}
     \end{subfigure}
     \hfill
     \begin{subfigure}[b]{0.32\textwidth}
         \centering
         \includegraphics[width=\textwidth]{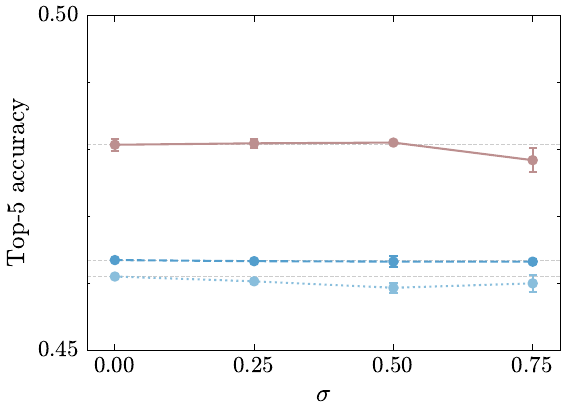}
         \caption{High School}
         \label{fig:sigmas-highschool}
     \end{subfigure}
     \hfill
    \caption{Detection performance, measured by top-5 accuracy, across all six networks and for the three methods GCN~\cite{Shah:2020}, SME, and MCMF. All performance measures are averaged over three runs, each using a different random seed. Results are shown for four values of $\sigma$, where $\sigma$ is the standard deviation of the underlying normal distribution. Horizontal dashed lines indicate the baseline performance at $\sigma=0$ as a visual guide. Error bars represent 95\% confidence intervals, reflecting uncertainty across runs.}
    \label{fig:sigma}
\end{figure}

Figure~\ref{fig:sigma} presents the results. As expected, detection performance generally deteriorates with increasing $\sigma$, although the drops tend to be modest. This is consistent with the fact that the sampling distribution remains centered on the true $\beta$ used for testing. Some results are nonetheless counterintuitive: for instance, MCMF performance on the Karate network improves with increasing $\sigma$, the reasons for which are unclear to us.

\section{Computational Complexity}\label{sec:complexity}

The time complexity of source detection methods can be divided into three main components: (i)~the generation of training data through simulation, (ii)~the model training (only applicable to GNNs), and (iii)~inference, i.e., applying one of the methods to detect the source.

\subsection{Simulation of Training Data}  
This step is required by GNNs, SME, and MCMF, all of which rely on a set of simulated outbreak scenarios. In our paper, we employ the static-network variant of the fast (event-driven) simulation algorithm proposed by Holme~\cite{Holme:2021} (see Appendix). Its worst-case complexity for generating one simulation is $\mathcal{O}(N^2 \log N)$: $\mathcal{O}(N \log N)$ arises from operating the priority queue for the algorithm, while an additional factor of $N$ accounts for the worst-case cost of iterating over all neighbors of each node in the queue.

\subsection{Model Training}  
As we use a convolutional message-passing architecture, we adapt the complexity analysis of Wu et al.~\cite{Wu:2021}. The cost of a full forward pass over all $L$ layers is $\mathcal{O}\left(L \cdot (Es + Ns^2)\right)$, where $s$ denotes the embedding dimension. The first term in the sum captures the neighborhood aggregation where the typically sparse adjacency matrix (representing the $E$ edges) is multiplied with the node feature matrix of dimensions $N \times s$, while the second term accounts for the linear transformation in the update equation (see equation~\ref{eq:updateShah}), where the $N \times s$ aggregation matrix is multiplied by an $s \times s$ weight matrix. 

Following Bajaj et al.~\cite{Bajaj:2025}, we approximate the cost of a combined forward and backward pass as a multiplicative factor $(1 + c)$ of the forward-pass cost. The total training cost must be further scaled by the batch size $B$ (since each pass processes $B$ identical graphs), the number of batches, and the number of training epochs.

\subsection{Inference}  
The inference complexity for a single outbreak snapshot varies by method. For the GCN architecture, inference consists of a single forward pass, as described above. For MCMF, inference requires summing (logarithmic) node state probabilities over all nodes for each potential source node, yielding a complexity of order $\mathcal{O}(N^2)$. This, however, assumes that these node state probabilities are already computed: obtaining them incurs an additional (one-time) cost of order $\mathcal{O}(N^2)$. For SME, the bottleneck lies in computing the Jaccard distance between the observed outbreak and each simulated outbreak, resulting in a complexity of order $\mathcal{O}(n N)$ (see Supplemental Material of~\cite{Antulov-Fantulin:2015}).

\subsection{Empirical Results}
Table~\ref{tab:runtimes} reports the average run times for the outbreak simulations, GCN training, and inference for the GCN, MCMF, and SME. The empirical cost of generating the simulations is negligible across all networks. As expected, these run times generally increase with the network size $N$, with the Highschool network ($N=327$) exhibiting the largest value. Notably, the Fraternity network ($N=58$) is an exception to this trend: despite being one of the smaller networks, it incurs run times comparable to the larger Workplace network ($N=92$), likely due to its high density.

\begin{table*}
    \caption{Average run times in seconds. For the simulations, the reported values correspond to the average run time per simulation. GCN training times are reported on a per-batch basis. Inference times represent the average time per test instance. For reference, all experiments were conducted on a machine equipped with an AMD Ryzen Threadripper PRO (24 cores), 128 GB of RAM, and a single RTX 5080 GPU with 16 GB of VRAM.}
    \centering
    \label{tab:runtimes}
    \begin{tabular}{l|c|cc|c|c}
        \toprule
        \textbf{Network} 
        & Simulations
        & \multicolumn{2}{c|}{GCN~\cite{Shah:2020}}
        & SME 
        & MCMF \\
        \cmidrule(lr){2-6}
        & 
        & Training 
        & Inference 
        & Inference
        & Inference \\
        \midrule
        Karate & $1.40\mathrm{e}{-06}$ & 0.0048 & 0.0014 & 0.0022 & 0.0013 \\
        Iceland & $2.54\mathrm{e}{-06}$ & 0.0046 & 0.0013 & 0.0033 & 0.0013 \\
        Dolphin & $2.53\mathrm{e}{-06}$ & 0.0048 & 0.0014 & 0.0029 & 0.0014 \\
        Fraternity & $5.23\mathrm{e}{-06}$ & 0.0050 & 0.0018 & 0.0031 & 0.0017 \\
        Workplace & $5.71\mathrm{e}{-06}$ & 0.0050 & 0.0017 & 0.0040 & 0.0017 \\
        Highschool & $3.24\mathrm{e}{-05}$ & 0.0199 & 0.0041 & 0.0186 & 0.0059 \\
        \bottomrule
    \end{tabular}
\end{table*}

Although all three methods incur the cost of generating simulated training data, only the GCN requires an additional, typically substantial, computational cost for model training. Table~\ref{tab:runtimes} shows that the average per-batch training time is practically identical for all networks but Highschool, which exhibits a cost roughly one order of magnitude higher. This suggests that for small graphs the training cost is dominated by a fixed overhead, and that the theoretically predicted scaling with $E$ and $N$ only becomes visible at larger network sizes.

Finally, the theoretical analysis above indicates that the dominant inference cost for SME and MCMF scales as $\mathcal{O}(nN)$ and $\mathcal{O}(N^2)$, respectively. Since $n > N$ for all networks in our experimental setting, SME incurs a substantially higher empirical inference cost than MCMF, with the gap widening for larger networks: the SME/MCMF inference ratio grows from approximately 1.7 for Karate to 3.2 for Highschool. The GCN's dominant cost term depends on the sparsity of the graph, more concretely whether $E \gg Ns$ or not. Table~\ref{tab:runtimes} shows that the GCN's inference cost is closely comparable to MCMF for small networks, but for the Highschool network the GCN is roughly 30\% faster than MCMF, suggesting it may become the most efficient inference method for larger, sparser networks.

\section{Real-World Case}\label{sec:real-world}
As a real-world demonstration, we consider the global spread of the 2009 H1N1 (``swine flu'') pandemic. It is well established that the outbreak originated in Mexico~\cite{Brockmann:2013}. The observed snapshot can be created by considering the arrival times of the virus in 91 countries obtained from Table S4 of Brockmann and Helbing's paper~\cite{Brockmann:2013}. We re-centered these reported arrival times such that Mexico has arrival time zero, and all other countries’ values represent the number of weeks after initial detection in Mexico.

\subsection{Network}
To construct a data-driven mobility network, we used the global airline transportation network derived from OpenFlights data (\url{https://openflights.org/data}) from 2014. Although data for 2009 were unavailable, we assume that large-scale global airline connectivity patterns did not change in ways that would invalidate the present analysis. We aggregated the airline data at the country level, yielding a network in which nodes correspond to countries. Following~\cite{Brockmann:2013}, we treated all connections as undirected and defined the weight of an edge as the total number of airline connections between the corresponding pair of countries. The strongest connection in the network is between Spain and the United Kingdom, with 1030 daily flights. The resulting network is undirected and weighted, comprising 225 nodes and 2317 edges. It consists of a single connected component, with an average degree of 20.60, a diameter of 5, an average shortest-path length of 2.33, and an average clustering coefficient of 0.65 (Figure~\ref{fig:airline-network}).

\subsection{Epidemic Model}
Instead of employing a metapopulation framework as is common in such use cases~\cite{Colizza:2006,Brockmann:2013}, we generated synthetic training data using a simple SI model with no recovery. The base infection rate was set to 1 and was, for each edge, adjusted proportional to the edge weight. For each potential source node, we simulated 500 outbreak realizations and the duration of the simulated outbreaks was chosen such that approximately 45 countries (roughly half of the 92 countries reported to have been affected during the actual pandemic) were infected, on average.

\subsection{Results}
\begin{figure}[ht]
  \centering
    \begin{subfigure}[b]{0.95\textwidth}
         \centering
         \includegraphics[width=\textwidth]{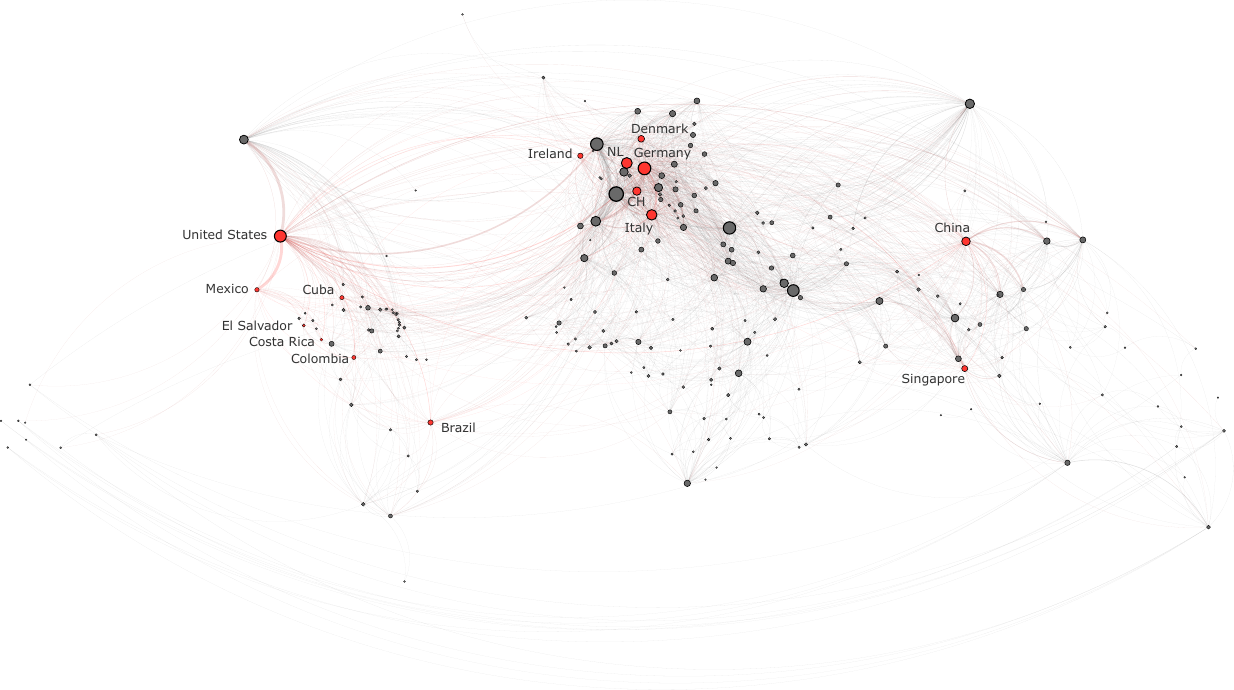}
         \caption{Global airline network.}
         \label{fig:airline-network}
     \end{subfigure}
     \hfill
     \begin{subfigure}[b]{0.6\textwidth}
         \centering
         \includegraphics[width=\textwidth]{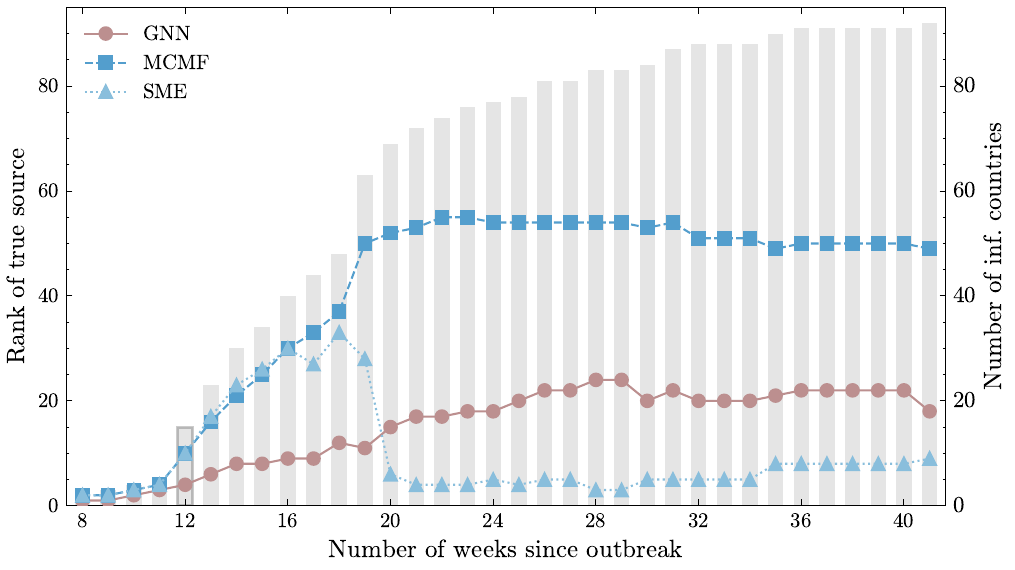}
         \caption{Rank of true source over time.}
         \label{fig:ranks}
     \end{subfigure}
     \hfill
     \begin{subfigure}[b]{0.38\textwidth}
         \centering
         \includegraphics[width=\textwidth]{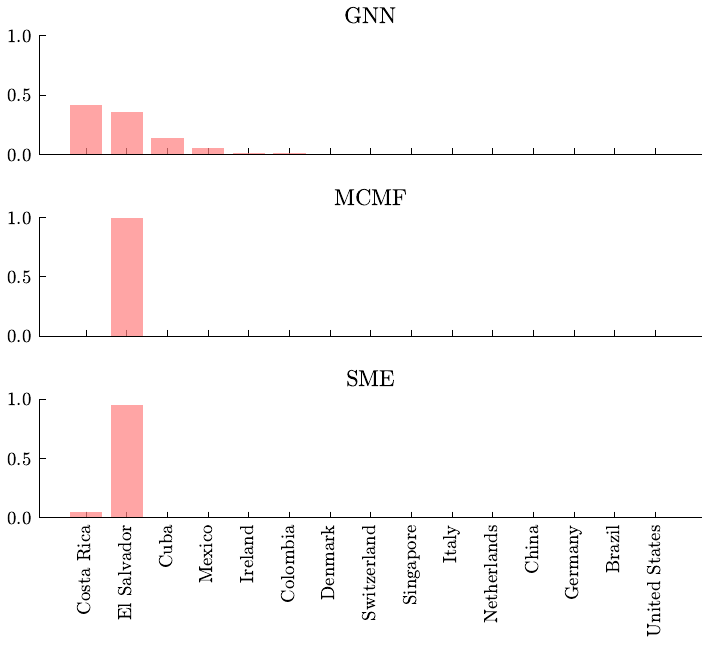}
         \caption{Source distributions after 12 weeks.}
         \label{fig:distr-real}
     \end{subfigure}
     \hfill
  \caption{Source detection results for the 2009 H1N1 (swine flu) pandemic. (a)~Global airline transportation network (in 2014), aggregated at the country level. Mexico is the known origin of the outbreak. The 15 countries shown in red (and labeled) correspond to those reported as infected 12 weeks after the initial detection in Mexico. (b)~Rank of the true source (Mexico) over the course of the pandemic for the three source-detection methods considered. Bars (right y-axis) denote the number of infected countries at each point in time. (c)~Source probability distributions inferred by the GNN, MCMF, and SME at week 12, corresponding to the set of 15 infected countries highlighted in panel (a). Countries along the x-axis are sorted in decreasing order of GNN source probabilities.}
  \label{fig:h1n1}
\end{figure}

We trained the GCN architecture of Shah et al.~\cite{Shah:2020} as outlined in Section~\ref{sec:setting}, but with an embedding dimension of 16 instead of 128, as this proved sufficient in the ablation study. After training, the GNN was applied to epidemic snapshots at multiple stages of the spreading process. For comparison, we also computed source probability distributions using SME and MCMF, applied to the same simulation data as the GNN. The results are presented in Figure~\ref{fig:h1n1}.

Figure~\ref{fig:ranks} shows the inferred rank of Mexico in the source distribution over time. The GNN assigns Mexico lower ranks than the other two methods for up to 19 weeks after the outbreak, after which SME achieves a better performance. Figure~\ref{fig:distr-real} displays the full source distributions at week 12. Across all three methods, probability mass concentrates on a small set of countries: Costa Rica, El Salvador, Mexico, and Cuba, with SME and MCMF concentrating almost all probability mass on El Salvador. Inspection of the airline network (Figure~\ref{fig:airline-network}) suggests two plausible source clusters, one in Central America and the other in Western Europe. Notably, all three methods assign overwhelmingly higher probability to the Central American cluster.

\section{Conclusions}\label{conclusion}

This study set out (i) to review existing work on the use of GNNs for epidemic source detection and (ii) to assess their effectiveness for this task. Our analysis highlights some gaps in the related literature. In particular, comparisons with classical methods are sometimes incomplete or not straightforward to interpret. Against this backdrop, our empirical results demonstrate that GNNs perform remarkably well at detecting the source of an epidemic across a diverse set of network topologies. In all considered settings, the main contenders from traditional work on source detection, MCMF and SME, are clearly outperformed. For MCMF, the underlying independence assumption results in source distributions that fail to adequately capture uncertainty across candidate sources. For SME, achieving competitive performance requires a prohibitively large number of simulated outbreaks, rendering the method computationally inefficient. A~simple MLP architecture often outperforms these traditional methods and in some cases comes surprisingly close to the GNN performance. Finally, our results indicate that performance is not overly sensitive to specific architectural choices with the notable exception of residual connections, which substantially stabilize training.

We hope that the present work provides useful guidance for future research on epidemic source detection. First, we caution against describing GNN-based source detection methods as (spreading-) model-agnostic. Instead, we advocate for further systematic investigations of robustness to model misspecification in order to meaningfully assess sensitivity to incorrect spreading dynamics or parameters. Second, our results suggest that the number of simulated outbreaks should be determined on a per-node basis rather than globally, as this choice has important implications for both performance and computational cost and should take the graph size into account. Finally, we argue that epidemic source detection constitutes an interesting benchmark task for evaluating GNN architectures. In its single-source formulation, the problem naturally corresponds to a multi-class graph prediction task. An attractive feature of this task is that training and test data can be generated synthetically and efficiently for common spreading models such as SI, SIR, or SEIR, allowing precise control over data scale and structure. While we release code and network data, future work will focus on providing additional resources and evaluation protocols to further establish epidemic source detection as a benchmark task.

Several directions for future research remain open. First, a more comprehensive understanding of how training set size affects GNN performance, as well as that of competing methods, is still needed. Recent work on scaling laws, largely motivated by the success of large language models~\cite{Kaplan:2020}, suggests that jointly increasing training data and model capacity can lead to substantial performance improvements. While our study explores the effects of scaling the training data alone, we leave a systematic investigation of simultaneous architectural scaling for future work. Second, source detection performance should be studied as a function of graph size and structural properties, potentially using synthetic graph models to disentangle these effects. Finally, the more challenging multi-source detection problem remains underexplored, with open questions ranging from the inference of the number of sources to the development of appropriate evaluation metrics.

While the present study is primarily academic in nature, our application to a real-world case study yields promising and practically relevant results. We therefore view this work as a step toward making GNN-based epidemic source detection methods applicable in real-world settings and intend to pursue this direction in future research.

\bibliographystyle{unsrt}  
\bibliography{references}

\appendix

\renewcommand{\thefigure}{A.\arabic{figure}}
\renewcommand{\theequation}{A.\arabic{equation}}
\setcounter{figure}{0}
\setcounter{equation}{0}

\section{Benchmark Methods}

\subsection{Monte Carlo Simulations of SIR on Static Networks}
We employ a fast, event-driven simulation method for static networks in continuous time, implemented in C by Holme~\cite{Holme:2018}. An extension of this algorithm to temporal networks is described in detail in a paper~\cite{Holme:2021}.

The key idea of the algorithm is to manage infection events using a priority queue (implemented in practice as a binary heap). The core simulation loop proceeds by removing the node $v$ associated with the earliest scheduled event from the priority queue and sampling its recovery time. Subsequently, for each neighbor of $v$, a potential infection time is sampled. Recovery and infection times are drawn from exponential distributions with rates $\mu$ and $\beta$, respectively. If a sampled infection time occurs before the recovery of $v$ and before any previously scheduled infection event involving the neighbor, the corresponding infection event is inserted into the priority queue. The queue is then reordered, and the algorithm continues by processing the next event at the top of the queue.

\subsection{Soft Margin Estimator (SME)}
The Soft Margin Estimator (SME) was introduced by Antulov-Fantulin et al.~\cite{Antulov-Fantulin:2015} in 2015. Its core idea is to compare the $n$ simulated outbreaks $r_{q,i}$ generated from a candidate source $q$ with the observed snapshot, denoted here by $r_*$. The closer the simulated outbreaks are to the observed one, the higher the likelihood of the corresponding source $q$.

Antulov-Fantulin et al.\ propose measuring similarity between the observed and simulated outbreaks using the Jaccard similarity $\varphi (r_*, r_{q,i})$. Treating $\varphi (r_*, r_{q,i})$ as a random variable, its empirical probability density function is given by
\begin{equation}
\label{eq:empdens}
\hat{f}_q(x) = \frac{1}{n} \sum_{i=1}^n \delta\!\left(x - \varphi(r_*, r_{q,i})\right),
\end{equation}
where $\delta(\cdot)$ denotes the Dirac delta function. Using this empirical density together with a Gaussian weighting function $w_a(x) = \exp\!\left(-(x-1)^2/a^2\right)$, the likelihood is estimated as
\begin{equation}
\label{eq:smelik}
\begin{split}
\hat{P}(r_* \mid q)
&= \int_0^1 w_a(x)\, \hat{f}_q(x)\, dx \\
&= \int_0^1 w_a(x)\, \frac{1}{n} \sum_{i=1}^n \delta\!\left(x - \varphi(r_*, r_{q,i})\right)\, dx \\
&= \frac{1}{n} \sum_{i=1}^n \int_0^1 w_a(x)\, \delta\!\left(x - \varphi(r_*, r_{q,i})\right)\, dx \\
&= \frac{1}{n} \sum_{i=1}^n w_a\!\left(\varphi(r_*, r_{q,i})\right) \\
&= \frac{1}{n} \sum_{i=1}^n \exp\!\left(-(\varphi(r_*, r_{q,i}) - 1)^2/a^2\right).
\end{split}
\end{equation}

The convergence parameter $a$ controls the extent to which the estimator approaches direct Monte Carlo estimation, where only exact matches between $r_*$ and $r_{q,i}$ contribute to the likelihood. In the limit $a \to 0$, the weighting function satisfies $w_{a \to 0}(x) = 0$ for $x \neq 1$ and $w_{a \to 0}(1) = 1$. Following the procedure described in the Supplemental Material of~\cite{Antulov-Fantulin:2015}, we determine the optimal value of $a$ as the smallest value that satisfies the convergence criterion
\begin{equation}
\label{eq:converg}
\left|\hat{P}_a^{n}(q_{\mathrm{MAP}} \mid r_*) - \hat{P}_a^{2n}(q_{\mathrm{MAP}} \mid r_*)\right| \leq 0.05.
\end{equation}
Here, only the most probable source candidate $q_{\mathrm{MAP}}$ is considered when assessing convergence. The quantity $\hat{P}_a^{2n}(q_{\mathrm{MAP}} \mid r_*)$ is estimated via bootstrap resampling of the $n$ simulated outbreaks.

Finally, we note that for the SIR model, SME based on Jaccard similarity does not distinguish between infectious and recovered nodes. This limitation may place it at a disadvantage relative to methods that fully exploit the available node state information.

\subsection{Monte Carlo Mean Field (MCMF)}
The Monte Carlo Mean Field (MCMF) method builds on prior work by Sterchi et al.~\cite{Sterchi:2023} and proceeds in two stages. In the first stage, we estimate conditional node-state probabilities $P(X_v(t_1)\mid q)$ for every node $v$ and each candidate source node $q$ at the inference time $t_1$ using the Monte Carlo simulations. For the SIR model, these probabilities are obtained by computing, for each source $q$, the fraction of simulation runs in which node $v$ is in the susceptible, infectious, or recovered state at time $t_1$.

In the second stage, we compute the likelihood of each candidate source $q$ given the observed node states $E_{t_1}$ (which, for SME, we denoted as $r_*$). To make this computation tractable, we adopt a strong mean-field-like assumption of conditional independence across nodes, which yields the factorized likelihood
\begin{equation}
\label{eq:meanfield}
\hat{P}(E_{t_1}\mid q) = \prod_{v,\, x_v(t_1)\in E_{t_1}} P\!\left(X_v(t_1)=x_v(t_1)\mid q\right).
\end{equation}
In practice, we evaluate the log-likelihood, thereby converting the product of node-state probabilities into a sum of their logarithms. When a normalized posterior over source nodes is required, we employ the standard log-sum-exp trick for numerical stability.

Finally, when the number of Monte Carlo simulations is small, some estimated node-state probabilities may be zero for a given source $q$, potentially resulting in a zero likelihood even if the source is a plausible candidate. To mitigate this issue, we apply add-one smoothing to the estimated probabilities, while enforcing that a source node $q$ cannot have a nonzero probability of being susceptible at the inference time.

\subsection{Multi Layer Perceptron (MLP)}
We consider two MLP variants, MLP$_{\text{node}}$ and MLP$_{\text{snapshot}}$, which differ in how the initial node feature matrix $\mathbf{H}^{(0)}$ is processed. 

\begin{figure}[ht]
  \centering
  \includegraphics[width=0.8\linewidth]{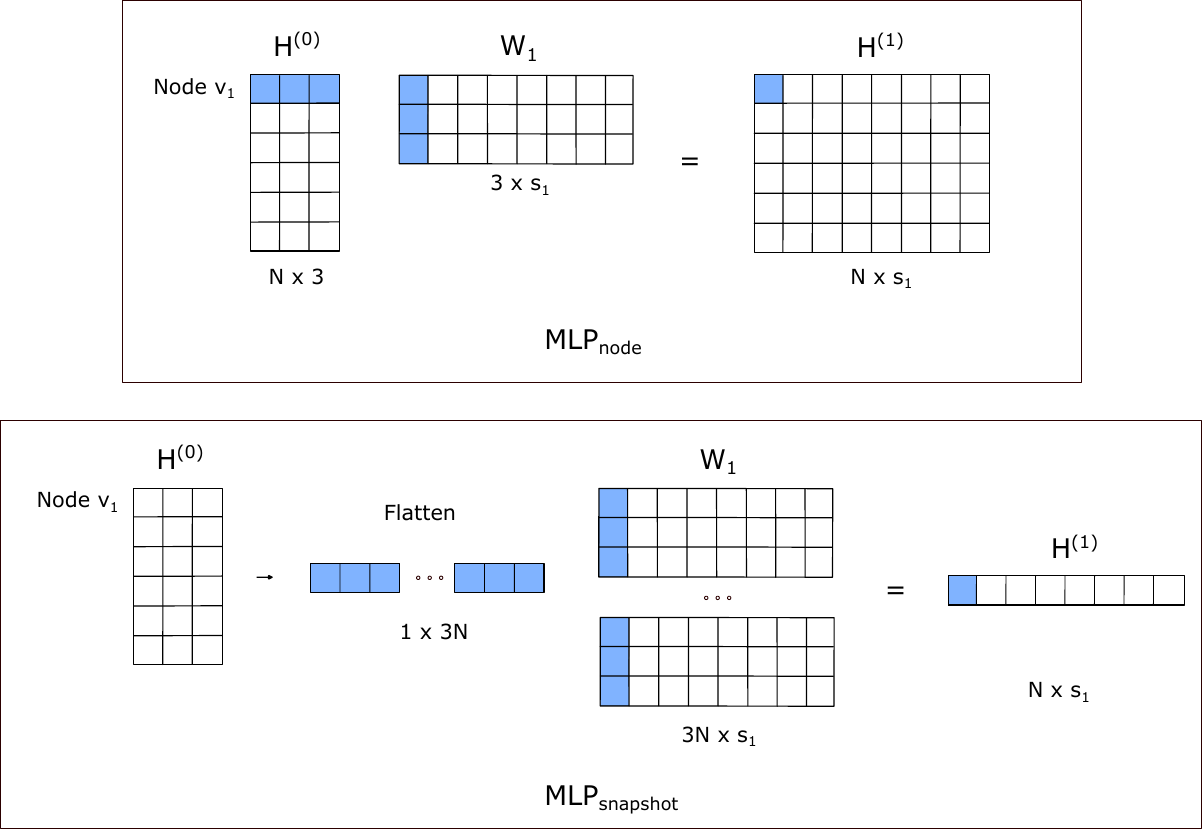}
  \caption{Illustration of the first-layer computation of MLP$_{\text{node}}$ (top) and MLP$_{\text{snapshot}}$ (bottom), illustrated for $N=6$ nodes and embedding dimensionality $s_1=8$. The key distinction is the scope of the input: MLP$_{\text{node}}$ operates on individual node feature vectors, while MLP$_{\text{snapshot}}$ takes the full snapshot as a single input.}
  \label{fig:mlps}
\end{figure}

MLP$_{\text{node}}$ transforms the one-hot encoded node states independently for each node (Figure~\ref{fig:mlps}, top). Since no message-passing is involved, nodes sharing the same epidemic state produce identical outputs, and the model has no access to the broader network context. As a consequence, the model can only learn a ranking over epidemic states. In our experiments, it learns to assign higher source probability to recovered nodes than to infectious ones.

MLP$_{\text{snapshot}}$, by contrast, first flattens the full node feature matrix before passing it through the network (Figure~\ref{fig:mlps}, bottom). This allows the model to implicitly capture structural information about the snapshot as a whole, rather than treating each node in isolation.

Both variants share the same hyperparameters as the GNN architectures: ReLU activation, dropout (rate 0.1), 5~hidden layers, and 128~hidden units. MLP$_{\text{snapshot}}$ additionally uses batch normalization and a residual connection following~\cite{Shah:2020}, ensuring a fair comparison with the top GCN architecture.

\section{Further Results}
This Appendix contains further results that were not included in the main text.

\subsection{Ablation Study}
\begin{figure}[ht]
  \centering
  \includegraphics[width=0.95\linewidth]{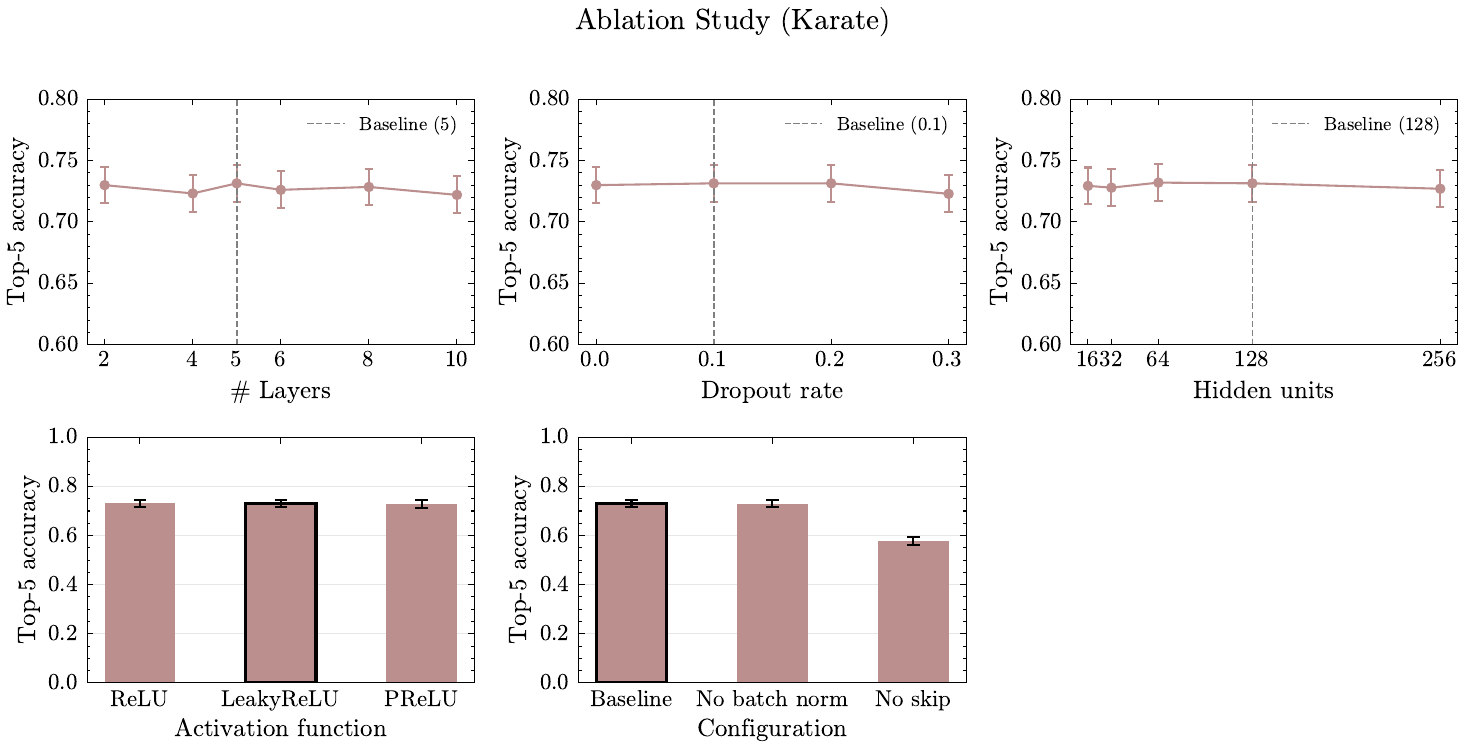}
  \caption{Ablation study results for the Karate network. All plots show top-5 accuracy evaluated on a test set of 100 simulated outbreaks per node. Error bars represent 95\% confidence intervals. In the top row, the baseline configuration is indicated by a vertical line; in the bottom row, it is marked by a dark border around the corresponding bar.}
  \label{fig:karate-ablation}
\end{figure}

\begin{figure}[ht]
  \centering
  \includegraphics[width=0.95\linewidth]{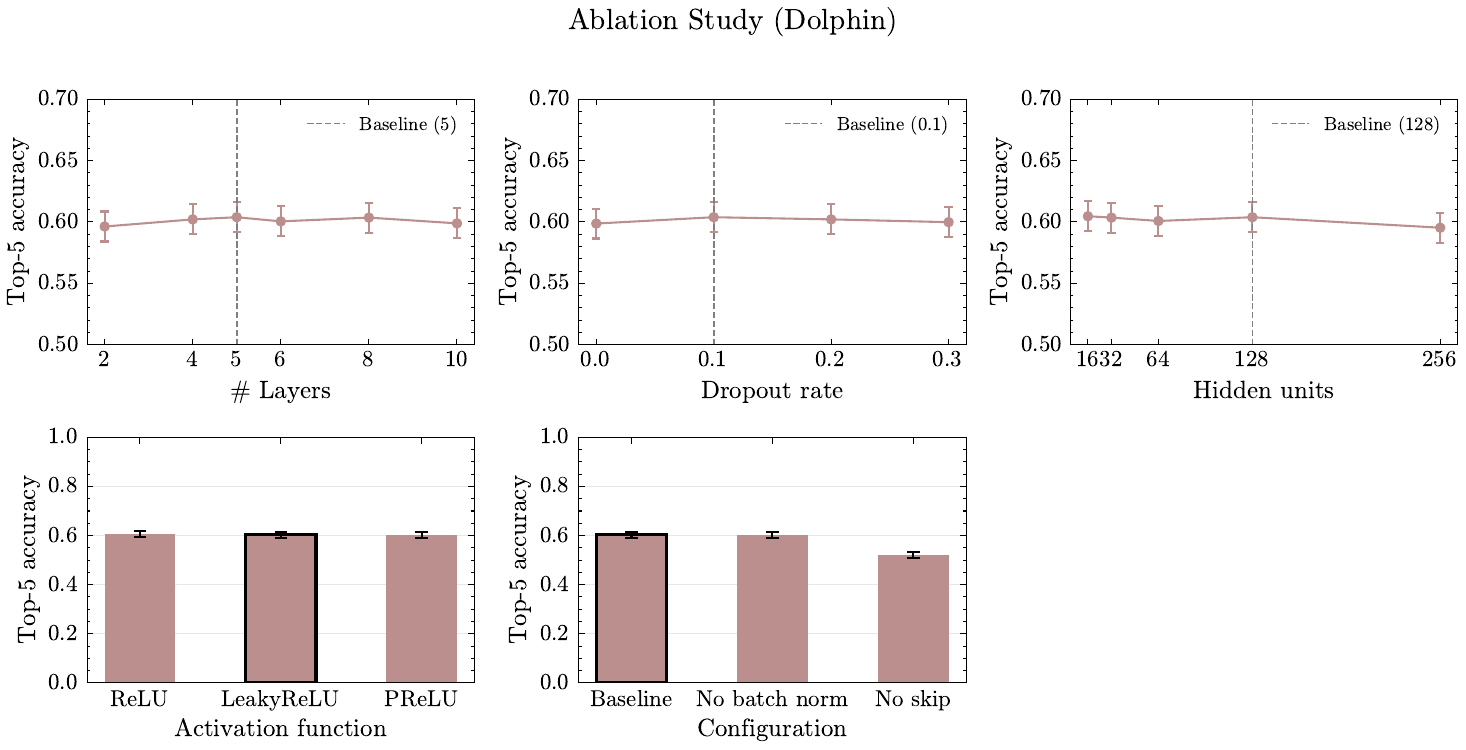}
  \caption{Ablation study results for the Dolphin network. See Figure~\ref{fig:karate-ablation} for a description of the plot elements.}
  \label{fig:dolphin-ablation}
\end{figure}

\begin{figure}[ht]
  \centering
  \includegraphics[width=0.95\linewidth]{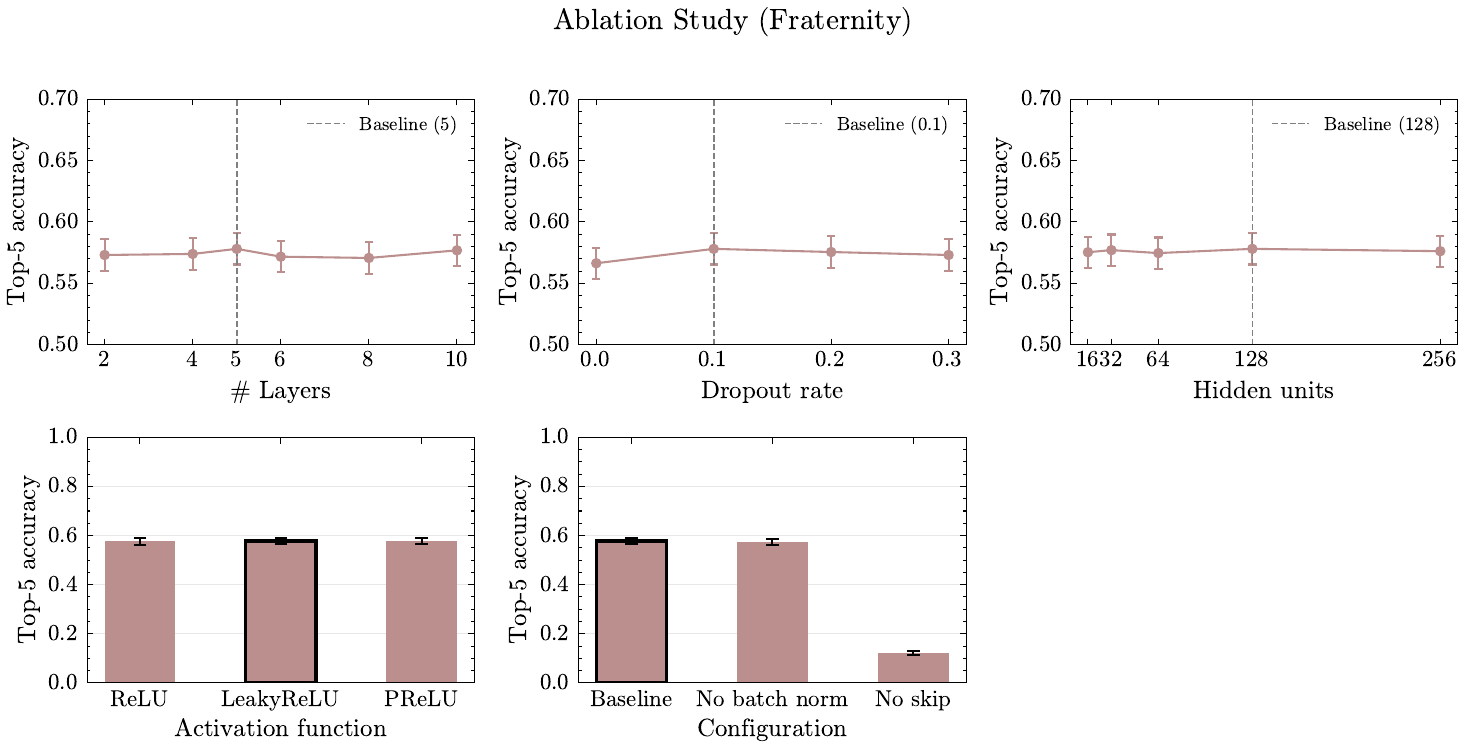}
  \caption{Ablation study results for the Fraternity network. See Figure~\ref{fig:karate-ablation} for a description of the plot elements.}
  \label{fig:fraternity-ablation}
\end{figure}

\begin{figure}[ht]
  \centering
  \includegraphics[width=0.95\linewidth]{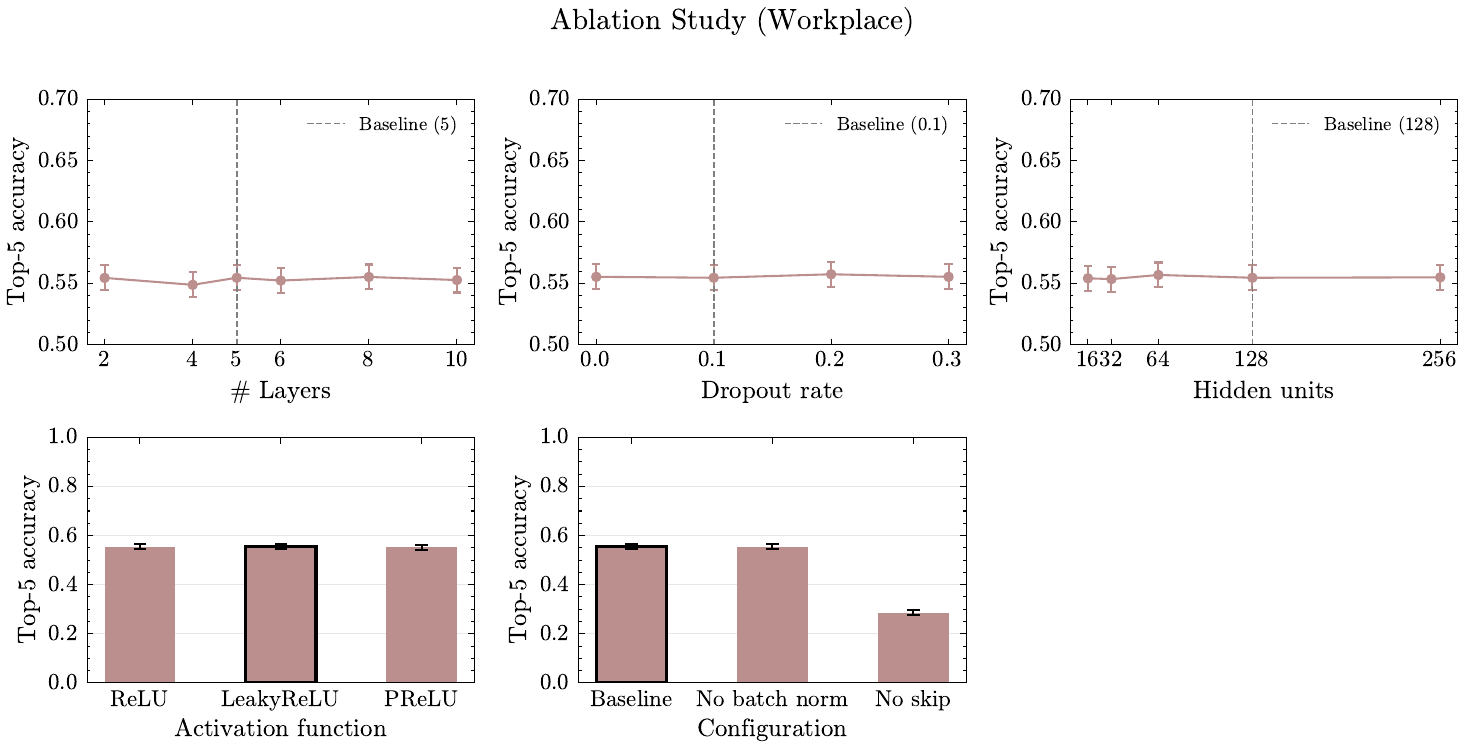}
  \caption{Ablation study results for the Workplace network. See Figure~\ref{fig:karate-ablation} for a description of the plot elements.}
  \label{fig:workplace-ablation}
\end{figure}

\begin{figure}[ht]
  \centering
  \includegraphics[width=0.95\linewidth]{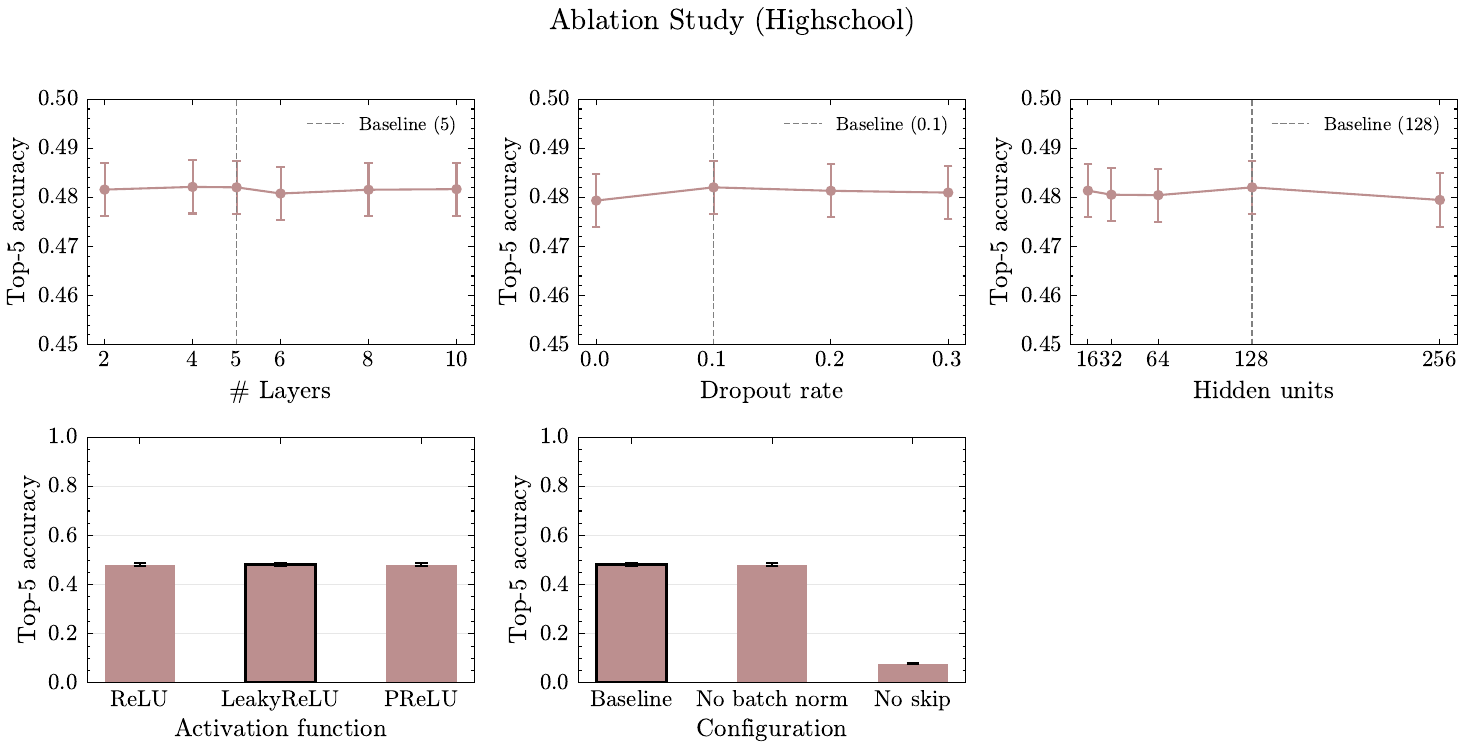}
  \caption{Ablation study results for the Highschool network. See Figure~\ref{fig:karate-ablation} for a description of the plot elements.}
  \label{fig:highschool-ablation}
\end{figure}

This subsection contains the results of the ablation study for the Karate network (Figure~\ref{fig:karate-ablation}), the Dolphin network (Figure~\ref{fig:dolphin-ablation}), the Fraternity network (Figure~\ref{fig:fraternity-ablation}), the Workplace network (Figure~\ref{fig:workplace-ablation}), and the Highschool network (Figure~\ref{fig:highschool-ablation}).

\subsection{Learning Curves}
\begin{figure}[ht]
    \centering
    \begin{subfigure}[b]{0.32\textwidth}
         \centering
         \includegraphics[width=\textwidth]{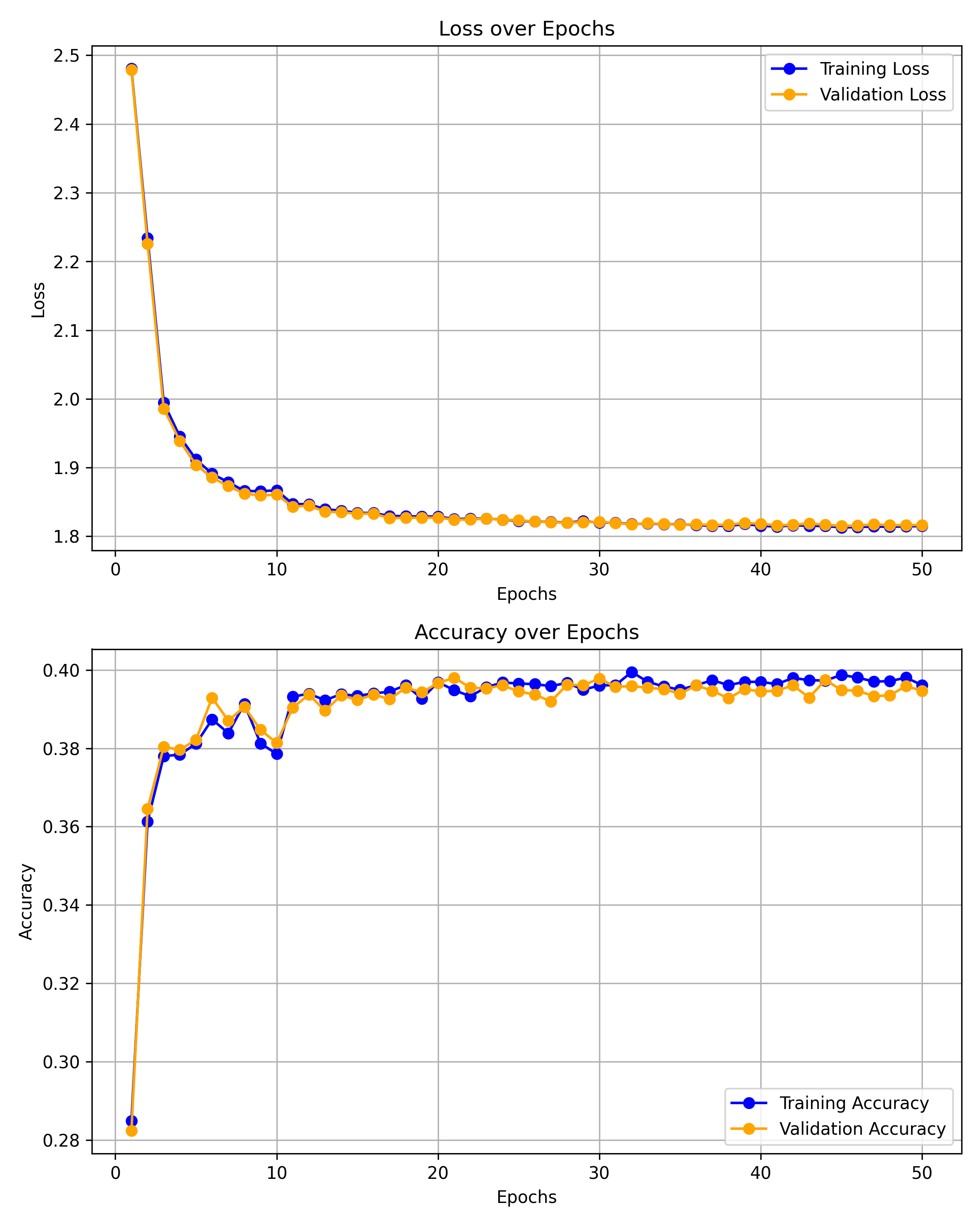}
         \caption{Karate}
         \label{fig:learn-karate}
     \end{subfigure}
     \hfill
     \begin{subfigure}[b]{0.32\textwidth}
         \centering
         \includegraphics[width=\textwidth]{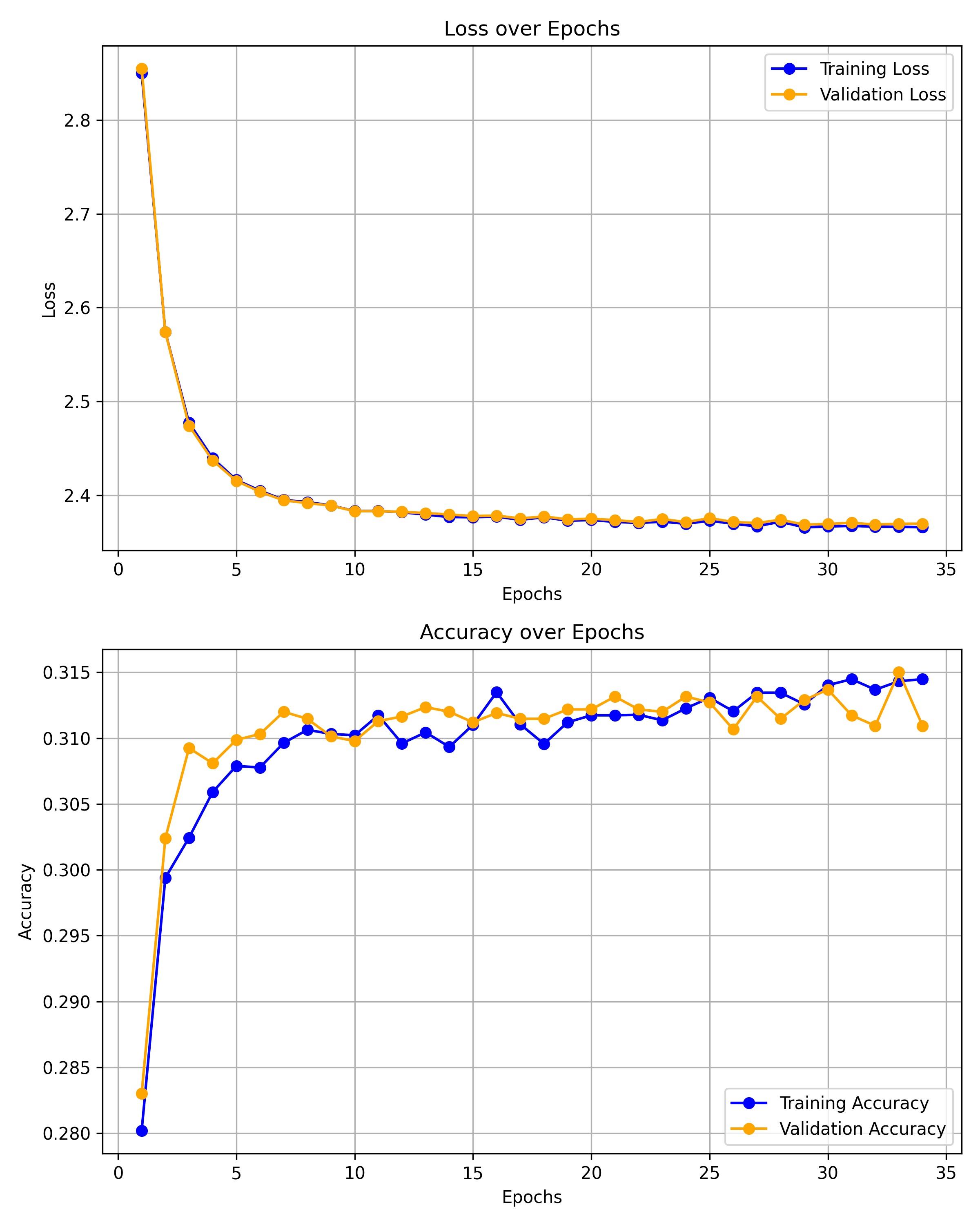}
         \caption{Iceland}
         \label{fig:learn-iceland}
     \end{subfigure}
     \hfill
     \begin{subfigure}[b]{0.32\textwidth}
         \centering
         \includegraphics[width=\textwidth]{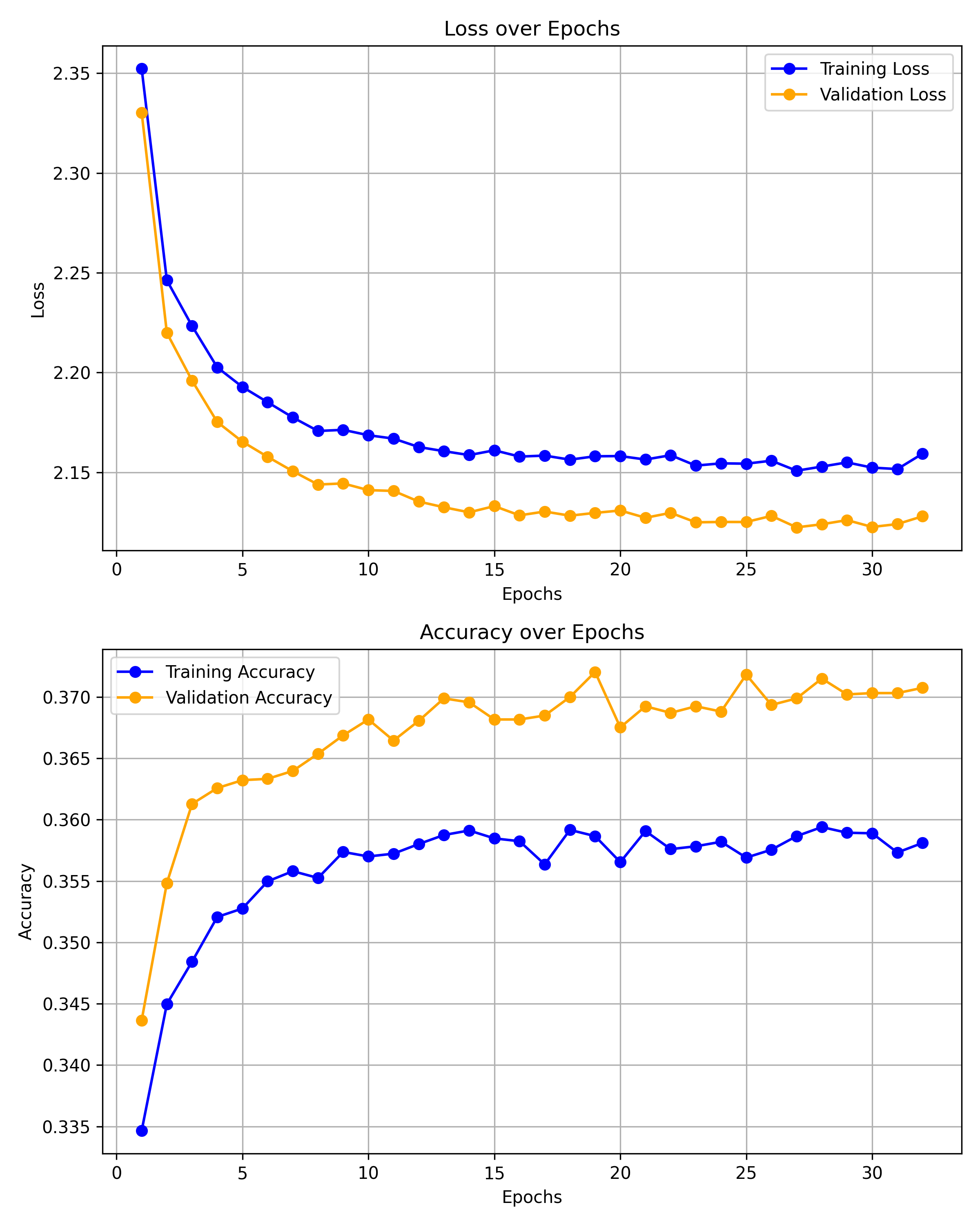}
         \caption{Dolphin}
         \label{fig:learn-dolphin}
     \end{subfigure}
     \hfill
     \begin{subfigure}[b]{0.32\textwidth}
         \centering
         \includegraphics[width=\textwidth]{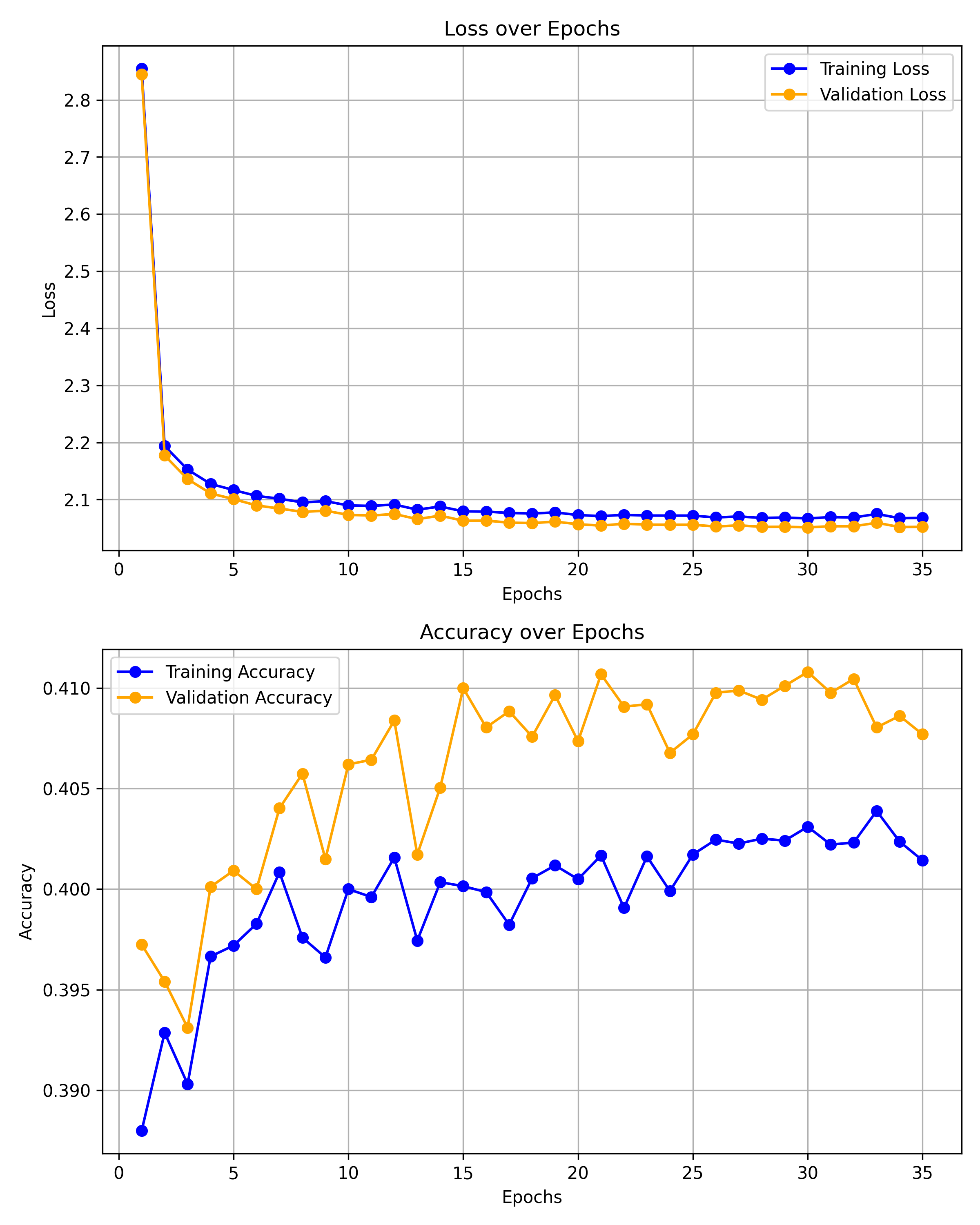}
         \caption{Fraternity}
         \label{fig:learn-fraternity}
     \end{subfigure}
     \hfill
     \begin{subfigure}[b]{0.32\textwidth}
         \centering
         \includegraphics[width=\textwidth]{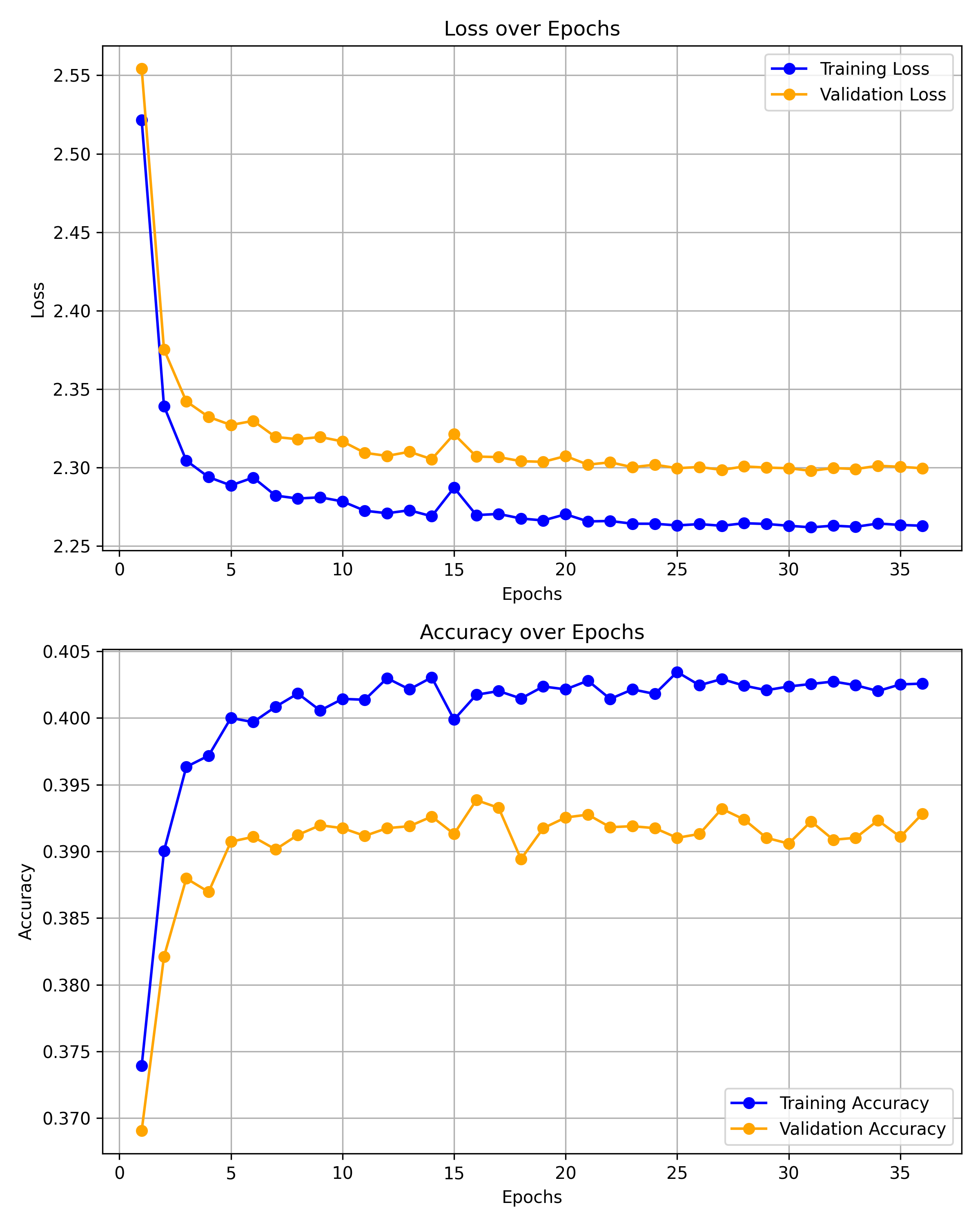}
         \caption{Workplace}
         \label{fig:learn-workplace}
     \end{subfigure}
     \hfill
     \begin{subfigure}[b]{0.32\textwidth}
         \centering
         \includegraphics[width=\textwidth]{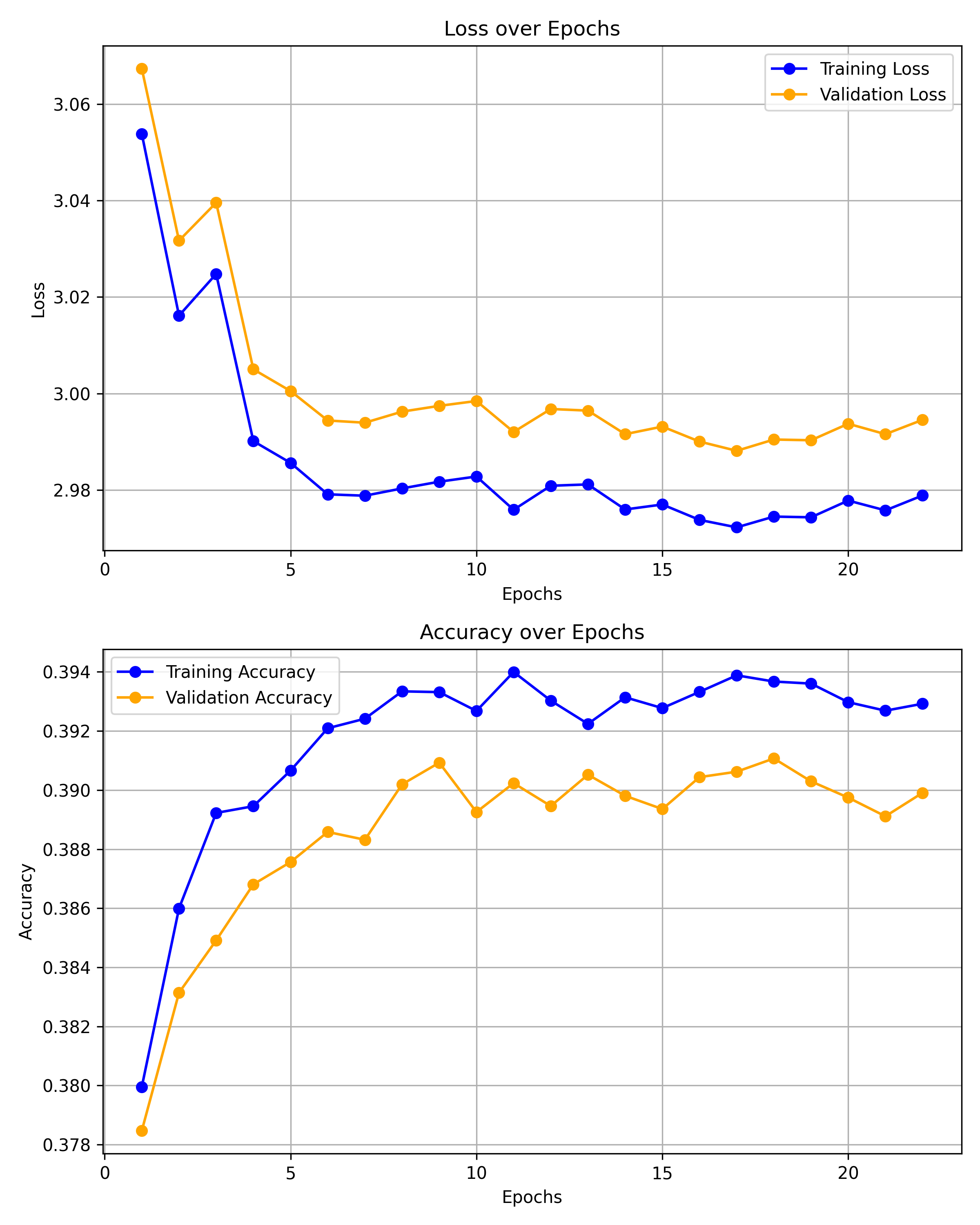}
         \caption{High School (2013)}
         \label{fig:learn-highschool2013}
     \end{subfigure}
     \hfill
    \caption{Learning curves for the six empirical networks. For each network, the top panel reports the training and validation loss across epochs, and the bottom panel shows the corresponding training and validation top-1 accuracy. The curves shown correspond to a representative run of the GCN architecture by Shah et al.~\cite{Shah:2020}, using 16 hidden channels.}
    \label{fig:learning-curves}
\end{figure}

This subsection shows the learning curves (Figure~\ref{fig:learning-curves}) for the six networks used in the main text.

\subsection{Powergrid Network}\label{sec:powergrid}
Finally, we present results on a larger, relatively sparse network: the Powergrid network, used for example in~\cite{Lokhov:2014}. The network consists of $N=4{,}941$ nodes and $E=6{,}594$ edges, with an average degree of 2.67 and a diameter of 46.

We generated 50 simulations per node, using epidemic parameters yielding $R_0 \approx 2$ and approximately 40\% of nodes infected at the time of inference. The number of simulations per node was limited to 50 to avoid exhausting the available GPU memory. The average time per simulation was 1.77e-04 seconds.

We trained one GCN~\cite{Shah:2020} instance with 16 hidden units and compared it against the random baseline. The results are shown in Table~\ref{tab:powergrid}. Source inference is considerably harder on this network, where on average nearly 2{,}000 nodes are infected at the time of inference, leaving relatively little information to discriminate among candidate sources. Nevertheless, the GCN markedly outperforms the random baseline on all metrics. In particular, the GCN roughly halves the error distance relative to the random baseline (5.89 vs.\ 11.85), and improves the top-5 accuracy by approximately 66\% (24.97\% vs.\ 15.06\%), demonstrating that the GCN extracts meaningful patterns even under these difficult conditions.

\begin{table}[ht]
\small
\centering
\caption{Performance of the random baseline and the GCN by Shah et al.~\cite{Shah:2020} for the Powergrid network. Test sets are generated by simulating 10 outbreaks from each node in the network. The 95\% confidence intervals shown in parentheses reflect the uncertainty for one run.}
\begin{tabular}{l|l|rrrrr}
\toprule
\textbf{Dataset} & \textbf{Model} & \textbf{Top-5 accuracy} & \textbf{Error dist.} & \textbf{Rec. rank} & \textbf{|90\% CSS|} & \textbf{Resistance} \\
\midrule
Powergrid & Random & 15.06\% ($\pm 0.32\%$) & 11.8526 ($\pm 0.0731$) & 0.1503 ($\pm 0.0031$) & - & -\\
& \cellcolor[HTML]{BC8F8F}GCN~\cite{Shah:2020} & \cellcolor[HTML]{BC8F8F}24.97\% ($\pm 0.38\%$) & \cellcolor[HTML]{BC8F8F}5.8867 ($\pm 0.0425$) & \cellcolor[HTML]{BC8F8F}0.2108 ($\pm 0.0032$) & \cellcolor[HTML]{BC8F8F}507.0269 ($\pm 3.7383$) & \cellcolor[HTML]{BC8F8F}1.1594 ($\pm 0.0077$) \\
\bottomrule
\end{tabular}
\label{tab:powergrid}
\end{table}

\begin{figure}[ht]
  \centering
  \includegraphics[width=0.6\linewidth]{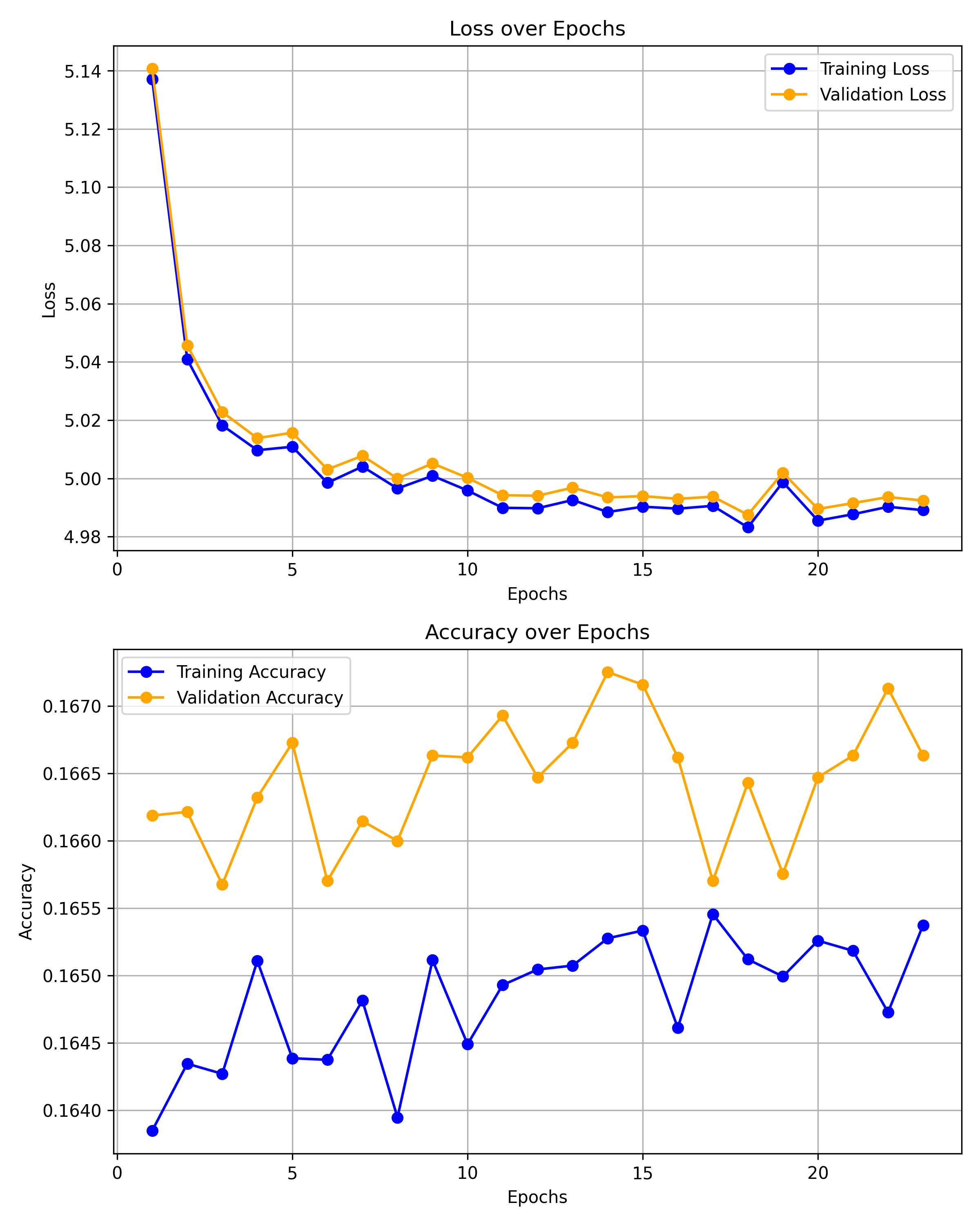}
  \caption{Learning curves for the Powergrid network. The top panel reports the training and validation loss across epochs, and the bottom panel shows the corresponding training and validation top-1 accuracy.}
  \label{fig:powergrid}
\end{figure}

Figure~\ref{fig:powergrid} shows the learning curves for the Powergrid network. The loss curves indicate a stable and relatively smooth training process. The average per-batch training time was 0.0379 seconds and the average inference time per instance was 0.0305 seconds. Comparing these to the Highschool network ($N=327$), the training time increases by a factor of roughly 1.9 and inference by a factor of roughly 7.4, which is broadly consistent with the theoretically predicted scaling with $E$ and $N$

\end{document}